\documentclass[showpacs,showkeys,preprintnumbers,twocolumn,eqsecnum,prd]{revtex4-1}  %
\usepackage{graphicx}
\usepackage{latexsym}
\usepackage{amsmath}
\usepackage{amssymb}
\usepackage{bm}
\usepackage[utf8]{inputenc}

\def\beq{\begin{equation}}
\def\eeq{\end{equation}}
\def\HPhi{\breve \Phi}

\def\pmb#1{\setbox0=\hbox{$#1$}%
  \kern-.025em\copy0\kern-\wd0
  \kern.05em\copy0\kern-\wd0
  \kern-.025em\raise.0433em\box0}

\allowdisplaybreaks

\begin{document}

\title{Gravitational radiation reaction along general orbits \\ in the effective one-body formalism}

\author{Donato Bini}
\affiliation{
Istituto per le Applicazioni del Calcolo ``M. Picone,'' CNR, I-00185 Rome,
Italy}

\author{Thibault Damour}
  \affiliation{Institut des Hautes \'{E}tudes Scientifiques, F-91440
    Bures-sur-Yvette, France}

\date{\today}

\begin{abstract}
We derive the gravitational radiation-reaction   force modifying the Effective One Body (EOB) description of the conservative dynamics of binary systems. Our result is applicable to general orbits (elliptic or hyperbolic) and keeps terms of fractional second post-Newtonian order  (but does not include tail effects). Our derivation of radiation-reaction 
is based on a new way of requiring energy and angular momentum balance. We give several applications of our results, notably the value of the (minimal) \lq\lq Schott" contribution to the energy, the radial component of the radiation-reaction force, and the radiative contribution to the angle of scattering during hyperbolic encounters. We present also new results about the conservative relativistic dynamics of hyperbolic motions.
\end{abstract}

\pacs{04.30.-w, 04.25.Nx}

\maketitle

\section{Introduction}\label{sec1}

The Effective One Body (EOB) formalism \cite{Buonanno:1998gg,Buonanno:2000ef,Damour:2000we,Damour:2001tu} is an approach to the relativistic dynamics of gravitationally interacting binary systems which was originally proposed as a way to extend the validity of the usual post-Newtonian (PN) 
formalism beyond the slow-motion $(v^2/c^2\ll 1)$ and weak-field $(GM/(c^2r)\ll 1)$ regime. The EOB approach is made of three, basic building blocks:

\begin{enumerate}
  \item a description of the {\it conservative} (Hamiltonian) part of the dynamics of two compact bodies;
  \item an expression for the {\it radiation-reaction} force ${\pmb {\mathcal F}}$ which  must be added to the conservative, Hamiltonian equations of motion;
  \item a description of the asymptotic {\it gravitational waveform} emitted by the binary system.
\end{enumerate}

The building  block 1, i.e., the EOB Hamiltonian, has been analytically computed with an increasing accuracy in a sequence of papers, both for non-spinning black holes \cite{Buonanno:1998gg,Damour:2000we},  for spinning black holes \cite{Damour:2001tu,Damour:2008qf,Barausse:2009xi,Nagar:2011fx,Taracchini:2012ig}
and for systems involving tidally-deformed bodies \cite{Damour:2009wj,Bini:2012gu}. In addition, the comparison between the EOB dynamics and numerical simulations of binary systems has allowed one to improve the knowledge of some of the functions entering the EOB Hamiltonian (see Ref. \cite{Damour:2009ic} for a review). More recently, results from gravitational self-force theory \cite{Barack:2009ux} have also allowed one to learn new information about the EOB formalism (See Ref. \cite{Akcay:2012ea} for recent progress and references).
The description of the second building block, i.e. the radiation-reaction force ${\pmb {\mathcal F}}$ has also improved over the years, both through the conception of new resummation methods \cite{Damour:2008gu} and from the comparison with numerical simulations (both in the comparable-mass case \cite{Damour:2009kr,Pan:2011gk}, and in the extreme-mass-ratio case \cite{Nagar:2006xv,Yunes:2009ef,Bernuzzi:2010xj}). The same remarks apply to the third building block, i.e., the gravitational waveform.

While the EOB Hamiltonian is able to describe the conservative dynamics of {\it general} binary orbits (quasi-circular, elliptic-like or hyperbolic-like), the currently existing accurate implementations of the radiation-reaction force and of the emitted waveform are limited to the case of quasi-circular, inspiralling orbits. The main reason behind this limitation is that the EOB program was originally motivated as a tool for computing accurate waveforms from the type of circularized binary systems that are likely sources for ground-based interferometric gravitational wave detectors. 
However, the progress in numerical relativity simulations has opened the possibility of numerically exploring the dynamics of binary systems in more exotic configurations. For instance, Refs. \cite{Sperhake:2008ga,Shibata:2008rq} have considered high-velocity encounters of black holes and other bodies, and Ref. \cite{Sperhake:2007gu} has considered eccentric orbits of binary black holes.
We anticipate that more simulations of general orbits will become routinely possible in the near future. See Ref.  \cite{Gold:2012tk} for a recent example, and more references.

This perspective motivates the main aim of the present work, namely, to provide an expression of the radiation-reaction force ${\pmb {\mathcal F}}$ along general orbits (elliptic or hyperbolic) within the EOB formalism. [We leave to future work a corresponding generalization of the EOB gravitational waveform.]

Gravitational radiation-reaction, notably in binary systems, has a  long history. Let us only recall that three general different approaches have been used. The first approach derives the full equations of motion of matter 
(including both conservative and radiative effects)
from a direct integration of the retarded field generated by the 
source. Because of its difficulty, this approach has been implemented essentially only up to the next-to-leading order in ${\pmb {\mathcal F}}$, i.e., at the {\it fractional} 1PN accuracy \cite{chandra-esp,Damour:1981bh,Damour:1983tz,Schaefer:1986rd,Jaranowski:1996nv}.

A second approach focuses on the radiation-reaction piece in the 
equations of motion and derives it
by using a matching between between the near-zone field and the 
wave-zone field.  This approach has been also implemented only up to the next-to-leading order in ${\pmb {\mathcal F}}$ \cite{Burke,Thorne,Blanchet:1984wm,Blanchet:1993ng,Iyer:1993xi,Iyer:1995rn,Blanchet:1996vx, Pati:2002ux,Konigsdorffer:2003ue,Nissanke:2004er}, 
with some vistas on the effect of tails \cite{Blanchet:1987wq}.

Finally, a third approach is based on requiring a {\it balance} between the losses of mechanical energy and angular momentum radiated by gravitational waves at infinity. This \lq\lq third" balance approach has been particularly developed by Iyer and Will and their collaborators \cite{Iyer:1993xi,Iyer:1995rn,Gopakumar:1997ng} 
and has been implemented to a higher PN accuracy than the other approaches, namely the next-to-next-to-leading order in ${\pmb {\mathcal F}}$, i.e., the {\it fractional} 2PN accuracy \cite{Gopakumar:1997ng}.

Note, however, that Ref. \cite{Gopakumar:1997ng} does not include the effect of tails. We shall, similarly, postpone the inclusion of tails (entering at the fractional 1.5PN, $v^3/c^3$, level) to future work. We note that Ref. \cite{Blanchet:1987wq} has shown that the tail contribution to ${\pmb {\mathcal F}}$ satisfies the balance requirement.

In view of the technical efficiency of the balance approach (and of the direct proof by several authors of the consistency between this approach and other ones \cite{Iyer:1993xi,Iyer:1995rn}), we shall also base our work on this approach. However, instead of attempting to \lq\lq translate" the radiation-reaction force ${\pmb {\mathcal F}}$ derived in Refs.
\cite{Iyer:1993xi,Iyer:1995rn,Gopakumar:1997ng} (which was derived in harmonic coordinates, and was expressed in terms of quasi-Newtonian equations of motion) into the EOB formalism (which uses different coordinates, and Hamiltonian equations of motion), we found  more efficient to develop a new way of using the balance approach. We shall explain in detail below our new way of implementing the balance approach.

Let us only say here that it is based on three essential ingredients: (i) we start from the 2PN-accurate expressions of the fluxes of energy and angular momentum,
$\Phi_E$ and $\Phi_J$, that have been derived in the PN literature \cite{Blanchet:1995ez,Blanchet:1995fg,Blanchet:1996pi,Will:1996zj,Gopakumar:1997bs} (see references \cite{Blanchet:2001ax,Blanchet:2004ek,Arun:2007rg,Arun:2007sg,Arun:2009mc} for recent higher PN accuracy results). 
These fluxes are expressed in terms of three scalars ${\mathbf v}_h^2$, $\dot r_h^2$ and $GM/r_h$, where ${\mathbf x}_h$ and ${\mathbf v}_h$ denote harmonic coordinate and velocities (of the relative orbit). Then, (ii) we derive the transformation connecting 
the three scalars ${\mathbf v}_h^2$, $\dot r_h^2$ and $GM/r_h$ to the three scalars that are natural within the EOB formalism, namely ${\mathbf p}_{e}{}^2$, 
$p_{e\, r}^2$ and $GM/r_e$, where ${\mathbf x}_e$ and ${\mathbf p}_e$ denote EOB coordinates and momenta. Finally, (iii) we introduce a new way of using the two EOB-expressed fluxes $\Phi_E({\mathbf x}_e,{\mathbf p}_e)$ and $\Phi_J({\mathbf x}_e,{\mathbf p}_e)$ to derive the two independent components of the radiation-reaction force
${\mathcal F}_r^{\rm (eob)}({\mathbf x}_e,{\mathbf p}_e)={\pmb {\mathcal F}}^{\rm (eob)}\cdot {\mathbf n}_e$, ${\mathcal F}_\phi^{\rm (eob)}({\mathbf x}_e,{\mathbf p}_e)=({\mathbf x}_e\times {\pmb {\mathcal F}}^{\rm (eob)})\cdot {\mathbf e}_z$.

The structure of this paper is as follows. We present in Sec. II our new way of implementing the balance approach. Then, in Sec. III, after presenting a brief review of the EOB formalism, we apply our method to the 2PN-accurate EOB-variables forms of $\Phi_E$ and $\Phi_J$, and derive explicit expressions for 
${\mathcal F}_r^{\rm (eob)}$ and ${\mathcal F}_\phi^{\rm (eob)}$. 
We also obtain the explicit expressions of the associated \lq\lq Schott" energy contribution.
Sec. IV discusses the gauge freedom in ${\pmb {\mathcal F}}$ and explains how it is related to the freedom in defining the Schott contributions to the energy and angular momentum. Then, Sec. V gives some applications of our results, and discusses notably the scattering angle during hyperbolic encounters, and its modification by radiation-reaction effects. We summarize our main results in Sec. VI, and discuss future directions. Finally, to relieve the tedium we have relegated several explicit technical details to various appendices.

\section{A new approach to radiation-reaction}

Here, we introduce a new approach to the computation of radiation reaction by the balance method. Let us consider the effect of adding a radiation-reaction force, say ${\mathcal F}_i$, to the Hamiltonian form of the equations describing the relative motion of a binary system (with masses 
$m_1$ and $m_2$)

\beq
\label{eqmot}
\dot x^i=\frac{\partial {\mathcal H}({\mathbf x},{\mathbf p})}{\partial p_i}\,,\qquad 
\dot p_i=-\frac{\partial {\mathcal H}({\mathbf x},{\mathbf p})}{\partial x^i}+{\mathcal F}_i\,.
\eeq
Here ${\mathcal H}({\mathbf x},{\mathbf p})$ denotes the Hamiltonian and a dot denotes differentiation with respect to time. When considering the motion within the orbital plane, we can take as coordinate and momenta
$x^i=(r,\phi)$ and $p_i=(p_r,p_\phi)$.
Correspondingly, the radiation-reaction will have two independent components: ${\mathcal F}_r$ and ${\mathcal F}_\phi$. 

Let us see how one can determine the two force components ${\mathcal F}_r$ and ${\mathcal F}_\phi$ by writing balance equations for the energy and the angular momentum of the {\it binary system}, namely
\begin{eqnarray}
E_{\rm (system)}(t)&=&{\mathcal H}({\mathbf x}(t),{\mathbf p}(t))-(m_1+m_2)c^2\nonumber\\
J_{\rm (system)}(t)&=&p_\phi(t)\,.
\end{eqnarray}

On the one hand, the equations of motion (\ref{eqmot}) yield the following time changes for $E_{\rm (system)}(t)$ and $J_{\rm (system)}(t)$
\begin{eqnarray}
\frac{d E_{\rm (system)}(t)}{dt}&=&\frac{d {\mathcal H}}{dt}=\dot x^i \frac{\partial {\mathcal H}}{\partial x^i}+\dot p_i \frac{\partial {\mathcal H}}{\partial p_i}=\dot x^i{\mathcal F}_i\nonumber\\
\frac{d J_{\rm (system)}(t)}{dt}&=&\frac{d p_\phi}{dt}=-\frac{\partial {\mathcal H}}{\partial \phi}+{\mathcal F}_\phi\,.
\end{eqnarray}
The explicit form of these two equations read (when using the fact that ${\mathcal H}$ does not depend on $\phi$)
\begin{eqnarray}
\label{n1}
\dot E_{\rm (system)}(t)&=&\dot r{\mathcal F}_r+\dot \phi{\mathcal F}_\phi
\end{eqnarray}
and
\begin{eqnarray}
\label{n2}
\dot J_{\rm (system)}(t)&=&\frac{d p_\phi}{dt}={\mathcal F}_\phi\,.
\end{eqnarray}
It will also be useful to consider the following combination of these two equations
\beq
\label{n3}
\dot E_{\rm (system)}-\dot \phi \dot J_{\rm (system)}=\dot r{\mathcal F}_r\,.
\eeq
Formally speaking, Eqs. (\ref{n1}) and (\ref{n2}) provide two equations relating the two unknowns ${\mathcal F}_r$ and ${\mathcal F}_\phi$ to the losses of energy and angular momentum.

On the other hand, we require that there is a balance between the energy and angular momentum losses of the {\it system}, and the corresponding energy and angular momentum {\it fluxes} (in the form of gravitational waves) at infinity, say $\Phi_E$ and $\Phi_J$. As was pointed out by Schott long ago \cite{Schott}, one cannot, however, simply equate 
$\dot E_{\rm (system)}$ and  $\dot J_{\rm (system)}$ to, respectively, $-\Phi_E$ and $-\Phi_J$. One must allow for the existence of {
\it Schott terms} that represent additional contributions to the energy and angular momentum of the system, due to its interaction with the radiation field, say
$E_{\rm (schott)}(t)=E_{\rm (schott)}({\mathbf x}(t),{\mathbf p}(t))$ and $J_{\rm (schott)}(t)=J_{\rm (schott)}({\mathbf x}(t),{\mathbf p}(t))$.
The correspondingly modified balance equations then read
\begin{eqnarray}
\label{n4}
&& \dot E_{\rm (system)}+\dot E_{\rm (schott)}+\Phi_E=0\nonumber\\
&& \dot J_{\rm (system)}+\dot J_{\rm (schott)}+\Phi_J=0\,.
\end{eqnarray}
Inserting the identities (\ref{n1}), (\ref{n2}) into (\ref{n4}) leads to the following two conditions on the two components of the radiation-reaction force
\begin{eqnarray}
\label{n5}
&& \dot r{\mathcal F}_r+\dot \phi{\mathcal F}_\phi+\dot E_{\rm (schott)}+\Phi_E=0\nonumber\\
&& {\mathcal F}_\phi+\dot J_{\rm (schott)}+\Phi_J=0\,.
\end{eqnarray}

Up to now, all the equations we have written down are equivalent to the standard \lq\lq balance approach" to radiation-reaction, as used, in particular, by Iyer, Will and collaborators \cite{Iyer:1993xi,Iyer:1995rn,Gopakumar:1997ng,Gopakumar:1997bs}, except for the fact that we are working within a Hamiltonian framework.
Let us now explain the new, simplifying features of our approach.

The first simplifying feature is to note that it is always possible to impose the condition that the Schott-contribution to the angular momentum {\it vanishes}:
\beq
\label{n6}
J_{\rm (schott)}({\mathbf x}(t),{\mathbf p}(t))=0\,.
\eeq
The proof that this is possible is simply that, after imposing Eq. (\ref{n6}), we shall be able to find a solution to the general balance equations (\ref{n5}).
Indeed, after making the assumption (\ref{n6}), we can use the second Eq. (\ref{n5}) to determine the instantaneous value of the $\phi$-component of the radiation-reaction force, in terms of the corresponding instantaneous $J$-flux:
\beq
\label{n7}
{\mathcal F}_\phi =-\Phi_J({\mathbf x}(t),{\mathbf p}(t))\,.
\eeq

Let us note in passing that the result (\ref{n7}) for ${\mathcal F}_\phi$ is standardly used in the current implementations of the EOB equations of motion \cite{Damour:2009ic}. Then, by inserting the result (\ref{n7}) into the first equation (\ref{n5}), we get an equation involving only ${\mathcal F}_r$ and $\dot E_{\rm (schott)}$, namely
\beq
\label{n8}
\dot r{\mathcal F}_r+\dot E_{\rm (schott)}=-\Phi_{EJ}\,,
\eeq
where we introduced the notation
\beq
\label{n9} 
\Phi_{EJ}({\mathbf x},{\mathbf p})=\Phi_{E}({\mathbf x},{\mathbf p})-\dot \phi({\mathbf x},{\mathbf p}) \Phi_{J}({\mathbf x},{\mathbf p})\,.
\eeq
As we shall discuss in detail in the next section, we assume here that we have in hands explicit expressions for $\Phi_E$, $\Phi_J$ (as well as for the \lq\lq combined flux" $\Phi_{EJ}$) as functions of the instantaneous dynamical state of the system. Within a Hamiltonian framework it means 
$\Phi_E=\Phi_E ({\mathbf x},{\mathbf p})$, $\Phi_J=\Phi_J ({\mathbf x},{\mathbf p})$ and $\Phi_{EJ}=\Phi_{EJ} ({\mathbf x},{\mathbf p})$. [Note that, by Hamilton's equations, the instantaneous orbital frequency $\dot \phi$ entering $\Phi_{EJ}$ is a function of position and momenta, given by 
$\dot \phi({\mathbf x},{\mathbf p})=\partial {\mathcal H}({\mathbf x},{\mathbf p})/\partial p_\phi$. As we shall further discuss below, contrary to $\Phi_E$ and  $\Phi_J$, $\dot \phi$ is not a gauge invariant quantity; we shall only consider its explicit expression  $\dot \phi({\mathbf x},{\mathbf p})$
in EOB coordinates.]

While Eq. (\ref{n7}) provides an explicit expression for ${\mathcal F}_\phi$ in terms of the instantaneous state of the system, our remaining problem is to show how Eq. (\ref{n8}) can be used to determine both ${\mathcal F}_r({\mathbf x},{\mathbf p})$ and $E_{\rm (schott)}({\mathbf x},{\mathbf p})$.
Let us now explain how this can be done.

The basic idea is that the specific combination $\Phi_{EJ}$ has the property of vanishing along circular motions. Indeed, it is well known that (because of the monochromatic nature of the emitted radiation) one has $\Phi_E=\Omega \Phi_J$ along a circular motion with orbital frequency $\Omega$. As a consequence, when considering general, noncircular motions, $\Phi_{EJ}$ will necessarily be given by an expression which can be written as a combination of the {\it two independent} quantities that vanish along circular motions, namely
\beq
Z_1({\mathbf x},{\mathbf p})=p_r^2
\eeq
and
\beq
\label{n10}
Z_2({\mathbf x},{\mathbf p})=r \frac{\partial {\mathcal H}}{\partial r}=- r \dot p_r\,,
\eeq
where the factor $r$ in $Z_2$ is introduced for later convenience. [See Sec. IIIC where we will work with rescaled versions of $Z_1$ and $Z_2$ that have the same dimensions.]

Here, we are availing ourselves of several simplifications that are allowed at the PN accuracy at which we shall be working. First, as we shall explicitly check, the combination $\Phi_{EJ}({\mathbf x},{\mathbf p})$ is invariant under time reversal, and can therefore be expressed as a function of $p_r^2 \sim \dot r^2$, rather than simply of $p_r \sim \dot r$. Second, modulo terms of 5PN order (i.e., $O(1/c^{10})$) one can neglect the ${\mathcal F}_r$ contribution to the link between $\dot p_r$ and $-\partial {\mathcal H}/\partial r$.

We can then write
\begin{eqnarray}
\label{n11}
\Phi_{EJ}({\mathbf x},{\mathbf p})&=&\Phi_1({\mathbf x},{\mathbf p})Z_1({\mathbf x},{\mathbf p})+\Phi_2 ({\mathbf x},{\mathbf p}) Z_2 ({\mathbf x},{\mathbf p})\nonumber\\
&=& \Phi_1 p_r^2 -r \Phi_2 \frac{d p_r}{dt}\,,
\end{eqnarray}
where $\Phi_1$ and $\Phi_2$ exist but are not uniquely defined.
For instance, we can move a term $\propto Z_2$ in $\Phi_1$ to $\Phi_2$, and reciprocally a term $\propto Z_1$ in $\Phi_2$ to $\Phi_1$. We shall discuss below the effect of these ambiguities in the definition of $\Phi_1$ and $\Phi_2$.

Operating by parts on the second expression (\ref{n11}) (which involves $\dot p_r$), we can then write
\begin{eqnarray}
\label{n11bis}
\Phi_{EJ}({\mathbf x},{\mathbf p})&=&p_r\left[p_r\Phi_1+\frac{d}{dt}(r\Phi_2)  \right] \nonumber\\
&& -\frac{d}{dt}(r\Phi_2p_r)\,,
\end{eqnarray}
which is a decomposition of $\Phi_{EJ}$ in a part proportional to $p_r$ (and therefore to $\dot r$, in view of $\dot r=\partial {\mathcal H}/\partial p_r$), and a total derivative. But, such a decomposition is precisely the content of the balance requirement (\ref{n8}).

We therefore see that, given any choice of $\Phi_1$ and $\Phi_2$ such that Eq. (\ref{n11}) holds, we can obtain one particular corresponding solution to Eq. (\ref{n8}), namely
\begin{eqnarray}
\label{n12}
{\mathcal F}_r({\mathbf x},{\mathbf p})&=&-\frac{p_r}{\dot r}\left[p_r\Phi_1+\frac{d}{dt}(r\Phi_2)  \right] \nonumber\\
E_{\rm (schott)}({\mathbf x},{\mathbf p})&=& rp_r\Phi_2\,.
\end{eqnarray}

In keeping with our approximations, the time derivative of $r\Phi_2({\mathbf x},{\mathbf p})$ in the first Eq. (\ref{n12}) should be evaluated along the (conservative) Hamiltonian dynamics, so that ${\mathcal F}_r$ can be explicitly expressed in terms of the instantaneous dynamical state of the system.

The results (\ref{n12}), together with Eqs. (\ref{n6}) and (\ref{n7}), give a {\it constructive} algorithm for determining the two components ${\mathcal F}_r$ and 
${\mathcal F}_\phi$ of radiation-reaction, as well as the Schott contributions to energy and angular momentum. [This contrasts with Refs. \cite{Iyer:1993xi,Iyer:1995rn,Gopakumar:1997ng} which had to use the method of undetermined coefficients.] This proves our claim that is indeed possible to define a radiation-reaction force such that the Schott contribution to the angular momentum vanishes. [By contrast,  one can show that it is generally impossible to define ${\mathcal F}_i$ such that  $E_{\rm (schott)}$ vanishes.] We shall discuss later, while implementing our construction, the impact of the non uniqueness in the decomposition (\ref{n11}), as well as a simple, algorithmic way of fixing it.
Let us only note here that, in keeping with the analysis of Iyer and Will \cite{Iyer:1993xi,Iyer:1995rn} and later developments by Gopakumar et al \cite{Gopakumar:1997ng}, all the non uniqueness in the definition of the radiation-reaction ${\pmb {\mathcal F}}$ has the character of a gauge freedom (and is actually related to possible coordinate changes). This also applies to the freedom of setting $J_{\rm (schott)}$ to zero, that we have used here to simplify the search for ${\pmb {\mathcal F}}$.

\section{Radiation reaction force in the EOB formalism}

Let us now apply the method explained in the previous section to the construction of the radiation-reaction force in the EOB formalism. To do that, we need the following items:
\begin{enumerate}
  \item The expressions of the various flux functions $\Phi_E$, $\Phi_J$ and $\Phi_{EJ}$ in terms of the positions and momenta of the EOB formalism; 
  \item An algorithmic way of decomposing $\Phi_{EJ}({\mathbf x},{\mathbf p})$ in the form (\ref{n11}).
\end{enumerate}
Before considering these items, let us recall the structure we shall need of the EOB formalism.

\subsection{EOB formalism: a short review}

At the 2PN accuracy that we shall consider here, the EOB Hamiltonian for the relative dynamics of two masses $m_1$ and $m_2$, is completely described by the following {\it effective metric} 
\begin{eqnarray}
ds^2_{\rm (eob)}&=&-A(r) c^2 dt^2+B(r)dr^2\nonumber\\
&& +r^2(d\theta^2+\sin^2\theta d\phi^2)
\end{eqnarray}
where
\begin{eqnarray}
A(r)&=& 1-2 \left(\frac{GM}{c^2r}\right)+2\nu\left(\frac{GM}{c^2r}\right)^3+...\nonumber\\
B(r)&=& 1+2 \left(\frac{GM}{c^2r}\right)+2(2-3\nu)\left(\frac{GM}{c^2r}\right)^2+...\nonumber\\
A(r)B(r)&\equiv &D(r)=1-6\nu \left(\frac{GM}{c^2r}\right)^2+... \,.
\end{eqnarray}
Our notation is
\beq
M=m_1+m_2\,,\quad \mu=\frac{m_1m_2}{M}\,,\quad \nu=\frac{\mu}{M}\,.  
\eeq
It will often be convenient to work with 
\beq
u\equiv \frac{GM}{r}\,.
\eeq
With an abuse of notation we will then write
\begin{eqnarray}
A(u) &=& 1-2\frac{u}{c^2}+2\nu \frac{u^3}{c^6}+...\nonumber\\
B(u) &=& 1+2 \frac{u}{c^2}+2(2-3\nu)\frac{u^2}{c^4}+...\nonumber\\
D(u)&=& 1-6\nu \frac{u^2}{c^4}+...\,.
\end{eqnarray}
The EOB Hamiltonian ${\mathcal H}_{\rm (eob)}$ is then defined as the following  function of the EOB coordinates $(r, \phi)$ and momenta $(p_r,p_\phi)$ in the plane of the relative trajectory
\begin{eqnarray}
\label{H_EOB}
{\mathcal H}_{\rm (eob)}&=& Mc^2\sqrt{1+2\nu \left( \frac{{\mathcal H}_{\rm (eff)}}{\mu c^2}-1 \right)}\nonumber\\
&\equiv & Mc^2 h\,,
\end{eqnarray}
where
\begin{eqnarray}
\left(\frac{{\mathcal H}_{\rm (eff)}}{\mu c^2}\right)^2&=&A(r)\left[1+\frac{({\mathbf n}_e\cdot {\mathbf p}_e)^2}{\mu^2 c^2 B(r)} +
\frac{({\mathbf n}_e\times {\mathbf p}_e)^2}{\mu^2 c^2}
 \right]\nonumber\\
&=&A(r)\left[1+\frac{\tilde p_r^2}{c^2 B(r)} +
\frac{\tilde p_\phi^2}{c^2r^2 }
 \right]\,,
\end{eqnarray}
that is
\begin{eqnarray}
\left(\frac{{\mathcal H}_{\rm (eff)}}{\mu c^2}\right)^2&=&
A(u)
\left(1+\frac{ A(u)\tilde p_r^2}
{D(u) c^2}+\frac{\tilde p_\phi^2}{c^2r^2}\right)\nonumber\\
&=& A(u)
\left(1+\frac{A(u)\tilde p_r^2}
{D(u)c^2}+\frac{u^2j^2}{c^2}\right)\,,
\end{eqnarray}
and
\beq
\label{smallh}
h=\sqrt{1+2\nu \left( \frac{{\mathcal H}_{\rm (eff)}}{\mu c^2}-1 \right)}\,.
\eeq
Here we have introduced a tilde to denote the result of a rescaling by the reduced mass $\mu$, e.g. $\tilde p=p/\mu$ and
\beq
\tilde E\equiv \frac{ {\mathcal H}_{\rm (eob)}-Mc^2 }{\mu}\,,
\eeq
where we subtract the rest mass contribution to the energy before scaling by $\mu$.
In addition it is convenient to introduce a special notation for some useful rescalings by $GM$, namely 
\beq
j\equiv \frac{\tilde p_\phi}{GM}\equiv \frac{ p_\phi}{\mu GM} \,,\qquad {\mathbf q}\equiv \frac{{\mathbf r}}{GM}\,,\quad
\hat t =\frac{t}{GM}\,.
\eeq
If we denote by $V$ any quantity having the dimension of a velocity, we note that the dimensions of the $GM$-rescaled quantities $u$, $j$, $q$ and $\hat t$ is 
$u \sim V^2$, $j\sim V^{-1}$, $q\equiv |{\mathbf q}|=u^{-1}\sim V^{-2}$ and $\hat t \sim V^{-3}$.
In the following, we shall often find convenient to work with the Hamiltonian pair of variables $q, \tilde p_r$; $\phi, j$. These variables are canonically conjugated with respect to the $\mu$-scaled Hamiltonian  $\tilde {\mathcal H}_{\rm (eob)}={\mathcal H}_{\rm (eob)}/\mu$, and correspond to an evolution with respect to the $GM$-scaled time $\hat t$. For instance, we have
\begin{eqnarray}
\frac{dq}{d \hat t}&=&\frac{\partial \tilde {\mathcal H}_{\rm (eob)}}{\partial \tilde p_r}\,,\quad 
\frac{d\tilde p_r}{d \hat t}=-\frac{\partial \tilde {\mathcal H}_{\rm (eob)}}{\partial q}+GM \tilde {\mathcal F}_r\,,\nonumber\\
\frac{d\phi}{d \hat t}&=&GM \frac{d\phi}{d  t}=\frac{\partial \tilde {\mathcal H}_{\rm (eob)}}{\partial j}\,,\quad
\frac{dj}{d \hat t}=\tilde {\mathcal F}_\phi\,.
\end{eqnarray}
Note also the vectorial relation
\beq
{\mathbf j}={\mathbf q}\times \tilde {\mathbf p}=\frac{{\mathbf r}}{GM}\,\times \frac{{\mathbf p}}{\mu}=\frac{{\mathbf J}}{GM\mu}\,,
\eeq
where ${\mathbf J}=  {\mathbf r}\times {\mathbf p}$ is the orbital angular momentum of the system.
Let us also note the following relations
\beq
\label{derivHEOB}
\frac{\partial \tilde {\mathcal H}_{\rm (eob)}}{\partial \tilde p_a}=\frac{1}{h}\frac{\partial \tilde {\mathcal H}_{\rm (eff)}}{\partial \tilde p_a}\,,
\eeq
with
\begin{eqnarray}
\frac{\partial \tilde {\mathcal H}_{\rm (eff)}}{\partial \tilde p_r}&=&\frac{c^2A}{B \tilde {\mathcal H}_{\rm (eff)}} \tilde p_r \nonumber\\
\frac{\partial \tilde {\mathcal H}_{\rm (eff)}}{\partial \tilde p_\phi}&=& \frac{c^2A}{r^2 \tilde {\mathcal H}_{\rm (eff)}} \tilde p_\phi\,.
\end{eqnarray}

\subsection{$\Phi_E$, $\Phi_J$ and $\Phi_{EJ}$ in EOB variables}

Let us now indicate how one can express the flux functions $\Phi_E$, $\Phi_J$ and $\Phi_{EJ}$ in terms of EOB variables. The first, crucial remark is that $\Phi_E$ and $\Phi_J$ are gauge-invariant quantities, and are scalars. [Note, however, that this is not true for  $\Phi_{EJ}= \Phi_E - \dot \phi \Phi_J$, because $\dot \phi$  is {\it not} a gauge invariant quantity (along non-circular orbits), but depends on the chosen coordinate system.  Here, we shall only consider the value of the
{\it combined} flux in EOB coordinates: $\Phi_{EJ}^{EOB}= \Phi_E - {\dot \phi}^{EOB} \Phi_J$.] This implies that the numerical values of  $\Phi_E$ and $\Phi_J$ are  independent of the choice of coordinates, and of any related choice of dynamical variables. We can therefore start from the results in the literature that have computed  $\Phi_E$ and $\Phi_J$, say at 2PN accuracy, in terms of, e.g. harmonic relative coordinates and velocities, ${\mathbf x}_h$ and ${\mathbf v}_h$, and transform these expressions in terms of EOB coordinates and momenta. This transformation is facilitated by the fact that $\Phi_E$ and $\Phi_J$ being scalars, are actually expressed in terms of a basis of scalar combinations of ${\mathbf x}_h$ and ${\mathbf v}_h$. [Here ${\mathbf x}_h={\mathbf x}_1^h-{\mathbf x}_2^h$, ${\mathbf v}_h=d {\mathbf x}_h/dt={\mathbf v}_1^h-{\mathbf v}_2^h$ are the relative, harmonic positions and velocities, considered in the center of mass system.]

Let us use the notation
\begin{eqnarray}
X_1^h&\equiv &v_h^2\nonumber\\
X_2^h&\equiv &({\mathbf n}_h\cdot {\mathbf v}_h)^2=\dot r_h^2\nonumber\\
X_3^h&\equiv &\frac{GM}{r_h}\,,
\end{eqnarray}
and introduce $X_A^h$, $(A=1,2,3)$ to refer collectively to these three scalars.
The corresponding, natural EOB scalars are $X_A^e$, $A=1,2,3$ with
\beq
\label{Xevariabili}
X_1^e\equiv \tilde {\mathbf p}^2\,,\quad X_2^e\equiv \tilde p_r^2\,,\quad X_3^e\equiv \frac{GM}{r_e}=\frac{1}{q}=u\,,
\eeq
where, as above, $\tilde {\mathbf p}={\mathbf p}_e/\mu$ and $q=r_e/(GM)$.
Note that all the scalars $X_A^h$, $X_A^e$ have the dimensions of a squared velocity. In other words $X_A^h/c^2$, $X_A^e/c^2$ are dimensionless.

In terms of this notation we have simply
\begin{eqnarray}
\Phi_E^e(X_A^e)&=&\Phi_E^h(X_A^h)\,,\nonumber\\
\Phi_J^e(X_A^e)&=&\Phi_J^h(X_A^h)\,.
\end{eqnarray}
Therefore, starting from the known results for $\Phi_E^h(X_A^h)$, $\Phi_J^h(X_A^h)$, it is enough to derive the transformation (taken at a fixed, common dynamical time $t^h=t^e$)
\beq
\label{n13}
X_A^h=f(X_A^e)
\eeq
to get the fluxes expressed in EOB variables.
When PN-expanded the transformation (\ref{n13}) has a polynomial structure, namely,
\begin{eqnarray}
\label{n14}
X_A^h&=&\xi_{AB_1}X_{B_1}^e+\epsilon^2 \xi_{AB_1B_2}X_{B_1}^eX_{B_2}^e\nonumber\\
&& +\epsilon^4 \xi_{AB_1B_2B_3}X_{B_1}^eX_{B_2}^eX_{B_3}^e+O(\epsilon^6)
\end{eqnarray}
Here $\epsilon\equiv 1/c$ is the PN expansion parameter and the structure of the 2PN-accurate expansion follows from the fact that $X_A/c^2\sim V^2/c^2$ is dimensionless.

Actually, we have derived the transformation (\ref{n13}) by combining the two transformations that have been explicitly worked out in the literature: (i) the transformation between EOB $({\mathbf q}_e, {\mathbf p}_e)$ and ADM $({\mathbf q}_a, {\mathbf p}_a)$ phase-space variables \cite{Buonanno:1998gg,Damour:2000we, Nagar:2011fx}; and (ii) the transformation between the ADM phase-space variables  $({\mathbf q}_a, {\mathbf p}_a)$ and the harmonic positions and velocities  $({\mathbf q}_h, {\mathbf v}_h)$  \cite{DamourSchaefer85,Damour:2000ni,Gopakumar:1997bs}.

We give in appendix \ref{coord_trans} the explicit forms of the various transformations
$({\mathbf q}_e, {\mathbf p}_e)\leftrightarrow ({\mathbf q}_a, {\mathbf p}_a) \leftrightarrow ({\mathbf q}_h, {\mathbf v}_h)$ we used, together with the explicit form of the resulting transformation (\ref{n13}), (\ref{n14}) between the corresponding scalars.

By inserting the latter transformation in the results of Refs. \cite{Gopakumar:1997ng} for the 2PN-accurate $\Phi_E^h$, $\Phi_J^h$ we get the explicit expressions of 
$\Phi_E^e$, $\Phi_J^e$ in EOB variables.
In order to better comprehend the structure of these results it is convenient to introduce a special notation for a general polynomial in $X_A^e$.

Given a collection of (symmetric) multi-index coefficients
$C_{A_1A_2\ldots A_p}$, $C_{A_1A_2\ldots A_p A_{p+1}}, \ldots , C_{A_1A_2\ldots A_q}$, 
where $0\le p <q$ and $A_i=1,2,3$, we denote
\begin{widetext}
\beq
C_{p,q}(X_A)=C_{A_1\ldots A_p} X_{A_1}\ldots X_{A_p}+\epsilon^2 C_{A_1\ldots A_p A_{p+1}} X_{A_1}\ldots X_{A_p}X_{A_{p+1}}+\ldots +
\epsilon^{2(q-p)}C_{A_1\ldots A_q} X_{A_1}\ldots X_{A_q}\,,
\eeq
\end{widetext}
where we have $q-p+1$ contributions, each one (using Einstein's summation convention) is a sum over all the indices $A_1\ldots A_n$ it involves.
Also the short-hand notation
\beq
\label{shorthand}
X_AX_BX_C\ldots =X_{ABC\ldots}
\eeq
will be adopted hereafter, when convenient. 
Note that in the multisummation 
\beq
C_{A_1\ldots   A_{n}} X_{A_1}\ldots X_{A_n}
\eeq
the coefficient of $(X_1)^{n_1}(X_2)^{n_2}(X_3)^{n_3}$  (with $n_1+n_2+n_3=n $) is
\beq
S(n_1,n_2,n_3)\, C_{\underbrace{11}_{n_1 \,\rm {times}} \ldots \underbrace{22}_{n_2\,\rm {times}} \ldots \underbrace{33}_{n_3\,\rm {times}} \ldots}
\eeq
where  the symmetry factor $S(n_1,n_2,n_3)$ is given by
\beq \label{symfactor}
S(n_1,n_2,n_3)=\frac{(n_1+n_2+n_3)!}{n_1! \, n_2! \, n_3!} \,.
\eeq
In addition, as our basic variables are the EOB ones, we shall often, for brevity, suppress the index $e$ (standing for EOB) on them: $X_A=X_A^e$.

Before considering the higher PN corrections to the energy and angular momentum fluxes it is useful to recall their leading order (\lq\lq Newtonian order")
expressions. They are easily deduced from the well known quadrupolar approximation (see e.g., \cite{LL}), namely
\beq
\Phi_E=\frac{G}{5c^5} \left(I_{ij}^{(3)}\right)^2 \,,\quad
\Phi_J=\frac{2G}{5c^5} \epsilon_{zij}I_{is}^{(2)}I_{js}^{(3)} \,,
\eeq 
with (in the center of mass)
\beq
I_{ij}=m_1 x_1^{<i}x_1^{j>}+m_2 x_2^{<i}x_2^{j>}=\mu x^{<i}x^{j>}\,,
\eeq
using standard notation for symmetric and tracefree tensors.
This yields 
\begin{eqnarray}
\label{scaledfluxes}
G\Phi_E &=&\frac{1}{c^5}\frac85  \nu^2 \left(\frac{GM}{r}  \right)^4 \left(4v^2-\frac{11}{3}\dot r^2\right)\nonumber\\
\frac{\Phi_J}{M} &=&\frac{1}{c^5} \frac85 \nu^2 \left(\frac{GM}{r}  \right)^3 \, j \left(2v^2-3\dot r^2+2\frac{GM}{r}\right)\,.\nonumber
\end{eqnarray}
Note that these fluxes are both proportional to $\nu^2$  and contain a factor $1/c^5$ (2.5PN order). In terms of a characteristic Newtonian velocity $V$ (with 
$GM/r\sim V^2, j=1/V$), we have
\beq
G\Phi_E \sim \nu^2 \frac{V^{10}}{c^5}\,,\quad \frac{\Phi_J}{M} \sim \nu^2 \frac{V^{7}}{c^5}\,.
\eeq
It will be also convenient to work with the quantities
\begin{eqnarray}
\widehat \Phi_E &=& \frac{c^5 }{\nu} G\Phi_E \sim \nu V^{10}\nonumber\\
\widehat \Phi_J &=& \frac{c^5}{\nu} \frac{\Phi_J}{ M} \sim \nu V^{7}\,,  
\end{eqnarray}
which have a finite limit when $c\to \infty$ and in which one power of $\nu$ has been factored out (so that they will be conveniently related to
${\pmb {\mathcal F}}/\mu$).

With this notation our 2PN-accurate results in EOB variables have the form
\begin{eqnarray}
\label{PhiEJ_eqsadim}
\widehat \Phi_E^e ({\mathbf X}^e)&=& \left(\frac{GM}{r_e} \right)^4 C_{1,3}(X_A^e)\nonumber\\
\widehat \Phi_J^e ({\mathbf X}^e)&=& j \left(\frac{GM}{r_e} \right)^3 B_{1,3}(X_A^e)\,,
\end{eqnarray}
where the explicit values of the coefficients entering
\begin{widetext}
\beq
C_{1,3}(X_A)=C_{A_1} X_{A_1}+\epsilon^2 C_{A_1 A_2} X_{A_1}X_{A_2}+\epsilon^4 C_{A_1 A_2 A_3} X_{A_1}X_{A_2}X_{A_3}
\eeq
\end{widetext}
and $B_{1,3}(X_A)$ are listed in Appendix \ref{en_flux} and \ref{am_flux}.
Let us, for illustration, explicitly display here the leading order contributions to $\widehat \Phi_E$ and $\widehat \Phi_J$ (\lq\lq Newtonian order"), namely
\begin{eqnarray}
\widehat \Phi_E^{e \rm {(Newt)}} &=&  \frac{8}{5}\nu \left( \frac{GM}{r_e}\right)^4 \left(4 \tilde p^2 -\frac{11}{3}\tilde p_r^2\right)\nonumber\\
&=& \frac{8}{5}\nu \left( \frac{GM}{r_e}\right)^4 \left(4 X_1^e -\frac{11}{3}X_2^e\right)\nonumber\\
\widehat \Phi_J^{e \rm {(Newt)}} &=& \frac85 \nu \left( \frac{GM}{r_e}\right)^3\, j\, \left(2\tilde p^2 -3 \tilde p_r^2+2\frac{GM}{r_e} \right)\nonumber\\
&=& \frac85 \nu \left( \frac{GM}{r_e}\right)^3 \, j \, (2X_1^e-3X_2^e+2X_3^e)\,.\nonumber
\end{eqnarray}

\subsection{Algorithm for decomposing $\Phi_{EJ}$}
Finally, we need to compute the  correspondingly rescaled version of the  combined flux 
$\Phi_{EJ}^{EOB}= \Phi_E -   \dot \phi^e \Phi_J$, Eq. (\ref{n9}) (with the EOB angular velocity
$ \dot \phi^e \equiv {\dot \phi}^{EOB}$), in terms of EOB variables, namely
\beq
\widehat \Phi_{EJ}^e \equiv \frac{c^5G}{\nu}  \Phi_{EJ}^e= \widehat \Phi_{E}^e  - GM \dot \phi^e \, \widehat \Phi_{J}^e\,.
\eeq
Combining Hamilton equations for the angular motion, $\dot \phi^e=\partial {\mathcal H}_{\rm (eob)}/\partial p_\phi$, whose explicit expression is obtained
from Eqs. (\ref{derivHEOB}), i.e.
\beq
\frac{d \phi^e}{d \hat t}=GM \dot \phi^e = \frac{c^2A}{\hat r^2 h\tilde {\mathcal H}_{\rm (eff)}}j\,,
\eeq
with our above explicit expressions for $\widehat \Phi_E$ and $\widehat \Phi_J$, Eqs. (\ref{PhiEJ_eqsadim}), yields the following expression
for $\widehat \Phi_{EJ}$,
\begin{widetext}
\beq
\label{m1}
\widehat \Phi_{EJ}^e=\left(\frac{GM}{r_e} \right)^3 \left[\frac{GM}{r_e} C_{1,3}(X_A^e)-\frac{c^2 A}{h\tilde {\mathcal H}_{\rm (eff)}}\left(\frac{GM}{r_e} \right)^2 \,j^2 B_{1,3}(X_A^e) \right]\,,
\eeq
\end{widetext}
where $h$ has been defined in Eq. (\ref{smallh}).
In Eq. (\ref{m1}), the factor $(GM/r_e)^2j^2$ can be expressed in terms of the $X_A^e$'s, Eq. (\ref{Xevariabili}), since $j=\tilde p_\phi/(GM)$ and
\begin{eqnarray}
\left(\frac{GM}{r_e}\right)^2j^2&=&
\frac{\tilde p_\phi^2}{r_e^2}=({\mathbf n}_e \times \tilde {\mathbf p}_e)^2\nonumber\\
&=&\tilde p_e^2-({\mathbf n}_e \cdot \tilde {\mathbf p}_e)^2\,, 
\end{eqnarray}
and hence
\beq
\label{eqjquad}
\left(\frac{GM}{r_e}\right)^2 \, j^2= X_1^e-X_2^e\,.
\eeq
Similarly, one can replace the remaining factor  $c^2 A/(h\tilde {\mathcal H}_{\rm (eff)})=1+O(\epsilon^2)$ in terms of the $X_A^e$'s, namely
\begin{widetext}
\begin{eqnarray}
\frac{r_e^2  \dot \phi^e}{GM j} = \frac{c^2 A}{h\tilde {\mathcal H}_{\rm (eff)}}&=&  
1+ \epsilon^2 \left(-\frac{\nu+1}{2}X_1^e+(\nu-1)X_3^e  \right)
+\epsilon^4 \left(\frac{3\nu^2-\nu-1}{2}(X_{33}^e-X_{13}^e )  \right. \nonumber\\
&&\left.+ (\nu+1) X_{23}^e+\frac38 (\nu^2+\nu+1)X_{11}^e \right) \,.
\end{eqnarray}

 For instance, the leading order contribution
(\lq\lq Newtonian order") to $\widehat \Phi_{EJ}^e$ reads

\begin{eqnarray}
\label{phiejnewtonian}
\widehat \Phi_{EJ}^{e \rm {(Newt)}} &=&   \widehat \Phi_{E}^{e \rm {(Newt)}}-\left( \frac{GM}{r_e} \right)^2j^2\, \widehat \Phi_{E}^{e \rm {(Newt)}}\nonumber\\
&=& \frac{8}{5}\nu \left(\frac{GM}{r_e}  \right)^3 \left[\frac{GM}{r_e}\left(4 \tilde p^2 -\frac{11}{3}\tilde p_r^2\right) -(\tilde p^2-\tilde p_r^2)\left(2\tilde p^2 -3 \tilde p_r^2+2\frac{GM}{r_e} \right)
  \right]\nonumber\\
&=& \frac{8}{5}\nu \left(\frac{GM}{r_e}  \right)^3 \left[X_3^e \left(4 X_1^e -\frac{11}{3}X_2^e\right) -(X_1^e-X_2^e)(2X_1^e-3X_2^e+2X_3^e) \right]\nonumber\\
&=& \frac{8}{5}\nu \left(\frac{GM}{r_e}  \right)^3 \left(-2(X_1^e)^2-3(X_2^e)^2 +5X_1^eX_2^e+2X_1^eX_3^e-\frac53 X_2^eX_3^e \right)\nonumber\\
&=& \frac{8}{5}\nu \left(\frac{GM}{r_e}  \right)^3 \left(-2 X_{11}^e-3X_{22}^e +5X_{12}^e+2X_{13}^e-\frac53 X_{23}^e  \right)\,.
\end{eqnarray}
\end{widetext}
Note that, while one could naturally factor $u^4=(GM/r)^4$ in front of $\widehat \Phi_{E}^{e \rm {(Newt)}}$, it is only the third power of $u=GM/r$ which one can naturally factor out of $\widehat \Phi_{EJ}^{e \rm {(Newt)}} $. This difference is linked to the fact that $\widehat \Phi_{E}^{e \rm {(Newt)}}/u^4$ was {\it linear} in $X_A$, while  $\widehat \Phi_{EJ}^{e \rm {(Newt)}}/u^3 $ is {\it quadratic} in the $X_A$'s.

When keeping the higher order PN corrections (which involve more powers of $X_A^e/c^2\sim v^2/c^2$), the adimensionalized combined flux has the structure
\beq
\label{eq:hatphiejformal}
\widehat \Phi_{EJ}^e ({\mathbf X}^e)=\left(\frac{GM}{r_e} \right)^3 Q_{2,4}(X_A^e) \,,
\eeq
where   
\begin{widetext}
\beq
Q_{2,4}(X_A)= Q_{A_1A_2} X_{A_1}X_{A_2}+\epsilon^2 Q_{A_1 A_2A_3} X_{A_1}X_{A_2}X_{A_3}+\epsilon^4 Q_{A_1 A_2 A_3A_4} X_{A_1}X_{A_2}X_{A_3}X_{A_4}\,,
\eeq
\end{widetext}
the coefficients of which are listed in Appendix \ref{enandam_flux}.

As indicated above, the first step of our new approach consists in separating out of $\widehat \Phi_{EJ}^e$ either a factor $Z_1=p_r^2$ or a factor $Z_2=r\partial {\mathcal H}_{\rm (eob)}/\partial r=-r\dot p_r$. As we are working in terms of $\tilde p_i=p_i/\mu$ and $GM/r_e=1/q$, we replace $Z_1$ and $Z_2$ respectively by
\begin{eqnarray}
\label{Z12}
\tilde Z_1 &\equiv& \tilde p_r^2 \equiv  X_2^e \nonumber\\
\tilde Z_2 &\equiv& -r_e \frac{d {\tilde p}_r^e}{dt}=r_e\frac{\partial \tilde {\mathcal H}_{\rm (eob)}}{\partial r_e} \equiv  X_4^e \,,
\end{eqnarray}
which both have the dimensions of a squared velocity.
In order to separate out a factor $\tilde Z_1=X_2^e$ or $\tilde Z_2=X_4^e$ from $\widehat \Phi_{EJ}^e$, Eq. (\ref{eq:hatphiejformal}), we need to replace our basic set of scalar variables $(X_1^e,X_2^e,X_3^e)$ by the new set of scalar variables $(X_2^e=\tilde Z_1 ,X_3^e,X_4^e=\tilde Z_2)$. This is done by first expressing
$X_4^e=r_e\partial \tilde {\mathcal H}_{\rm (eob)}/\partial r_e$ in terms of $(X_1^e,X_2^e,X_3^e)$ (by differentiating the EOB Hamiltonian (\ref{H_EOB}) with respect to the variable $r_e$) and then solving for $X_1^e$ as a function of $X_2^e$, $X_3^e$ and $X_4^e$. For instance, at the Newtonian order we have
\begin{eqnarray}
 \frac{1}{\mu}\left( {\mathcal H}_{\rm (eob)}-Mc^2\right)^{\rm (Newt)}&=&\frac12 \tilde {\mathbf p}^2-\frac{GM}{r} \nonumber\\
&=&\frac12 \tilde p_r^2+\frac12 \frac{\tilde p_\phi^2}{r^2} -\frac{GM}{r} \nonumber\\
&=& \frac12 \tilde p_r^2+\frac12 \frac{j^2}{q^2} -\frac{1}{q}
\end{eqnarray}
so that
\begin{eqnarray}
\tilde Z_2^{\rm (Newt)}&\equiv&X_4^{\rm (Newt)}=-\frac{j^2}{q^2} +\frac{1}{q}\nonumber\\
&=&-X_1+X_2+X_3\,.
\end{eqnarray}
Therefore, at the leading order, $X_1$ can be solved in terms of $X_2$, $X_3$ and $X_4$ according to
\beq
\label{m3}
X_1=X_2+X_3-X_4+O\left(\frac{1}{c^2}\right)\,.
\eeq
The extension of this result to 2PN accuracy is obtained by first computing $\tilde Z_2(X_1,X_2,X_3)$ to higher order, namely
\begin{eqnarray}
\tilde Z_2&\equiv&X_4^e(X_1^e,X_2^e,X_3^e)=r_e\frac{\partial \tilde {\mathcal H}_{\rm (eob)}}{\partial r_e}\nonumber\\
&=& -  \hat  C_{1,3}(X_A^e)\,,
\end{eqnarray}
where the coefficients of $\hat  C_{1,3}$ in 
\begin{widetext}
\beq
\hat C_{1,3}(X_A)=\hat C_{A_1} X_{A_1}+\epsilon^2 \hat C_{A_1 A_2} X_{A_1}X_{A_2}+\epsilon^4 \hat C_{A_1 A_2 A_3} X_{A_1}X_{A_2}X_{A_3}
\eeq
are listed in Appendix \ref{eq_re}. Then one  solves (perturbatively) for $X_1$ in terms of $X_2$, $X_3$ and $X_4$, starting with the Newtonian solution (\ref{m3}).
This yields
\begin{eqnarray}
\label{X1_expanso}
X_1^e&=&X_2^e+X_3^e -X_4^e +\epsilon^2\left(2X_{23}^e +3X_{33}^e +\frac{\nu-5}2  X_{34}^e -\frac{\nu +1}{2}  X_{24}^e   +\frac{1+\nu}{2}  X_{44}^e \right)\nonumber\\
&& +\epsilon^4\left((2-6\nu) X_{233}^e  -3(\nu-3) X_{333}^e  +\frac18 (\nu ^2+7\nu-63 )X_{334}^e +\frac18( \nu ^2-\nu+1)X_{224}^e \right.\nonumber\\
&& \left.+\frac{1}{4}(5\nu+8)  X_{344}^e  +\frac34 \nu X_{244}^e -\frac14 (\nu^2+\nu+3)X_{234}^e   -\frac{1}{8}(\nu^2+5\nu +1) X_{444}^e\right)\,,
\end{eqnarray}
\end{widetext}
where we have used the short-hand notation (already introduced in Eq. (\ref{shorthand}) for the variables $X_1$,$X_2$, $X_3$)
\beq
X^e_{IJK\ldots}=X^e_IX^e_JX^e_K\ldots\,\quad (I,J,K=2,3,4)\,.
\eeq
Here and below we find often convenient to use an explicit form for the polynomial expansion in powers of $X_I$'s (rather than a tensorial form $C_IX_I+\epsilon^2 C_{IJ}X_IX_J+\ldots$ where one must take into account the symmetry factors associated with each term in the multi summations).

Finally, by substituting the PN expansion of $X_1^e (X_2^e,X_3^e,X_4^e)$, Eq. (\ref{X1_expanso}), into the combined flux (\ref{phiejnewtonian}), we get the expression of $\widehat \Phi_{EJ}$ in terms of $X_I^e =(X_2^e,X_3^e,X_4^e)$. For example, at the Newtonian order, it suffices to replace Eq. (\ref{m3}) into Eq. (\ref{phiejnewtonian}). This yields
\begin{widetext}
\beq
\widehat \Phi_{EJ}^{\rm (Newt)}(X_I)=\frac{8\nu}{5}\left( \frac{GM}{r_e} \right)^3 \left(2X_3X_4 +\frac43 X_2X_3-2X_4^2-X_2X_4 \right)\,.
\eeq
\end{widetext}
As anticipated, each term in this expression contains either a factor $\tilde Z_1=X_2$ or $\tilde Z_2=X_4$. It can therefore be decomposed in the form (\ref{n11})
that we mentioned above. Actually, there are many ways in which such a decomposition can be performed because the term $-X_2X_4=-\tilde Z_1 \tilde Z_2$ can be considered either as a part of $\Phi_1 Z_1$ or of $\Phi_2 Z_2$.

We shall define the minimal decomposition (\ref{n11}) of a polynomial in the $X_I$'s (which vanishes when $X_2=0=X_4$) as the one of the form
\beq
X_2\widehat \Phi_2(X_2,X_3,X_4)+X_4\widehat \Phi_4(X_3,X_4)\,,
\eeq
in which the coefficient of $X_4$ does not contain any dependence on $X_2$. (In other words, all the terms $\propto X_2^n$ are shuffled into the $X_2 \widehat \Phi_2$ contribution.)

This minimal choice somewhat simplifies the expression of $\widehat\Phi_4$, i.e., the coefficient denoted as $\Phi_2$ in Eq. (\ref{n11})-(\ref{n12}). In turn this simplifies both the radial component of radiation reaction and the Schott energy, because, according to our above result (\ref{n12}), these contain respectively $d\widehat \Phi_4/dt$ and $\widehat \Phi_4$.
[Note the mnemonic rule that the indices get multiplied by a factor of two when passing from the notation of Sec. II to the notation here, $Z_1\to X_2$, $Z_2\to X_4$, $\Phi_1\to \widehat \Phi_2$ and $\Phi_2\to \widehat \Phi_4$.] 

For instance, at the Newtonian level, the minimal decomposition of $\widehat \Phi_{EJ}$ reads
\begin{widetext}
\begin{eqnarray}
\label{phiejnewtonian2}
\widehat \Phi_{EJ}^{e \rm {(Newt)}}(X_I) 
&=& \frac{8}{5}\nu \left(\frac{GM}{r_e}  \right)^3 \left[X_2 \left(\frac43 X_3-X_4 \right)+X_4 \left( 2X_3-2X_4\right)\right]\,,
\end{eqnarray}
while its 2PN-accurate generalization reads
\begin{eqnarray}
\label{m4}
\widehat \Phi_{EJ}^{e }(X_I) &=& X_2\widehat \Phi_2(X_2,X_3,X_4)+X_4\widehat \Phi_4(X_3,X_4) \nonumber\\
&\equiv &  \left(\frac{GM}{ r_e}  \right)^3 \left[X_2\HPhi_2(X_2,X_3,X_4)+X_4\HPhi_4(X_3,X_4)\right]
\end{eqnarray}
where we found it convenient to factorize the term $({GM}/{ r_e})^3$ in the above expression so that
\beq
\widehat \Phi_2(X_2,X_3,X_4)=(X_3^e)^3\HPhi_2\,,\qquad\quad \widehat \Phi_4(X_3,X_4)=(X_3^e)^3\HPhi_4\,,
\eeq
with
\begin{eqnarray}
\HPhi_2 &=& \frac{8}{5}\nu \left(\frac{4}{3}X_3^e-X_4^e\right) \nonumber\\
&&+\epsilon^2
\left(\frac{236}{105}\nu^2X_{24}^e -\frac{5252}{105}\nu X_{33}^e -\frac{608}{105}\nu X_{23}^e -\frac{256}{105}\nu X_{44}^e-\frac{484}{105}\nu^2 X_{34}^e \right.\nonumber\\
&& \left.+\frac{548}{105}\nu X_{24}^e +\frac{76}{21}\nu^2 X_{44}^e +\frac{24}{35}\nu^2 X_{33}^e-\frac{80}{21}\nu^2X_{23}^e +\frac{1300}{21}\nu X_{34}^e\right)\nonumber\\
&& + \epsilon^4\left(\frac{1756}{63}\nu^2 X_{333}^e+\frac{854948}{2835}\nu X_{333}^e +\frac{1112}{105} \nu^2 X_{223}^e +\frac{1378}{45}\nu X_{233}^e+\frac{120268}{945}\nu^2 X_{233}^e\right.\nonumber\\
&& +\frac{416}{105}\nu X_{223}^e -\frac{45916}{315}\nu^2 X_{234}^e -\frac{4066}{35}\nu X_{234}^e-\frac{32}{21}\nu X_{244}^e -\frac{1973}{315}\nu  X_{224}^e \nonumber\\ 
&& +\frac{398}{21}\nu^2X_{244}^e -\frac{499}{63}\nu^2 X_{224}^e+\frac{1496}{315}\nu X_{344}^e -\frac{14597}{35}\nu X_{334}^e-\frac{25442}{315}\nu^2 X_{344}^e\nonumber\\
&& 
-\frac{892}{105}\nu^2 X_{334}^e+\frac{668}{35}\nu^2 X_{444}^e -\frac{289}{35}\nu X_{444}^e +\frac{701}{35} \nu^3 X_{444}^e +\frac{9164}{945}\nu^3 X_{333}^e\nonumber\\
&& -\frac{2428}{63}\nu^3 X_{344}^e+\frac{1459}{315}\nu^3 X_{334}^e -\frac{176}{105}\nu^3 X_{244}^e -\frac{857}{315}\nu^3 X_{224}^e+\frac{16}{3}\nu^3 X_{223}^e\nonumber\\
&& \left.+\frac{1672}{945}\nu^3 X_{233}^e -\frac{4}{9}\nu^3 X_{234}^e \right)\,,
\end{eqnarray}
and
\begin{eqnarray}
\HPhi_4&=&\frac{16}{5}\nu (X_3^e-X_4^e)  \nonumber\\
&& +\epsilon^2\left(-\frac{704}{105}\nu ^2 X_{33}^e+\frac{278}{105}\nu X_{44}^e -\frac{256}{105}\nu ^2 X_{44}^e +\frac{568}{35}\nu X_{34}^e +\frac{1168}{105}\nu ^2 X_{34}^e -\frac{538}{35}\nu X_{33}^e\right)\nonumber\\
&& +\epsilon^4\left(-\frac{58}{45}\nu X_{444}^e-\frac{1135}{63}\nu ^2 X_{344}^e-\frac{14597}{315}\nu ^2 X_{334}^e-\frac{286}{315}\nu X_{334}^e
-\frac{9832}{315}\nu^3 X_{334}^e\right.\nonumber\\
&& \left.+\frac{44}{35}\nu^3 X_{444}^e+\frac{3272}{945}\nu X_{333}^e  +\frac{6082}{945}\nu ^3X_{333}^e
+\frac{1377}{35}\nu ^2 X_{333}^e+\frac{1363}{315}\nu ^2X_{444}^e-\frac{6536}{315}\nu  X_{344}^e\right.\nonumber\\
&& \left. +\frac{654}{35}\nu ^3 X_{344}^e\right)\,.
\end{eqnarray}
It should be noted that  $\HPhi_2$ and $\HPhi_4$ have the dimension of $V^2$.
Moreover, in the circular orbit limit $X_2=0=X_4$ (for a later use) the above expressions reduce to 
\begin{eqnarray}
\HPhi_2(0,X_3^e,0)&=& \frac{32}{15}\nu X_3^e \left[ 1+ \epsilon^2 X_{3}^e \left( \frac{9}{28}\nu -\frac{1313}{56} \right)+\epsilon^4 X_{33}^e \left( \frac{2291}{504}\nu^2 +\frac{2195}{168}\nu +\frac{213737}{1512} \right) \right]
\nonumber\\
\HPhi_4(0,X_3^e,0) &=& \frac{16}{5}\nu X_3^e \left[ 1+ \epsilon^2 X_{3}^e \left( -\frac{44}{21}\nu -\frac{269}{56} \right)+\epsilon^4 X_{33}^e \left( \frac{3041}{1512}\nu^2 +\frac{1377}{112}\nu +\frac{409}{378} \right) \right]\,.
\end{eqnarray}

\end{widetext}

\subsection{Minimal expressions of ${\mathcal F}_r$ and $E_{\rm (schott)}$}

Having obtained a particular, {\it minimal} decomposition of the 2PN-accurate combined flux $\Phi_{EJ}({\mathbf x},{\mathbf p})$ in the form (\ref{n11}), namely Eq. (\ref{m4}), we can now apply our general results (\ref{n12}), i.e., derive the corresponding minimal expressions of ${\mathcal F}_r$ and $E_{\rm (schott)}$.
Modulo the $\mu$-rescaling ($\tilde E=E/\mu$, $\tilde p=p/\mu$), the prefactor $(GM/r_e)^3$ and $(\Phi_1\to\widehat\Phi_2,\Phi_2\to\widehat\Phi_4)$, the second Eq. (\ref{n12}) yields the following minimal Schott energy per unit reduced mass, $\tilde E_{\rm schott}^{\rm (min)}={E_{\rm schott}^{\rm (min)}}/{\mu}$
\beq
\label{minimalschottenergy}
\tilde E_{\rm schott}^{\rm (min)}=\frac{1}{c^5} \left( \frac{GM}{r_e} \right)^2 \tilde p_r \HPhi_4(X_3^e,X_4^e)\,.
\eeq
Note that the Newtonian order approximation to the (rescaled) Schott energy reads
\begin{widetext}
\begin{eqnarray}
\label{schottnewtonian}
\tilde E_{\rm schott}^{\rm (min, Newt)} 
&=& \frac{1}{c^5}\frac{16\nu}{5} \tilde p_r \left(\frac{GM}{r_e}  \right)^2 (X_3-X_4)\nonumber\\
&=&  \frac{1}{c^5}\frac{16\nu}{5} \tilde p_r \left(\frac{GM}{r_e}  \right)^2 \left[\left(\frac{{\mathbf p}}{\mu}\right)^2 -\left(\frac{p_r}{\mu}\right)^2 +O\left(\frac{1}{c^2}\right)  \right]\,,
\end{eqnarray}
\end{widetext}
where we used Eq. (\ref{m3}) to write the second form. The corresponding {\it minimal expression} of the ($\mu$-scaled) radiation reaction is obtained from
the first Eq. (\ref{n12}). To write it explicitly, we first need to derive the value of the ratio $\tilde p_r/\dot r$. This is obtained from Hamilton's equation
\beq
\dot r_e=\frac{\partial {\mathcal H}_{\rm (eob)}}{\partial p_r^e}\equiv \tilde C({\mathbf x}, {\mathbf p})\tilde p_r\,,
\eeq
with
\beq
\label{m5}
\tilde C({\mathbf x}, {\mathbf p})=\frac{c^2}{h \tilde {\mathcal H}_{\rm (eff)}}\frac{A(r_e)}{B(r_e)}
\eeq
where $h$ is given by Eq. (\ref{smallh}) above.
The expression (\ref{m5}) for $\tilde C$ is exact. Here, we shall work with its 2PN-accurate expansion which is found to be
\begin{eqnarray}
\tilde C 
&=& 1+\epsilon^2  \tilde  C_{1,2}(X_A^e)\,,
\end{eqnarray}
and the coefficients of $\tilde  C_{1,2}$ are listed in Appendix \ref{eq_re}.
In terms of $\tilde C$, $\HPhi_2$ and $\HPhi_4$, the radial component of the minimal ($\mu$-scaled) radiation-reaction is given by
\begin{widetext}
\begin{eqnarray}
\label{Fr_EOB}
\tilde {\mathcal F}^{\rm (eob)}_r&=&-\frac{1}{c^5} \left[\frac{1}{\tilde C} \frac{(GM)^2}{r_e^3}\, \tilde p_r \HPhi_2(X_2^e,X_3^e,X_4^e)+\frac{d}{dt}\left(\left(\frac{GM}{r_e}\right)^2 \HPhi_4(X_3^e,X_4^e) \right)\right]\,.
\end{eqnarray}
\end{widetext}
Let us also recall that the azimuthal component of the minimal ($\mu$-scaled) radiation-reaction is simply given by
\begin{eqnarray}
\label{Fphieob}
\tilde {\mathcal F}_\phi^{\rm (eob)}&=&-\tilde \Phi_J^{\rm (eob)} =-\frac{1}{c^5} \widehat \Phi_J^e\nonumber\\
&=&\frac{1}{c^5}\left( \frac{GM}{r_e}\right)^3\, j\, B_{1,3}(X_A^e)\,.
\end{eqnarray}
For illustration, let us display the leading-order (\lq\lq Newtonian order") terms in these expressions. To get in explicit form the leading order expression of 
$\tilde {\mathcal F}^{\rm (eob)}_r({\mathbf x},{\mathbf p})$ we need to perform the time derivative in Eq. (\ref{Fr_EOB}) by using the unperturbed (conservative) equations of motion. Here, we get some simplifications from having chosen $\widehat \Phi_4$ as a function of $X_3$ and $X_4$ only. Indeed, as $X_3^e=GM/r_e$ and 
\beq
X_4^{\rm (Newt)}=-\frac{\tilde p_\phi^2}{r^2}+\frac{GM}{r}\,,
\eeq
(where $p_\phi$ is constant along the conservative dynamics), the time derivative of $X_3$ and $X_4$ are both proportional to $\dot r$, e.g.
\beq
\frac{d X_4^{\rm (Newt)}}{dt}=\frac{\dot r}{r}\left(2\frac{\tilde p_\phi^2}{r^2}-\frac{GM}{r}\right)+O\left(\frac{1}{c^5}\right)\,.
\eeq
Re-expressing the result in terms of $\tilde p_r=\dot r (1+O(c^{-2}))$ we get
\begin{widetext}
\beq
\tilde {\mathcal F}_r({\mathbf x},{\mathbf p})^{\rm (Newt)}=\frac{1}{c^5} \frac{8}{15}\nu \frac{(GM)^2}{r^3}\tilde p_r \left(21 \tilde p^2-21 \tilde p_r^2-\frac{GM}{r}\right)+O\left(\frac{1}{c^7}\right)
\eeq
which, at this order, could alternatively be written in terms of velocities
\beq
\tilde {\mathcal F}_r({\mathbf x},{\mathbf v})^{\rm (Newt)}=\frac{1}{c^5} \frac{8}{15}\nu \frac{(GM)^2}{r^3}\dot r \left(21 v^2-21 \dot r^2-\frac{GM}{r}\right)+O\left(\frac{1}{c^7}\right)\,.
\eeq
The corresponding, Newtonian order, results for $\tilde {\mathcal F}_\phi ({\mathbf x},{\mathbf p})^{\rm (Newt)}$ read
\begin{eqnarray}
\tilde {\mathcal F}_\phi ({\mathbf x},{\mathbf p})^{\rm (Newt)}&=&-\frac{1}{c^5} \frac{8}{5}\nu \left(\frac{GM}{r}\right)^3\frac{\tilde p_\phi}{GM} \left(2\tilde p^2-3 \tilde p_r^2+2\frac{GM}{r}\right)+O\left(\frac{1}{c^7}\right)\nonumber\\
\tilde {\mathcal F}_\phi({\mathbf x},{\mathbf v})^{\rm (Newt)}&=&-\frac{1}{c^5} \frac{8}{15}\nu \frac{(GM)^2}{r}\dot \phi \left(2v^2-3 \dot r^2+2\frac{GM}{r}\right)+O\left(\frac{1}{c^7}\right)\,.
\end{eqnarray}
The explicit 2PN-accurate versions of our minimal $\tilde E_{\rm (schott)}$, $\tilde {\mathcal F}_r$ and $\tilde {\mathcal F}_\phi$ are given in 
Appendix \ref{Schott_energy} and \ref{Fr_and_phi_eob}.
They are expressed there in terms of $X_A=(X_1^e,X_2^e,X_3^e)$ and have the forms
\begin{eqnarray}
\label{minimalschottenergy2}
\tilde E_{\rm (schott)}^{\rm (min)}({\mathbf x},{\mathbf p}) &=& \frac{1}{c^5}\tilde p_r \left( \frac{GM}{r_e} \right)^2 (C_AX_A^e+\epsilon^2 C_{AB}X_{AB}^e+\epsilon^4C_{ABC}X_{ABC}^e)\nonumber\\
\tilde {\mathcal F}_r({\mathbf x},{\mathbf p}) &=&\frac{1}{c^5} \frac{(GM)^2}{r^3}\tilde p_r (R_AX_A^e+\epsilon^2 R_{AB}X_{AB}^e+\epsilon^4R_{ABC}X_{ABC}^e)\nonumber\\
\tilde {\mathcal F}_\phi({\mathbf x},{\mathbf p}) &=&\frac{1}{c^5} \left(\frac{GM }{r}\right)^3 \frac{\tilde p_\phi}{GM} (S_AX_A^e+\epsilon^2 S_{AB}X_{AB}^e+\epsilon^4S_{ABC}X_{ABC}^e)\,,
\end{eqnarray}
where the coefficients $C_{A_1\ldots A_n}$, $R_{A_1\ldots A_n}$, $S_{A_1\ldots A_n}$ are explicitly displayed in Eqs. (\ref{C1app})-(\ref{D8app}).

The Schott energy  as a function  of $X_2$, $X_3$ and $X_4$ (especially useful to study their limiting values along circular orbits) is given by Eq. (\ref{minimalschottenergy}),  while the radial  and azimuthal components of the radiation-radiation force follow from Eqs. (\ref{Fr_EOB}) and (\ref{Fphieob}), i.e., 
\begin{eqnarray}
\label{Fr_X234}
\tilde {\mathcal F}_r (X_2^e,X_3^e,X_4^e) &=& \frac{1}{c^5}\left(\frac{GM }{r}\right)^3 \tilde p_r(T_IX_I^e+\epsilon^2 T_{IJ}X_{IJ}^e+\epsilon^4T_{IJK}X_{IJK}^e)\nonumber\\
\label{Fphi_X234}
\tilde {\mathcal F}_\phi (X_2^e,X_3^e,X_4^e) &=&\frac{1}{c^5} \left(\frac{GM }{r}\right)^3 j (V_IX_I^e+\epsilon^2 V_{IJ}X_{IJ}^e+\epsilon^4V_{IJK}X_{IJK}^e)
\end{eqnarray}
where $I=2,3,4$ and the coefficients $T_{I_1\ldots I_n}$, $V_{I_1\ldots I_n}$  are explicitly displayed in Eqs. (\ref{TI_coeffs})-(\ref{TIJK_coeffs}).
Note that if one wishes to express  $\tilde {\mathcal F}_\phi$ entirely in terms of $X_2$, $X_3$ and $X_4$,
the (rescaled) angular momentum term $j$ should also be expanded in terms of $X_2$, $X_3$ and $X_4$; the result is the following
\beq
\label{j234}
j=\frac{\sqrt{X_1^e-X_2^e}}{X_3^e}=\frac{ \sqrt{X_3^e-X_4^e}}{X_3^e} \left[1 +\frac{W_1}{X_3^e-X_4^e}\epsilon^2 +  \frac{W_2}{(X_3^e-X_4^e)^2} \epsilon^4\right]
\eeq 
where

\begin{eqnarray}
W_1 &=& \frac{\nu+1}{4} X_{44}^e+\frac{\nu-5}{4} X_{34}^e-\frac{\nu+1}{4}X_{24}^e+\frac32 X_{33}^e +X_{23}\nonumber \\
W_2 &=&  \frac{\nu^2+8\nu +1}{32}X_{4444}^e+\frac{-4\nu^2-22\nu-24}{32}X_{3444}+\frac{\nu^2-4\nu+1}{16}X_{2444}^e\nonumber\\
&& +\frac{-3\nu^2+4\nu +121}{32}X_{3344}^e+\frac{3(\nu^2-1)}{16}X_{2344}-\frac{3(\nu^2+1)}{32}X_{2244}^e\nonumber\\
&& +\frac{\nu^2+25\nu-105}{16}X_{3334}^e-\frac{\nu^2-24\nu-2}{8}X_{2334}^e+\frac{\nu^2+3\nu+5}{16}X_{2234}^e+\frac{-12\nu+27}{8}X_{3333}^e\nonumber\\
&& -\frac{96\nu+16}{32}X_{2333}^e-\frac12 X_{2233}^e\,.
\end{eqnarray}
In the circular orbit limit these quantities reduce to
\begin{eqnarray}
W_1(0,X_3^e,0) &=& \frac32 X_{33}^e\,,\qquad
W_2(0,X_3^e,0) =  \frac{-12\nu+27}{8} X_{3333}^e\,.
\end{eqnarray}

\end{widetext}

\section{Non minimal choices  and associated gauge freedom}

Iyer and Will \cite{Iyer:1993xi,Iyer:1995rn} and later Gopakumar et al. \cite{Gopakumar:1997ng} have shown that, at each order in the PN expansion, there is a multi-parameter arbitrariness in the construction of a radiation-reaction force by the balance method, and that this arbitrariness is linked to the freedom in the choice of  coordinate gauge. Let us briefly discuss how this arbitrariness enters our approach. First, it can be checked that our simplifying constraint (\ref{n6}) that the Schott contribution to the {\it angular momentum} vanishes, $J_{\rm (schott)}=0$, corresponds to part of the freedom found by Iyer and Will.

Indeed, one easily checks that within their approach, all the (non necessarily vanishing)parameters entering $J_{\rm (schott)}$  are linearly independent, i.e., are unconstrained by the set of linear equations they obtained. Within our approach, this is immediately clear as we have obtained a solution with $J_{\rm (schott)}=0$, so that by choosing some given, general (nonzero) expression for $J_{\rm (schott)}$ (such that $\dot J_{\rm (schott)}$ vanishes along circular motions) we will be able to straightforwardly construct a corresponding (minimal) radiation reaction force. [Indeed, the condition that $\dot J_{\rm (schott)}$ vanishes along circular motion will introduce extra source terms in the equation (\ref{n8}) for ${\mathcal F}_r$ and $E_{\rm (schott)}$, linked to extra terms linear in $Z_1$ and $Z_2$ in the right hand side of (\ref{n8}), coming from an extra $\dot \phi (\delta {\mathcal F}_\phi)$ contribution to $\Phi_{EJ}$, linked to 
$\delta {\mathcal F}_\phi=-\dot J_{\rm (schott)}$.] This freedom in the choice of $J_{\rm (schott)}$ is parametrized by: (i) one parameter $(\lambda_0^J)$ at the leading (\lq\lq Newtonian") order, (ii) three parameters $(\lambda_2^J,\lambda_3^J,\lambda_4^J)$ at the 1PN order, and (iii) six parameters $(\lambda_{22}^J,\lambda_{33}^J,\lambda_{44}^J,\lambda_{23}^J,\lambda_{24}^J,\lambda_{34}^J)$ at the 2PN order.
The general form of $\tilde J_{\rm (schott)}= J_{\rm (schott)}/\mu$  can be written as
\begin{eqnarray}
\label{Jshott_non_min}
\tilde J_{\rm (schott)}^{\rm (non\, min)}&=&\frac{1}{c^5}\tilde p_r \tilde p_\phi\left( \frac{GM}{r_e}\right)^2\cdot \left(\lambda_0^J \right.\nonumber\\
&& \left. +\epsilon^2 \lambda_I^JX_I+\epsilon^4 \lambda_{IJ}^J X_IX_J\right)\,,
\end{eqnarray}
where the free gauge parameters parametrize the coefficients of a general polynomial in $X_I=(X_2,X_3,X_4)$.

Note that these parameters were indicated differently in previous papers \cite{Iyer:1993xi,Iyer:1995rn,Buonanno:2000ef}. In particular, the single $J$-related parameter $\lambda_0$ at leading (\lq\lq Newtonian") order was previously notated as
\beq
\beta_2^{GII}=\alpha^{IW}=\bar \alpha^{BD}
\eeq
and was normalized so that $\lambda_0^J=(8/5)\nu\beta_2^{GII}$.

Besides the parameters associated with the (non minimal) choice of a non vanishing $J_{\rm (schott)}$, there are further arbitrary parameters which, in our approach, correspond to further non minimal choices in the construction of $E_{\rm (schott)}$. Indeed, our general result (\ref{n12}) shows that the arbitrariness in the coefficient $\Phi_2$ of $Z_2$ in the decomposition (\ref{n11}), will directly affect $E_{\rm (schott)}$, and then ${\mathcal F}_r$. [Given a choice of $\Phi_2$, compatible with (\ref{n11}), the corresponding $\Phi_1$ is uniquely determined.] As discussed above, the arbitrariness in $\Phi_2$ is parametrized by a general term $\propto Z_1=p_r^2=X_2$. In terms of the relevant basis $X_2,X_3,X_4$ (with $X_2\propto Z_1$ and $X_4\propto Z_2$) the arbitrariness in $\widehat \Phi_4\sim \Phi_2$ in Eq. (\ref{Z12}) is of the general form
\begin{eqnarray}
\widehat \Phi_4^{\rm (non\, min)}(X_I)&=&\nu X_2 \left(\lambda_0^J+\epsilon^2 \lambda_I^JX_I \right.\nonumber\\
&& \left. +\epsilon^4 \lambda_{IJ}^J X_IX_J \right)\,,
\end{eqnarray}
corresponding to an additional non minimal contribution to $E_{\rm (schott)}$ of the form
\begin{eqnarray}
\label{e_schott_non_min}
E_{\rm (schott)}^{\rm (non\, min)}&=&\frac{\nu}{c^5}\tilde p_r^3 \left( \frac{GM}{r_e}\right)^{2}\cdot \left(\lambda_0^E\right. \nonumber\\
&& \left. +\epsilon^2 \lambda_I^EX_I+\epsilon^4 \lambda_{IJ}^E X_IX_J\right)\,.
\end{eqnarray}
This expression shows that the additional gauge-freedom associated with such non minimal choices in the Schott energy is parametrized by: (i) one parameter  
$\lambda_0^E$ at the leading (Newtonian) order, (ii) three parameters $(\lambda_2^E,\lambda_3^E,\lambda_4^E)$ at the 1PN order, and (iii) six parameters $(\lambda_{22}^E,\lambda_{33}^E,\lambda_{44}^E,\lambda_{23}^E,\lambda_{24}^E,\lambda_{34}^E)$ at the 2PN order.

In terms of the notation of \cite{Gopakumar:1997ng} (if we approximately identify their Lagrangian framework with our Hamiltonian one) these parameters correspond, respectively,  to: (i) $\alpha_3$, (ii) $\xi_2,\xi_4,\xi_5$ and (iii) $\psi_2,\psi_4,\psi_6,\psi_7,\psi_8,\psi_9$, i.e. to the following contributions
($\propto \dot r^3$) to the Schott energy considered in \cite{Gopakumar:1997ng}):
\begin{eqnarray}
E_{\rm (schott)}^{\rm (non\, min, Newt)}&=&\alpha_3 \frac{16}{5}\nu  \frac{G^2M^2}{c^5r^2}\dot r^3 \nonumber\\
E_{\rm (schott)}^{\rm (non\, min,1PN)}&=&\frac{16}{5}\nu  \frac{G^2M^2}{c^5r^2}\dot r^3\left(\xi_2 v^2 +\xi_4\dot r^2\right.\nonumber\\
&& \left. +\xi_5 \frac{GM}{r}\right)\nonumber\\
E_{\rm (schott)}^{\rm (non\, min,2PN)}&=&\frac{16}{5}\nu  \frac{G^2M^2}{c^5r^2}\dot r^3
\left(\psi_2 v^4+\psi_4v^2 \dot r^2\right.\nonumber\\
&&+\psi_6 v^2  \frac{GM}{r}+\psi_7\dot r^4+\psi_8\dot r^2 \frac{GM}{r}\nonumber\\
&&\left. +\psi_9 \left(\frac{GM}{r}\right)^2\right)\,.
\end{eqnarray}
Summarizing: the arbitrariness in the construction of a radiation-reaction force is parametrized by the parameters $\lambda_0^J$, $\lambda_{I_1}^J$, $\lambda_{I_1I_2}^J \ldots$ entering the (non-minimal) Schott angular momentum (\ref{Jshott_non_min}), together with the parameters $\lambda_0^E$, $\lambda_{I_1}^E$, $\lambda_{I_1I_2}^E \ldots$ entering the (non-minimal) $O(\tilde p_r^3)$ Schott energy (\ref{e_schott_non_min}) (expressed as a function of $X_2=\tilde p_r^2$, $X_3$ and $X_4$). It is easy to see that the number of arbitrary parameters entering the $n$PN order is equal to 
\beq
a_n=\left(\begin{array}{c}
n+2 \cr
n
\end{array} 
 \right)
= \left(\begin{array}{c}
n+2 \cr
2
\end{array}
 \right)= \frac{(n+1)(n+2)}{2}
 \,,
\eeq
for {\it each one} of these sources, with $a_0=1$, $a_1=3$, $a_2=6$, $a_3=10$, etc.

\section{Some applications of our results}

\subsection{Schott energy along quasi-circular inspirals}
Recently, Damour, Nagar, Pollney and Reisswig \cite{Damour:2011fu} have compared several different functional  relations $E(J)$ between the energy $E$ and the angular momentum $J$  of a binary system evolving along a radiation-reaction driven sequence of quasi-circular orbits. In particular, they compared a relation
$E^{\rm NR}(J)$ obtained from accurate numerical relativity (NR) simulations, to several of the relations $E^{\rm EOB}(J)$ that can be derived from EOB theory (under various approximations). Actually, the NR relation $E^{\rm NR}(J)$ computed in Ref. \cite{Damour:2011fu} was obtained by defining the NR energy $E^{\rm NR}$ and the NR angular momentum as being their initial values minus the time integral of their respective NR fluxes, $\Phi_E^{\rm NR}$ and $\Phi_J^{\rm NR}$ (as recorded at infinity).
In view of our general balance equations (\ref{n4}), we see that (modulo numerical errors) the NR energies and angular momenta can be identified with the {\it sum} of the system plus Schott contributions:
\begin{widetext}
\begin{eqnarray}
\label{EJ_NR}
E^{\rm NR}(t)&=& E_{\rm (system)}({\mathbf x}(t),{\mathbf p}(t)) +E_{\rm (schott)}({\mathbf x}(t),{\mathbf p}(t))\nonumber\\
J^{\rm NR}(t)&=& J_{\rm (system)}({\mathbf x}(t),{\mathbf p}(t)) +J_{\rm (schott)}({\mathbf x}(t),{\mathbf p}(t))\,.
\end{eqnarray}
\end{widetext}
On the other hand, one of the tenets of the current implementation of the EOB formalism is to require that the $\phi$-component of the radiation-reaction force be equal, at any moment, to minus the angular momentum flux $\Phi_J$.

In view of the second Eq. (\ref{n5}), this means that the EOB formalism has chosen a \lq\lq gauge" where
\beq
J_{\rm (schott)}^{\rm EOB}({\mathbf x}(t),{\mathbf p}(t))=0\,.
\eeq
In view of this, it is consistent to identify the instantaneous NR angular momentum  $J^{\rm NR}(t)$ with the EOB one $J^{\rm EOB}$, which indeed measures the angular momentum of the system, $J_{\rm (system)}$):
\beq
J^{\rm NR}(t)=J^{\rm EOB}({\mathbf x}(t),{\mathbf p}(t))\,.
\eeq
By contrast, in view of the first equation (\ref{EJ_NR}), the EOB measure of the total energy of the system, defined as
\begin{eqnarray}
E^{\rm EOB}({\mathbf x}(t),{\mathbf p}(t))&=&{\mathcal H}_{\rm (eob)}({\mathbf x}(t),{\mathbf p}(t))-Mc^2\nonumber\\
&=&E_{\rm (system)}({\mathbf x}(t),{\mathbf p}(t))\,,
\end{eqnarray}
cannot be simply identified with the NR computed energy $E^{\rm NR}$. Indeed, one expects the relation
\beq
E^{\rm NR}(t)=E^{\rm EOB}({\mathbf x}(t),{\mathbf p}(t))+E_{\rm (schott)}^{\rm EOB}({\mathbf x}(t),{\mathbf p}(t))\,.
\eeq 
In conclusion, as was already pointed out in Ref. \cite{Damour:2011fu}, the NR-derived functional relation $E^{\rm NR}(J)$ should differ from the EOB derived one 
$E^{\rm EOB}(J)$ by the quantity $E_{\rm (schott)}(t)$, re-expressed in terms of the corresponding instantaneous angular momentum
$J(t)=J^{\rm NR}(t)=J^{\rm EOB}(t)$.

Our results provide, for the first time, the explicit analytical value of $E_{\rm (schott)}$, namely the first of Eqs. (\ref{minimalschottenergy2}) (see Appendix \ref{Schott_energy}).
Note that $E_{\rm (schott)}$ is proportional to $\tilde p_r$, which stays rather small all along the radiation-reaction driven sequence of quasi-circular inspiralling orbits, including most of the subsequent plunge phase (see Fig. 1 of \cite{Buonanno:2000ef}). The smallness of $\tilde p_r$ further implies that the numerical value of $E_{\rm (schott)}$ is approximately gauge-invariant during the inspiral and the plunge. Indeed, Eq. (\ref{e_schott_non_min}) above shows that the general non-minimal contribution to $E_{\rm (schott)}$ contains an overall factor $\tilde p_r^3$, instead of the corresponding factor $\tilde p_r$ in $E_{\rm (schott)}^{\rm (min)}$. The ratio $E_{\rm (schott)}^{\rm (non\, min)}/E_{\rm (schott)}^{\rm (min)}$ is therefore generally expected to be numerically of order $
\tilde p_r^2$, and hence small during the inspiral (and even the plunge).

In addition, during the inspiral, i.e., before crossing the Last Stable (circular) Orbit (LSO), the dimensionless scalar 
$X_4^e/c^2=\tilde Z_2/c^2=(r_e/c^2) \partial {\tilde {\mathcal H}}_{\rm (eob)}/\partial r_e$ will also be numerically small. [Indeed, the orbital radius $r_e(t)$ approximately stays at the bottom of the effective radial potential $\tilde{\mathcal H}_{\rm (eob)}(j,r_e)$ during the inspiral.] The numerical value of $E_{\rm (schott)}$ during the inspiral can then be approximately evaluated by neglecting $X_4^e$ in $\widehat \Phi_4(X_3^e,X_4^e)$. This leads to an approximate, simplified expression for $E_{\rm (schott)}$, along the inspiral, as a function of the EOB radius $r_e$
\beq
\tilde E_{\rm (schott)}^{\rm (inspiral)}\simeq \tilde E_{\rm (schott)}^{\rm (min)}\simeq \frac{1}{c^5}\tilde p_r \widehat \Phi_4 (X_3^e,0)\,,
\eeq
i.e., explicitly
\begin{widetext}
\begin{eqnarray}
\label{minimalschottenergy2_circ}
&&\tilde E_{\rm (schott)}^{\rm (inspiral)}(t)\simeq  \frac{\nu}{c^5} \frac{16}{5} \tilde p_r \left(\frac{GM}{r_e}\right)^3\left[1-\frac{1}{168}(807+352\nu)
\left(\frac{GM}{r_e}  \right)\epsilon^2 \right.\nonumber\\
&& \left. \frac{1}{3024} (6082\nu^2+37179\nu+3272)\left(\frac{GM}{r_e}  \right)^2 \epsilon^4 \right]\,,
\end{eqnarray}
\end{widetext}

Note that $\tilde E_{\rm (schott)}^{\rm (inspiral)}(t)$ is {\it negative} (because $\tilde p_r\sim \dot r<0$ during the inspiral). It would be interesting to take into account the modifications of the EOB/NR comparison of Ref. \cite{Damour:2011fu} introduced  by the presence of the Schott contribution to the energy (especially during the late inspiral and the plunge). This might allow one to refine the conclusions of Ref. \cite{Damour:2011fu} and to extract some information about the exact form of the EOB Hamiltonian.

\subsection{About the radial component of radiation-reaction}

When Buonanno and Damour \cite{Buonanno:2000ef} incorporated radiation-reaction effects in the EOB formalism, they suggested that it is possible to use the radiative gauge freedom to put the radiation-reaction force in the simplified form
\begin{eqnarray}
\label{b1}
{\mathcal F}_r&=&0\\
\label{b2}
{\mathcal F}_\phi&=&-\Phi_J\,.
\end{eqnarray}
For instance, at the Newtonian order they argued that the choice
\beq
\label{b3}
\bar \alpha_{\rm BD}\equiv \alpha_{\rm IW}\equiv \beta_2{}_{\rm GII}=-\frac{10}{3}
\eeq
of one of the two free gauge parameters entering ${\mathcal F}_i^{\rm (Newt)}$ ensured the vanishing of the radial component ${\mathcal F}_r^{\rm (Newt)}$.
This statement is correct. However, this specific choice of $\bar \alpha_{\rm BD}\equiv \alpha_{\rm IW}\equiv \beta_2{}_{\rm GII}$ conflicts with the second requirement (\ref{b2}) that ${\mathcal F}_\phi$ be identified with minus the angular momentum flux.
Indeed, our results above (as well as the previous results of Iyer and Will) show that the simplifying requirement (\ref{b2}) actually determines the value of half of the free gauge parameters entering ${\mathcal F}_i$. More precisely, they determine the values of the parameters $\lambda^J_{I_1\ldots I_n}$ ($n=0,1,2$)
entering $J_{\rm (schott)}^{\rm (non\, min)}$, Eq. (\ref{Jshott_non_min}) (namely $\lambda_{I_1\ldots I_n}^J=0$).
One the other hand, as pointed out in Sec. IV above, the Newtonian order $J_{\rm (schott)}$-related parameter $\lambda_0^J$ 
happens to be proportional to the parameter $\bar \alpha_{\rm BD}=\alpha_{\rm IW}=\beta_2{}_{GII}$ which needed to take the nonzero value (\ref{b3}).
We see therefore that the choice (\ref{b3}) corresponds to a non-minimal (i.e., non vanishing) value for $J_{\rm (schott)}$, in conflict with the second, simplifying requirement (\ref{b2}).

In view of this result, we henceforth advocate to incorporate radiation-reaction in the EOB formalism by consistently enforcing the minimal choice
\beq
{\mathcal F}_\phi= -\Phi_J\,,
\eeq
corresponding to
\beq
J_{\rm (schott)}^{\rm EOB}({\mathbf x}(t),{\mathbf p}(t))=0\,,
\eeq
i.e., $\lambda^J_{I_1\ldots I_n}=0$.  This choice necessarily implies a {\it nonzero value} for ${\mathcal F}_r$. In particular, if we also require the second minimal choice,  
\beq
E_{\rm (schott)}^{\rm EOB}({\mathbf x}(t),{\mathbf p}(t))=E_{\rm (schott)}^{\rm EOB\, min}({\mathbf x}(t),{\mathbf p}(t))\,,
\eeq
we have seen above that ${\mathcal F}_r$ is completely determined, and has the form
\begin{widetext}
\begin{eqnarray}
\label{Fr_long}
\tilde {\mathcal F}_r&=&\frac{\nu}{c^5}\tilde p_r \frac{(GM)^2}{r_e^3} (R_A X_A^e+\epsilon^2 R_{AB}X_{AB}^e+\epsilon^4 R_{ABC}X_{ABC}^e)\,,
\end{eqnarray}
\end{widetext}
where the coefficients $R_{A_1\ldots A_n}$ are listed in Appendix \ref{Fr_and_phi_eob}. If we consider the case of a quasi-circular inspiral
we can neglect $X_2^e=\tilde Z_1=\tilde p_r^2$, and replace $X_1^e=\tilde p^2$ by the expression obtained by setting $X_2^e$ and $X_4^e$ to zero in the relation (\ref{X1_expanso}).

Specialized along circular orbits, relation (\ref{X1_expanso}) becomes
\beq
\label{X_1circ_orbit}
X_1^{\rm (circ)}=X_3^e +3\epsilon^2 X_{33}^e-  3(\nu-3) \epsilon^4 X_{333}^e \,.
\eeq

This leads to the following approximated form of $\tilde {\mathcal F}_r$
\begin{widetext}
\begin{eqnarray}
\label{Fr_inspiral}
\tilde {\mathcal F}^{\rm (inspiral)}_r &\simeq & \frac{\nu}{c^5} \frac{32}{3}\tilde p_r \frac{(GM)^3}{r_e^4} \left[1-\frac{1}{280} (1133+944\nu)\left(\frac{GM}{r_e}\right)\epsilon^2 \right.\nonumber\\
&& \left.
+\frac{1}{15120} (-175549+322623\nu+70794\nu^2)\left(\frac{GM}{r_e}\right)^2\epsilon^4 \right]\,.
\end{eqnarray}
\end{widetext}
It might be useful to record the value of the ratio between $\tilde {\mathcal F}_r$ and $\tilde {\mathcal F}_\phi$ during inspiral.
To this end, we first note that the inspiral value of $\tilde {\mathcal F}_\phi$ [obtained by replacing $X_2^e\to 0$ and $X_4^e\to 0$ in Eq. (\ref{Fphieob})]
reads
 
\begin{widetext}
\begin{eqnarray}
\label{Fphi_inspiral}
\tilde {\mathcal F}^{\rm (inspiral)}_\phi &\simeq& -\frac{\nu}{c^5} \frac{32}{5} \left(\frac{GM}{r_e}\right)^{7/2}\left[
1-\frac{1}{336} (588\nu+1247)\left(\frac{GM}{r_e}\right)\epsilon^2\right.\nonumber\\
&& \left. +\frac{1}{18144}(-89422+153369\nu+9072\nu^2)\left(\frac{GM}{r_e}\right)^2\epsilon^4\right]\,.
\end{eqnarray}
so that we have the ratio
\beq
\frac{\tilde {\mathcal F}^{\rm (inspiral)}_r}{\tilde {\mathcal F}^{\rm (inspiral)}_\phi}=-\frac53 GM \frac{\tilde p_r}{\tilde p_\phi}  
\left[1+\left(-\frac{227}{140}\nu+\frac{1957}{1680}  \right)\left(\frac{GM}{r_e}\right)\epsilon^2 +
\left(\frac{753}{560} \nu^2 +\frac{165703}{70560}\nu-\frac{25672541}{5080320} \right)\left(\frac{GM}{r_e}\right)^2\epsilon^4
 \right]\,.
\eeq
\end{widetext}

This result is consistent with Eqs. (3.14), (3.18) of \cite{Buonanno:2000ef} with the value $\bar \alpha_{BD}=0$ (i.e., $\lambda_0^J=0$).
We leave to future work a detailed study of the consequences of incorporating in the EOB formalism the non-vanishing value of $\tilde {\mathcal F}_r$ advocated here. The preliminary comparison performed at the end of Sec. V in Ref. \cite{Buonanno:2000ef} (between using $\tilde {\mathcal F}_r/\tilde {\mathcal F}_\phi=0$
and $\tilde {\mathcal F}_r/\tilde {\mathcal F}_\phi=\dot r/(r^2\dot \phi)$) indicates that the effect of the more consistent value of $\tilde {\mathcal F}_r/\tilde {\mathcal F}_\phi$ found here will be small. However, modern use of  EOB theory aims at a very high accuracy in the phasing, for which the new value of $\tilde {\mathcal F}_r$ will probably have a significant effect.
Let us also recall that along circular orbits, one finds  (at 2PN order),
using  $X_1^{\rm (circ)} = {\tilde p_\phi^2}/{r^2}$ and Eq. (\ref{X_1circ_orbit}), 
\begin{widetext}
\beq
\tilde p_\phi=\sqrt{GMr_e}\left[1+\frac32 \frac{GM}{r_e}\epsilon^2 -\frac38 (4\nu-9)\left(\frac{GM}{r_e}\right)^2 \epsilon^4 \right]
\eeq
and hence, for $\Omega^{\rm (circ)} =\partial {\mathcal H}_{\rm (eob)}/\partial p_\phi |_{\rm circ}$,
\beq
\frac{GM \Omega ^{\rm (circ)}}{c^3}=\left( \frac{GM}{c^2 r_e}\right)^{3/2}\left[ 1+\frac{\nu}{2}  \frac{GM}{r_e} \epsilon^2+ \frac38 (\nu-5) \left( \frac{GM}{r_e}\right)^2\epsilon^4 \right]\,.
\eeq
The latter equation implies the following expression for the  dimensionless frequency parameter  $x$, i.e.,
\beq
x\equiv\left(\frac{GM \Omega^{\rm (circ)}}{c^3} \right)^{2/3}=\left( \frac{GM}{c^2 r_e}\right)\left[ 1+\frac{\nu}{3} \frac{GM}{c^2 r_e} +\frac{\nu}{36}(-45+8\nu)\left( \frac{GM}{c^2r_e}\right)^2  
 \right]\,;
\eeq
inverting (perturbatively) this relation yields
\beq
 \frac{GM}{c^2 r_e}= x-\frac13 \nu x^2 +\frac54 \nu x^3\,,
\eeq
so that, in terms of $x$ we have
\begin{eqnarray}
\tilde E_{\rm (schott)}^{\rm (inspiral)} &=&\frac{16}{5}\nu x^3 \tilde p_r  \left[1+\left(-\frac{65}{21}\nu -\frac{269}{56} \right)x+
\left(\frac{7769}{1512}\nu^2 +\frac{7543}{336}\nu +\frac{409}{378} \right) x^2 \right]\, \nonumber\\
\tilde {\mathcal F}^{\rm (inspiral)}_r&=& \frac{32}{3}\frac{c^3}{GM}\nu x^4 \tilde p_r\left[1+\left(-\frac{494}{105}\nu-\frac{1133}{280}  \right)x
+\left(\frac{3071}{280}\nu^2 +\frac{55577}{1680}\nu -\frac{175549}{15120}\right) x^2 \right]\, \nonumber\\
\tilde {\mathcal F}^{\rm (inspiral)}_\phi &=& -\frac{32}{5}c^2 \nu x^{7/2}\left[1+\left(-\frac{35}{12}\nu -\frac{1247}{336}\right) x+\left(\frac{65}{18}\nu^2 +\frac{9271}{504}\nu -\frac{44711}{9072}\right) x^2 \right]\,.
\end{eqnarray}
The latter expression of  $\tilde {\mathcal F}^{\rm (inspiral)}_\phi $ as a function of the frequency parameter $x$ agrees with well-known previously derived results (see, e.g.,  Eq. (4.18) in \cite{Blanchet:1995fg}).

\end{widetext}

\subsection{Hyperbolic orbits: conservative aspects}
Up to now, the EOB formalism has been applied only to the description of radiation-reaction driven quasi-circular orbits, because these are the orbits of greatest relevance for the current network of ground based gravitational wave detectors. However, we anticipate that it will be useful to apply the EOB approach to other orbits, such as elliptic orbits, but also hyperbolic ones.
 It is now possible to do so because we have provided above a description of radiation-reaction along general motions. Here, we shall consider the case of hyperbolic motions, and focus on the effect of radiation-reaction on the angle of scattering of a gravitationally interacting binary system (considered in the center of mass system).

Before taking into account the additional effects of the radiation-reaction force ${\mathcal F}_i$, let us consider the conservative dynamics of hyperbolic encounters (at the 2PN accuracy). We recall that, at the 2PN accuracy, the relative motion in the orbital plane, $r(t),\phi(t)$ is described
(in {\it any} PN gauge; harmonic, ADM or EOB) by equations of the form \cite{Damour:1983tz,Damourtorino1985,Damour:1988mr,schaferwex}
\begin{eqnarray}
\label{c1}
\left(\frac{d\hat r}{d \hat t}  \right)^2&=&2\tilde E' +\frac{2'}{\hat r}-\frac{(j^{2})'}{\hat r^2} \nonumber\\
&& + \epsilon^2 \frac{R_3}{\hat r^3}+\epsilon^4 \frac{R_4}{\hat r^4}+\epsilon^4 \frac{R_5}{\hat r^5}\\
\label{c2}
\hat r^2 \frac{d\phi}{d\hat t}&=&j'\left(1+\epsilon^2 \frac{G_1}{\hat r}+\epsilon^4 \frac{G_2}{\hat r^2}+  \epsilon^4 \frac{G_3}{\hat r^3}\right)\,.
\end{eqnarray}
Here we have used the scaled variables $(\hat r=r/(GM)), \hat t=t/(GM))$, and the prime on any quantity $Q$ denotes a multiplicative modification
by higher PN terms of the type $Q'=Q(1+q_1\epsilon^2+q_2\epsilon^4)$, where $q_1\epsilon^2$, $q_2\epsilon^4$ (as well as the coefficients $R_p\epsilon^q$,
$G_p\epsilon^q$ above) are polynomials (with $\nu$-dependent coefficients) in the dimensionless quantities $\tilde E/c^2$ and $1/(jc)^2$. For instance, at the 1PN accuracy, and in harmonic coordinates \cite{DDI}
\begin{eqnarray}
2\tilde E' &=& 2\tilde E \left(1+\frac32 (3\nu-1)\frac{\tilde E}{c^2}  +O\left(\frac{1}{c^4} \right)\right)\nonumber\\
2'&=& 2\left(1+(7\nu-6)\frac{\tilde E}{c^2}  +O\left(\frac{1}{c^4} \right)  \right)\nonumber\\
(j^{2})'&=&j^2 \left( 1+2(3\nu-1)\frac{\tilde E}{c^2} -(5\nu-10)\frac{1}{(cj)^2} \right. \nonumber\\
&& \left. +O\left(\frac{1}{c^4} \right)  \right) \nonumber\\
j'&=& j  \left( 1+(3\nu-1)\frac{\tilde E}{c^2}+O\left(\frac{1}{c^4} \right)  \right)\,.
\end{eqnarray}
Note that $(j^{2})'$ is not the square of $j'$.
Many previous investigations \cite{Damour:1983tz,DDI,Damour:1988mr,schaferwex} were interested in describing the motion as a function of time. Here, we shall instead focus on the shape of the orbit, i.e., $\hat r$ as a function of $\phi$. This is obtained by eliminating $d\hat t$ between Eqs. (\ref{c1}) and(\ref{c2}). 
Introducing the dimensionless variable
\beq
\hat u= \frac{(j')^2}{\hat r}
\eeq
leads to a first-order differential equation for $\hat u(\phi)$ of the form
\begin{widetext}
\beq
\left( \frac{d\hat u}{d\phi}\right)^2= 2\tilde E' (j')^2 + 2' \hat u -1' \hat u^2 +\epsilon^2 \hat U_3 \hat u^3+\epsilon^4 \hat U_4 \hat u^4+
\epsilon^4 \hat U_5 \hat u^5
\eeq
\end{widetext}
where all coefficients $(\tilde E'$, $(j')^2$, $2'$, $1'$, $\epsilon^2 \hat U_3$, $\epsilon^4 \hat U_4$, $\epsilon^4 \hat U_5)$ are dimensionless. One can then reduce the above equation to a Newtonian-looking equation by a suitable change of (inverse) radial coordinate. Indeed, by appropriately choosing  the dimensionless coefficients
$\epsilon^2 \hat c_1,\epsilon^4 \hat c_2, \epsilon^4 \hat c_3$ in
\beq
\label{c3}
\bar u=\hat u (1+\epsilon^2 \hat c_1\hat u+\epsilon^4 \hat c_2 \hat u^2+ \epsilon^4 \hat c_3\hat u^3)
\eeq
one can get (modulo 3PN terms) an equation for $\bar u(\phi)$ of the simple form
\beq
\label{c4}
\left( \frac{d\bar u}{d\phi}\right)^2= 2(\tilde E j^2)''+2''\bar  u -1'' \bar u^2\,,
\eeq
where the double prime indicates further multiplicative modifications by higher-PN terms of the usual coefficients entering the Newtonian-order equation for
$u^{(N)}\equiv j^2/r$, namely
\beq
\left( \frac{d  u^{(N)}}{d\phi}\right)^2= 2 \tilde E j^2 +2  u^{(N)} -(u^{(N)})^2\,.
\eeq
The general solution of the latter (Newtonian-order) equation is the well known polar equation of a conic,
\beq
u^{(N)}(\phi)=1+e^{(N)}\cos \phi
\eeq
with $e^{(N)}=\sqrt{1+2\tilde E j^2}$. By contrast, the general solution of the modified Eq. (\ref{c4}) will be of the form
\beq
\label{c5}
\bar u (\phi) =C \left(1+\bar e \cos \frac{\phi-\phi_0}{K} \right)\,,
\eeq
where $\phi_0$ is an arbitrary integration constant and where $\bar e$, $C$ and $K$ are functions of $\tilde Ej^2$, $\tilde E/c^2$ and $1/(cj)^2$ which, respectively, reduce to $\sqrt{1+2\tilde Ej^2}$, $1$ and $1$ when $1/c^2\to 0$. Note that the quantity $K$ measures the periastron advance
\beq
K\equiv \frac{\Phi}{2\pi}\equiv 1+k\,,
\eeq
where $\Phi$ denotes the period of $\phi$ (i.e., $u(\phi+\Phi)=u(\phi)$ in the elliptic case; see below the definition of $\Phi$ in the hyperbolic case), and where $k$ is the usual notation for the relativistic contribution to periastron advance. It is given at 2PN by  \cite{Damour:1988mr}
\begin{eqnarray}
\label{kappapiccolo}
k (\tilde E, j)&=&\frac{3}{(cj)^2}\left[1+\left(\frac52 -\nu \right)\frac{\tilde E}{c^2} +\right. \nonumber\\
&& \left. \left(\frac{35}{4}-\frac52 \nu \right)\frac{1}{(jc)^2}+O\left(\frac{1}{c^4} \right)  \right]
\end{eqnarray}
[See Ref. \cite{Damour:1999cr} for the 3PN accurate value of $k$]. Here, we work with the {\it analytic continuation} (in $\tilde E$) of the function $k(\tilde E,j)$ from the elliptic-like case (where $\tilde E<0$) to the hyperbolic-like one ($\tilde E>0$). Note that we can further simplify the result (\ref{c5}) by modifying the leading-order coefficient 1 in the parenthesis appearing on the right hand side of Eq. (\ref{c3}) so as to absorb the coefficient $C=1+O(c^{-2})$ in a rescaling of $\bar u$. In other words, there exist coefficients $1'=1+O(\tilde E/c^2)+O(1/(jc)^2)$, and $\epsilon^2 \bar c_1$, $\epsilon^4 \bar c_2$, $\epsilon^4\bar c_3$ such that the polar equation $\hat r(\phi)$ of the orbit takes (at 2PN order) the form
\begin{widetext}
\beq
\label{c6}
\frac{j^2}{\hat r}\left(1'+ \epsilon^2 \bar c_1 \frac{j^2}{\hat r}+\epsilon^4 \bar c_2 \left(\frac{j^2}{\hat r}\right)^2+\epsilon^4\bar c_3 \left(\frac{j^2}{\hat r}\right)^3 \right)=1+\bar e \cos \frac{\phi-\phi_0}{K}\,.
\eeq
\end{widetext}
This form is valid in any PN gauge (harmonic, ADM or EOB). We will give below the explicit
values of its coefficients in the EOB case.
In this form the two coefficients, $\bar  e$ and $K$ entering the rhs acquire a gauge-invariant meaning. This is well known for the periastron advance parameter $K$ (when it is considered for the elliptic-like case), but this is also true (when considering asymptotically flat gauges) for the \lq\lq eccentricity" $\bar e$
(when  considering  the hyperbolic-like case). Indeed, when considering hyperbolic orbits the lhs will vanish both  in the infinite past (incoming state, $\hat r \to + \infty$) and in the infinite future (outgoing state, $\hat r \to + \infty$) so that (choosing the integration constant $\phi_0=0$; location of the periastron) $\phi$ will evolve from $\phi_-$ in the infinite past to $\phi_+$ in the infinite future, where $\phi_-(=-\phi_+)$ and $\phi_+$ are the two solutions of
\beq
1+\bar e \cos \frac{\phi}{K}=0\,,
\eeq
i.e. (we are in the hyperbolic case where $\bar e>1$)
\beq
\phi_\pm =\pm K {\rm arccos}\left(-\frac{1}{\bar e}  \right).
\eeq
The (center of mass) scattering angle, $\chi$ (taken with a positive sign) is related to $\phi_\pm$ via
\beq
\chi +\pi=\phi_+-\phi_-\equiv \Delta \phi
\eeq
so that we can write $\chi$ in terms of $K$ and $\bar e$ according to
\beq
\label{c7}
\chi +\pi=\Delta \phi=2K {\rm arccos}\left(-\frac{1}{\bar e}  \right)\,.
\eeq

Both the scattering angle $\chi$ and the periastron precession parameter $K$ are gauge-invariant physical quantities 
that can be expressed as functions of the two basic gauge-invariant dynamical parameters $\tilde E$ and $j^2$.
We see therefore from Eq. (\ref{c7}) that $\bar e$ can also be considered as a gauge-invariant quantity, and can be, in principle, expressed as a function of 
$\tilde E$ and $j^2$. [We shall give below some explicit integral definitions of the functions $\chi(\tilde E,j)$ and $K(\tilde E, j)$ from EOB theory.]

At the 1PN accuracy, the invariant eccentricity $\bar e$ coincides with the eccentricity denoted as $e_\theta$ in \cite{DDI} (see Eq. (5.7) there, which is of the form (\ref{c6})). The expression of $\bar e^2$ in terms of $\tilde E$ and $j^2$ is given by
(see Eq. (4.13) in \cite{DDI})
\begin{eqnarray}
\bar e^2&=& e_\theta^2=1+2\tilde Ej^2 \left[ 1+\left(\frac{\nu}{2}-\frac{15}{2}\right)\frac{\tilde E}{c^2} \right]\times \nonumber\\
&& \qquad \times \left(1-\frac{6}{(cj)^2}\right)+O\left(\frac{1}{c^4}\right)\,.
\end{eqnarray}
We have determined the extension of this relation to the 2PN accuracy by using results in the literature on the \lq\lq quasi-Keplerian" parametrization of the 2PN motion \cite{Damour:1988mr,schaferwex}, namely
\begin{widetext}
\begin{eqnarray}
n(t-t_0)&=& u-e_t \sin u +\frac{f_t}{c^4}\sin v+\frac{g_t}{c^4}(v-u)\nonumber\\
r&=& a_r(1-e_r\cos u)\nonumber \\
\frac{\phi-\phi_0}{K}&=& v+\frac{f_\phi}{c^4}\sin 2v+\frac{g_\phi}{c^4}\sin 3 v
\end{eqnarray}
\end{widetext}
where
\beq
v=2{\rm arctan}\left[ \left( \frac{1+e_\phi}{1-e_\phi} \right)^{1/2}\tan \frac{u}{2} \right]\,.
\eeq
Here the ``eccentric anomaly'' $u$ (and its analytic continuation $\bar u$ mentioned below) should
not be confused with the gravitational potential variables $u= GM/r_e$, $\bar u$, used above.

The form written here corresponds to elliptic-like motions ($\tilde E<0$). However, similarly to the Newtonian case (which is recalled in Appendix \ref{newtonian}) the corresponding parametrization of hyperbolic-like motion is obtained by the simple analytical continuation
\beq
u=i\bar u\,,
\eeq
which accompanies the continuation of $\tilde E$ from negative to positive values, as well as the continuation of the various eccentricities $e_t$, $e_r$, $e_\phi$ from $e_i<1$ to $e_i>1$.
In addition, $n^2\sim GM/a_r^3$ and $a_r\sim -GM/(2\tilde E)$ are continued from positive to negative values. In this continuation the angular variable $v$ remains real. The radial motion equation becomes $r=a_r (1-e_r \cosh \bar u)$, so that the outgoing and incoming states are described by $\bar u \to \pm \infty$.

This corresponds to finite values of the real angle $v$ given by
\beq
v_\pm =\pm 2 {\rm arctan} \left( \frac{e_\phi+1}{e_\phi-1} \right)^{1/2}\,,
\eeq
so that (choosing $\phi_0=0$ as above)
\beq
\frac{\phi_\pm }{K}= v_\pm +\frac{f_\phi}{c^4}\sin 2v_\pm+\frac{g_\phi}{c^4}\sin 3 v_\pm\,.
\eeq
Taking the cosine of this result, and using the 2PN accurate expressions of $e_\phi$, $f_\phi$, $g_\phi$ as functions of $\tilde E$ and $j^2$ \cite{schaferwex} then leads to the following explicit 2PN-accurate expression for $\bar e (\tilde E, j)$
\beq \label{e2PN}
\bar e^2 (\tilde E, j)=1+2\tilde Ej^2 [1+\epsilon^2B_1+\epsilon^4 B_2+O(\epsilon^6)]
\eeq
where
\begin{eqnarray}
B_1&=& \frac12 (\nu-15)\tilde E-\frac{6}{j^2}\nonumber\\
B_2&=& 5 \left(8-\frac32 \nu  \right)\tilde E^2+(23\nu-4)\frac{\tilde E}{j^2}\nonumber\\
&&+\frac32 (10\nu-17) \frac1{j^4} \,.
\end{eqnarray}

This leads to several possible ways of computing the scattering angle $\chi$ as a function of  $\tilde E$ and $j^2$, at the 2PN-accuracy. A first form would be obtained from Eq. (\ref{c7}) without any re-expansion, i.e.,
\beq
\label{c8}
\chi(\tilde E,j)+\pi=2K(\tilde E,j){\rm arcos}\left( -\frac{1}{\sqrt{\bar e^2 (\tilde E, j)}} \right)
\eeq
where $K(\tilde E, j)=1+k (\tilde E, j)$ is written in Eq. (\ref{kappapiccolo}) above, and $\bar e^2 (\tilde E, j)$ is the polynomial in  $\tilde E$ and $j^2$ written above.

Alternatively, one might consider re-expanding the result (\ref{c8}) as a straightforward expansion in $1/c^2$.
This leads to
\begin{widetext}
\beq
\frac12 \chi(\tilde E, j) =  \arctan \left( \frac{1}{\sqrt{2\tilde E j^2}} \right)+\epsilon^2 A_1+\epsilon^4 A_2 \,,
\eeq
where
\begin{eqnarray}
A_1&=&  \frac{3}{j^2}\phi_+^0(\tilde E, j)  -\frac{\sqrt{2\tilde E j^2}}{4j^2 (1+2\tilde E j^2)}[(\nu-15)\tilde E j^2 -12]\nonumber\\
A_2&=& \phi_+^0(\tilde E, j)  A_{2a} +A_{2b} \,,
\end{eqnarray}
\begin{eqnarray}
A_{2a}&=& \frac{3}{j^2}\left[-\frac{(5-2\nu)}{2}\tilde E+\frac{5(7-2\nu)}{4j^2}  \right]\nonumber\\
A_{2b}&=&  \frac{\sqrt{2\tilde E j^2}}{32 (1+2\tilde Ej^2)^2 j^4} \left[  2(3\nu^2+30 \nu +35)\tilde E^3 j^6 +(\nu^2-838 \nu +2593)\tilde E^2 j^4\right. \nonumber\\
&& \left. -32 (28 \nu-95)\tilde E  j^2-240 \nu +840\right] \, ,
\end{eqnarray}
and
\end{widetext}
\beq \label{phi0+}
\phi_+^0(\tilde E, j)={\rm arccos}\left(-\frac{1}{\sqrt{1+2\tilde E j^2}}  \right)
\eeq
One can also consider the PN expansion of $\tan \frac{\chi}{2}$, namely
\begin{widetext}
\beq
\tan \frac{\chi (\tilde E, j)}{2}=\frac{1}{\sqrt{2\tilde E j^2}}\left\{ 1+\frac{1+2\tilde E j^2}{\sqrt{2\tilde E j^2}}\left[\epsilon^2 A_1+\epsilon^4\left(A_2+\frac{1}{\sqrt{2\tilde E j^2}}A_1^2 \right)\right]\right\}\,.
\eeq
\end{widetext}

Beware that the straightforward PN expansions of $k (\tilde E, j)$ and $\chi(\tilde E,j)$ are badly convergent because of the presence of a singularity (where $k (\tilde E, j) \to \infty $ and $\chi(\tilde E,j) \to \infty$)  along the sequence of {\it unstable} circular orbits. Let us recall that, in the $(\tilde E , j)$ plane the sequence of circular orbits is defined by parametric equations of the type (when $\nu \to 0$)
$$
\frac{\tilde E (x)}{c^2} = \frac{1-2x}{\sqrt{1-3x}} - 1 + O(\nu)
$$
$$
cj(x) = \frac{1}{\sqrt{x(1-3x)}} + O(\nu) \, .
$$

The orbits we consider here (either elliptic-like or hyperbolic-like) lie between the two branches defined by the parametric equations above: the lower branch of stable circular orbits (corresponding to $0 \leq x \leq x_{\rm LSO} (\nu)$, with $x_{\rm LSO} (\nu) = \frac16 +O(\nu)$), and the upper branch of unstable circular orbits ($x_{\rm LSO} (\nu) \leq x < x_{\rm LR} (\nu)$, with $x_{\rm LR} (\nu) = \frac13 + O(\nu)$). [Both branches meet at a cusp corresponding to the LSO.] Many of the functions of $\tilde E$ and $j$ that we consider here (and notably $k (\tilde E, j)$ and $\chi(\tilde E,j)$) become singular on the upper branch. It might then be better to work with
the PN expansions of related variables that are regular on the upper branch, e.g. related
functions that {\it smoothly vanish} there instead of blowing up. [When considering the 
zero-eccentricity limit, this strategy was used in Refs. \cite{Damour:1999cr,Damour:2000we}, 
which replaced the singular function  $K(\tilde E, j)$ by the smoothly vanishing \cite{Damour:1988mr} function $K^{-4}(\tilde E, j)$.]

Let us finally note that the EOB formalism gives an exact integral form for the scattering angle. Indeed, applying the Hamilton-Jacobi method to the EOB Hamiltonian leads to a separated action of the type
\begin{eqnarray}
S_{\rm (eob)}(t,r,\phi; E, p_\phi)&=&-Et+p_\phi \phi \nonumber \\
&& + \int dr p_r(r; E, p_\phi)\,,
\end{eqnarray}
where $p_r(r; E, p_\phi)$ is obtained by solving the equation ${\mathcal H}_{\rm (eob)}=E$, or, in terms of the $\mu$-reduced effective energy $\tilde {\mathcal H}_{\rm (eff)}$ (using also $\hat r=r/(GM)$) 
\beq
\frac{\tilde {\mathcal H}_{\rm (eff)}^2}{c^4}= A(\hat r)\left(1+\frac{j^2}{c^2\hat r^2}+\frac{\tilde p_r^2}{c^2B(\hat r)} \right)\,.
\eeq
This yields
\beq
\frac{\tilde p_r}{c}=\pm \sqrt{\frac{B(\hat r)}{A(\hat r)}}\sqrt{\frac{\tilde {\mathcal H}_{\rm (eff)}^2}{c^4}-A(\hat r) \left(1+\frac{j^2}{c^2 \hat r^2} \right)}\,.
\eeq
The orbit $\phi(\hat r)$ is then obtained from using $\partial S_{\rm (eob)}/\partial p_\phi=\phi_0=$constant.
Setting $\phi_0=0$ yields
\beq
\phi(\hat r)=-(GM)\frac{\partial}{\partial \tilde p_\phi}\int d\hat r \tilde p_r =\pm \int d\hat r {\mathcal R}(\hat r; j,\tilde{\mathcal H}_{\rm (eff)})\,,
\eeq
where
\beq
{\mathcal R}(\hat r; j,\tilde{\mathcal H}_{\rm (eff)}) =\frac{j}{c\hat r^2}\frac{\sqrt{A(\hat r)B(\hat r)}}{\sqrt{\frac{\tilde{\mathcal H}_{\rm (eff)}^2}{c^4} -A(\hat r)\left( 1+\frac{j^2}{c^2 \hat r^2}\right)}}\,.
\eeq
It is useful to re-write this result in terms of the inverse radius $u=1/\hat r =GM/r$.  

Introducing
\beq
U(u; j,\tilde{\mathcal H}_{\rm (eff)})=j\frac{\sqrt{A(u)B(u)}}{\sqrt{\frac{\tilde{\mathcal H}_{\rm (eff)}^2}{c^2} -A(u)\left(c^2+ j^2u^2\right)}}
\eeq
we have
\beq
\phi(u)=\pm \int du U(u; j,\tilde{\mathcal H}_{\rm (eff)})\,.
\eeq
The function $U(u)$ is defined as a real function in the classical domain where the function appearing under the square root in its denominator, say ${\mathcal D}(u; j,\tilde{\mathcal H}_{\rm (eff)})\equiv {\mathcal D}(u)$
\beq
{\mathcal D}(u)= \frac{\tilde{\mathcal H}_{\rm (eff)}^2}{c^2} -A(u)\left(c^2+ j^2u^2\right)
\eeq
is positive.
In the elliptic-like case $(\tilde E<0)$ this is the case in an interval of the form $0<u_{\rm min}(\tilde E, j)\le u\le u_{\rm max}(\tilde E, j)$,
where $u_{\rm min}$ and $u_{\rm max}$ are two positive roots of ${\mathcal D}(u)$. 
In the Newtonian approximation ${\mathcal D}(u)^{\rm (Newt)}= 2\tilde E  +2u- j^2u^2$, these two positive roots are
\begin{eqnarray}
u_{\rm min}^{\rm (Newt)}(\tilde E, j)&=& \frac{1-\sqrt{1+2\tilde Ej^2}}{ j^2}\nonumber\\
u_{\rm max}^{\rm (Newt)}(\tilde E, j)&=& \frac{1+\sqrt{1+2\tilde Ej^2}}{ j^2}\,.
\end{eqnarray}
Then the angular period $\Phi=2\pi K$ is given by an integral over the interval $[u_{\rm min},u_{\rm max}]$, namely
\begin{eqnarray}
\label{eq_per_K}
\pi K(\tilde E, j)&=&\frac{\Phi (\tilde E, j)}{2}\nonumber\\
&=&\int_{u_{\rm min}(\tilde E, j)}^{u_{\rm max}(\tilde E, j)}du U(u;j,\tilde{\mathcal H}_{\rm (eff)})\,.
\end{eqnarray}
When one continues $\tilde E$ from negative to positive values, the analytic continuation of $u_{\rm min}(\tilde E, j)$ stays real, but becomes {\it negative}. However, nothing wrong happens to the integrand, and one can still consider that the real integral above defines $K(\tilde E, j)$ in the hyperbolic-like case ($\tilde E>0$, i.e., $\tilde {\mathcal H}_{\rm (eff)}/c^2>1$). [In terms of the usual radial variable $\hat r=1/u$ this means that one is taking an integral which goes {\it beyond} $\hat r=+\infty$ to formally extend to negative values of the variable $\hat r$.]

By contrast, the scattering angle $\chi$ is directly defined in the hyperbolic-like case $(\tilde E >0)$ by an integral over the interval $0\le u \le u_{\rm max}(\tilde E, j)$, namely
\begin{eqnarray}
\label{eq_per_chi}
&&\frac{\chi(\tilde E, j)}{2}+\frac{\pi}{2}=\frac{\Delta \phi (\tilde E, j)}{2}\nonumber\\
&&\qquad =\int_0^{u_{\rm max}(\tilde E, j)}du U(u;j,\tilde{\mathcal H}_{\rm (eff)})\,.
\end{eqnarray}
Here the interval $0\le u \le u_{\rm max}(\tilde E, j)$ corresponds to the radial interval $\hat r_{\rm min}\le \hat r\le +\infty$, where $\hat r_{\rm min}=1/u_{\rm max}$ is the minimum of $\hat r$ (periastron). 
By comparing Eq. (\ref{eq_per_chi}) with Eq. (\ref{eq_per_K}) we see that while $K$ is given by a {\it complete} integral (i.e., a {\it period} integral, between two successive roots of $U(u)$), $\chi$ is given by an {\it incomplete} version of the complete integral (going between a root and $u=0$, which is an intermediate point). This explains why the PN expansion of $\chi(\tilde E, j)$ has a more complicated analytical structure as a function of $\tilde E$ and $j^2$ [involving ${\rm arctan}(1/\sqrt{2\tilde Ej^2})$], than $K(\tilde E, j)$.

Let us finally indicate how one can rather easily compute the explicit quasi-conical equation (see Eq. (\ref{c6})) of the orbit in EOB coordinates. Let us consider the squared differential of the polar angle, $d\phi^2 = U^2 (u) \, du^2$. We wish to transform it, by a (2PN-accurate) change of $u$ variable of the form
\begin{equation}
\label{nn1}
u = \overline u + \epsilon^2 {\sf a} \, \overline u^2 + \epsilon^4 {\sf b} \, \overline u^3 + \epsilon^4 {\sf c} \, \overline u^4 + O(\epsilon^6) \, ,
\end{equation}
so that it simplifies (modulo $O(\epsilon^6)$) to a form involving a quadratic polynomial in $\overline u$ as denominator, i.e.
\begin{eqnarray}
\label{nn2}
d\phi^2 &=& j^2 \, \frac{D(u) \, (du)^2}{(c^2 + \varepsilon) - A(u)(c^2+j^2u^2)} \nonumber\\
&=& j^2 \, \frac{(d\overline u)^2}{\varepsilon + 2 \, \alpha \, \overline u - j^2 \beta \, \overline u^2} \, .
\end{eqnarray}
Here, $D(u) \equiv A(u) \, B(u) = 1 - \epsilon^2 \, 6 \, \nu \, u$, and we introduced the new energy measure  $\varepsilon$ (not to be confused with the PN ordering parameter $\epsilon \equiv 1/c$)
\begin{eqnarray}
\label{nn3}
\varepsilon &\equiv& \frac{\widetilde{\mathcal H}_{({\rm eff})}^2}{c^2} - c^2 = c^2 \left( 1 + \frac{\tilde E}{c^2} + \frac{1}{c^4} \, \frac{\nu}{2} \, \tilde E^2 \right)^2 - c^2 \nonumber \\
&= &2 \left( \tilde E + \frac{\nu}{2c^2} \, \tilde E^2 \right) + \frac{1}{c^2} \left( \tilde E + \frac{\nu}{2c^2} \, \tilde E^2 \right)^2 \, .
\end{eqnarray}
It is easy to check that the choice of coefficients
\begin{eqnarray}
\label{nn4}
{\sf a} &= &-1-\frac{1}{4 \, j^2 \, c^2} \, (17 - 10 \nu) + O \left( \frac{1}{c^4} \right) \nonumber \\
{\sf b} &= &\frac34 \, (1 + 2\nu) + O \left( \frac{1}{c^2} \right) \nonumber \\
{\sf c} &= &0 + O \left( \frac{1}{c^2} \right)
\end{eqnarray}
in Eq.~(\ref{nn1}) does yield the simple $\overline u$-form indicated in the second Eq.~(\ref{nn2}). The coefficients $\alpha$ and $\beta$ entering the quadratic $\overline u$-denominator 
$\varepsilon + 2 \, \alpha \, \overline u - j^2 \beta \, \overline u^2$ are then found to be (at 2PN accuracy) the following functions of $\tilde E$ and $j$:
$$
\alpha \, (\tilde E , j) = 1 - \frac{2}{c^2} \, {\sf a} \, \varepsilon + O \left( \frac{1}{c^6} \right) \, ,
$$
\begin{equation}
\label{nn5}
\beta \, (\tilde E , j) = \frac{1}{K^2} = \frac{1}{(1+k)^2} \, .
\end{equation}
[The latter result for $\beta$, that we explicitly checked at 2PN, must hold to all PN orders.] 

The integration of Eq.~(\ref{nn2}) then yields
$$
\overline u = \langle \overline u \rangle \left( 1 + \overline e \cos \frac{\phi}{K} \right)
$$
where, denoting by $\overline u_1$ and $\overline u_2$ ($\overline u_1 \leq \overline u_2$) the two roots of the quadratic $\overline u$-denominator, 
\beq
\varepsilon + 2 \, \alpha \, \overline u - j^2 \beta \, \overline u^2 \equiv j^2 \beta (\overline u -\overline u_1 )(\overline u_2 -\overline u )  \, ,
\eeq
we have
$$
\langle \overline u \rangle = \frac{\overline u_1 + \overline u_2}{2} \, , \quad \overline e = \frac{\overline u_2 - \overline u_1}{\overline u_1 + \overline u_2} \, .
$$
This yields
$$
\langle \overline u \rangle = \frac{\alpha}{j^2 \beta} = \frac{\alpha \, K^2}{j^2}
$$
and
\begin{equation}
\label{nn6}
\overline e^2 - 1 = \varepsilon \, j^2 \, \frac{\beta}{\alpha^2} = \frac{\varepsilon \, j^2}{K^2 \, \alpha^2} \, .
\end{equation}
When inserting in Eq.~(\ref{nn6}), the expressions of $\varepsilon$ (Eq.~(\ref{nn3})), $K$ (Eq.~(\ref{kappapiccolo})), and $\alpha$ (Eq.~(\ref{nn5}) with the first equation (\ref{nn4})), one finds, after PN reexpanding $\overline e^2 (\tilde E , j)$ the same result as Eq.~(\ref{e2PN}) above.

\subsection{Hyperbolic orbits: radiative effects}

Having explained the various ways in which one can compute the scattering angle $\chi$ as a function of $\tilde E$ and $j^2$, in the {\it conservative} case, let us now discuss the modification of $\chi$ brought by radiation-reaction. We define the supplementary contribution $\chi^{\rm (RR)}$ to $\chi$ entailed by radiation-reaction by decomposing the total $\chi$ as
\beq
\chi^{\rm (tot)}(\tilde E_-, j_-)=\chi^{\rm (conserv)}(\tilde E_-, j_-)+\chi^{\rm (RR)}(\tilde E_-, j_-).
\eeq

Here $\chi^{\rm (conserv)}(\tilde E, j)$ is the function defined above in the conservative case and we have denoted by $\tilde E_-$ and $j_-$  the energy and the angular momentum of the {\it incoming} state (considered in the infinite past, $t\to -\infty$). We are going to prove the following simple result concerning
$\chi^{\rm (RR)}$. When working linearly in the radiation-reaction ${\mathcal F}_i$, i.e. modulo terms that are formally quadratic in ${\mathcal F}_i$, we can write
\begin{eqnarray}
\label{deltachi}
\chi^{\rm (RR)} &=& \frac12 \left( \chi^{\rm (conserv)}(\tilde E_+, j_+)-\chi^{\rm (conserv)}(\tilde E_-, j_-) \right)\nonumber\\
&=& \frac12  \left( \frac{\partial \chi^{\rm (conserv)}}{\partial \tilde E}\delta^{\rm (RR)} \tilde E + \frac{\partial \chi^{\rm (conserv)}}{\partial j}\delta^{\rm (RR)} j\right)\nonumber\\
\end{eqnarray}
where $\delta^{\rm (RR)} \tilde E$ and $\delta^{\rm (RR)} j$ are the integrated  losses of energy and angular momentum, radiated (between $t=-\infty$ and $t=+\infty$) at infinity in the form of the corresponding fluxes $\Phi_E$ and $\Phi_J$. Note that (still modulo terms $O({\mathcal F}^2)$) the result (\ref{deltachi}) means that the total scattering angle $\chi^{\rm (tot)}$, in presence of radiation-reaction, can be written as
\begin{eqnarray}
\label{deltachi2}
\chi^{\rm (tot)}(\tilde E_-, j_-)&=&\frac12 \left(\chi^{\rm (conserv)}(\tilde E_+, j_+)\right.\nonumber\\
&& \left. +  \chi^{\rm (conserv)}(\tilde E_-, j_-)\right).
\end{eqnarray}
Moreover, it can also be written (modulo $O({\mathcal F}^2)$) as
\beq
\label{deltachi3}
\chi^{\rm (tot)}(\tilde E_-, j_-)=\chi^{\rm (conserv)}(\tilde E_0, j_0)\,,
\eeq
where
\begin{eqnarray}
\tilde E_0 &=& \frac12 (\tilde E_++\tilde E_-)\nonumber\\
j_0 &=& \frac12 (j_++ j_-)
\end{eqnarray}
are the average values of $\tilde E$ and $j$ over the incoming and outgoing states.
As the radiation-reaction is of PN order ${\mathcal F}=O(1/c^5)$, the accuracy of the results stated above is modulo corrections of PN order $O(1/c^{10})$.

To give a proof of the above statements, one should use the generalized method of variation of constants used in Refs. \cite{Damour:1983tz,Damourtorino1985,Damour:2004bz}, which considers the perturbation of the 2PN accurate conservative dynamics by the radiation-reaction force.
Moreover, one should extend the treatment of these references from the elliptic-like case they consider, to the hyperbolic-like one we are interested in here. 
This can be done, and yields a straightforward proof of the relations above. Here, for the benefits of simplicity, we shall content ourselves with presenting the proof of these relations in a simplified case where the unperturbed dynamics is treated as being Newtonian, while the perturbing force ${\mathcal F}_i$ is considered at the fractional 2PN accuracy.
We shall, however, indicate the essential reason why the result still holds in the case where {\it both} the conservative dynamics and the radiation-reaction are treated more exactly, i.e. with a Hamiltonian of the type
\beq
{\mathcal H}^{\rm (conserv)}={\mathcal H}^{\rm (Newt)}+\frac{1}{c^2}{\mathcal H}^{\rm (1PN)}+\frac{1}{c^4}{\mathcal H}^{\rm (2PN)}\,,
\eeq
and a radiation-reaction of the type
\beq
{\mathcal F}={\mathcal F}^{\rm (Newt)}+\frac{1}{c^2}{\mathcal F}^{\rm (1PN)}+\frac{1}{c^4}{\mathcal F}^{\rm (2PN)}\,.
\eeq
When considering the simple case where the unperturbed dynamics is Newtonian, we can simplify the calculations of $\chi^{\rm (RR)}$ by making use of the famous 
Laplace(-Lagrange-Runge-Lenz) conserved vector.
Using scaled variables, $\hat r=r/(GM)$, $j=J/(GM\mu)$, $\tilde {\mathbf p}={\mathbf p}/\mu$ (and, henceforth, dropping both the carets and the tilde's for easing the notation) we have the Laplace vector
\beq
\label{c8}
{\mathbf A}(t)={\mathbf p}\times {\mathbf j}-{\mathbf n}
\eeq
where ${\mathbf j}={\mathbf r}\times {\mathbf p}$ and ${\mathbf n}={\mathbf r}/r$.
Its time derivative is proportional to the perturbing force $\tilde {\pmb {\mathcal F}}$ (henceforth we shall also drop the tilde on ${\pmb {\mathcal F}}$) and is given by
\beq
\frac{d {\mathbf A}}{dt}={\pmb {\mathcal F}}\times {\mathbf j}+{\mathbf p}\times ({\mathbf r}\times {\pmb {\mathcal F}})\,.
\eeq
If we write ${\pmb {\mathcal F}}$ in vectorial form, it has the structure
\beq
{\pmb {\mathcal F}}=\alpha ({\mathbf r},{\mathbf p})p_r {\mathbf n}+\beta  ({\mathbf r},{\mathbf p})  {\mathbf p}
\eeq
where the crucial information is that the coefficients $\alpha$ and $\beta$ (which should not be confused
with the quantities introduced in the previous subsection) are {\it time-even} scalars, i.e., combinations of our usual scalars ${\mathbf p}^2$, $p_r^2$ and $1/r$.
[This holds for the 2PN-accurate $\alpha$'s and $\beta$'s.] Inserting this structure in the time derivative of ${\mathbf A}$ yields
\beq
\frac{d {\mathbf A}}{dt}=\alpha p_r  {\mathbf n}\times {\mathbf j}+2\beta {\mathbf p}\times {\mathbf j} \,.
\eeq
Let us now decompose all vectors with respect to an orthonormal basis ${\mathbf e}_x$, ${\mathbf e}_y$, ${\mathbf e}_z$, with the $x$ direction along the apsidal line (i.e. with ${\mathbf e}_x$ a unit vector directed from the origin towards the periastron) and with ${\mathbf e}_z$ being along the angular momentum: ${\mathbf j}=j{\mathbf e}_z$. We have
\begin{eqnarray}
{\mathbf n}&=& \cos \phi {\mathbf e}_x+ \sin \phi {\mathbf e}_y\nonumber\\
{\mathbf p}&=&\frac{1}{j}\left[ -\sin \phi {\mathbf e}_x+ (\cos \phi +e) {\mathbf e}_y\right]\,,
\end{eqnarray}
so that the two components of $\dot {\mathbf A}=\dot A_x {\mathbf e}_x+\dot A_ y{\mathbf e}_y$ read
\begin{eqnarray}
\dot A_x &=& \alpha e \sin^2 \phi +2\beta (\cos \phi +e)\nonumber\\{}
\dot A_ y &=& -\sin \phi (\alpha e \cos \phi +2\beta)\,,
\end{eqnarray}
where we used the fact that
\beq
p_r={\mathbf n}\cdot {\mathbf p}=\frac{e}{j}\sin \phi\,.
\eeq

The crucial fact we wish to stress is that $\dot A_x$ is an even function of $\phi$, while $\dot A_y$ is an odd function of $\phi$. [Recall that the scalars $\alpha$ and $\beta$ are functions of ${\mathbf p}^2$, $p_r^2$ and $1/r$ and are therefore even functions of $\phi$.] Remember that we have chosen the origin $\phi_0$ of $\phi$ at $\phi_0=0$, so that these parity properties of the vector $\dot {\mathbf A}$ correspond to simple symmetry properties between the first half of the motion (between infinity and the periastron) and the second half (from the periastron back to infinity). When integrating over time to get (at order $O({\mathcal F})$) the total radiation-reaction-induced change of ${\mathbf A}$ between $-\infty$ and $+\infty$, we deduce (using the fact that $\dot \phi=j/r^2=(1+e\cos\phi)^2/j^3$ is even in $\phi$)
\beq
\delta^{\rm (RR)}{\mathbf A}={\mathbf A}_+-{\mathbf A}_-
\eeq 
will be directed along the $x$ axis. As the unperturbed ${\mathbf A}$ vector is simply
\beq
{\mathbf A}^{\rm (conserv)}=e{\mathbf e}_x\,,
\eeq
we conclude that the effect of radiation-reaction on ${\mathbf A}$ amounts to changing{\it only the magnitude} of the eccentricity $e$, without introducing any further angular rotation in the apsidal line. More precisely, as the magnitude of the perturbed ${\mathbf A}^2(t)$ is given (at any moment) by
\beq
{\mathbf A}^2(t)={\mathbf p}^2 {\mathbf j}^2+1-\frac{2}{r} {\mathbf j}^2\equiv 1+2 \tilde E(t)j^2(t)\,,
\eeq
where $\tilde E(t)$ and $j(t)$ are the instantaneous (Newtonian) values of the energy and angular momentum along the perturbed motion, we conclude that an incoming ${\mathbf A}$ vector at $t=-\infty$ of the form
\beq
{\mathbf A}(t=-\infty)\equiv {\mathbf A}_-= \sqrt{1+2\tilde E_-j_-^2}{\mathbf e}_x
\eeq
will end up, at $t=+\infty$ with the value
\beq
{\mathbf A}(t=+\infty)\equiv {\mathbf A}_+= \sqrt{1+2\tilde E_+j_+^2}{\mathbf e}_x\,.
\eeq
Let us now use these asymptotic results to compute the value of the scattering angle $\chi^{\rm (tot)}$, including the cumulated effect of radiation-reaction. This is done by considering the limits $t\to \pm \infty$ in the defining expression (\ref{c8}) of ${\mathbf A}(t)$. Asymptotically, we have
\beq
{\mathbf p}(t=\pm \infty)\equiv {\mathbf p}_\pm= \pm \sqrt{2\tilde E_+}{\mathbf n}_\pm\,.
\eeq

Let us replace any vector ${\mathbf V}=V_x {\mathbf e}_x +V_y {\mathbf e}_y$ in the orbital plane by the corresponding complex number
$V=V_x +iV_y$. In particular, the unit vector ${\mathbf n}(t)$ becomes the complex number $n(t)=e^{i\phi(t)}$. Its limiting values are
$n_\pm=e^{i\phi_\pm}$, where $\phi_+=\phi(t\to +\infty)$ and $\phi_-=\phi(t\to -\infty)$. It is then easy to find that the asymptotic values of
$A(t)=A_x(t) +iA_y(t)$ are given by
\begin{eqnarray}
A_-&=& (-1+i \sqrt{2\tilde E_-j_-^2})n_-\nonumber\\
A_+&=& -(1+i \sqrt{2\tilde E_+j_+^2})n_+\,.
\end{eqnarray}
If we then define $\chi_\pm$ (and $e_\pm$) by
\beq
\tan \frac{\chi_\pm}{2}=\frac{1}{\sqrt{2\tilde E_\pm j^2_\pm}}=\frac{1}{\sqrt{e_\pm^2-1}}
\eeq
we conclude that
\begin{eqnarray}
A_- &=& i e_- e^{i\frac{\chi_-}{2}}n_-=i e_- e^{i\frac{\chi_-}{2}}e^{i\phi_-}\nonumber\\
A_+ &=& -i e_+ e^{-i\frac{\chi_+}{2}}n_+=-i e_+ e^{-i\frac{\chi_+}{2}} e^{i\phi_+}\,.
\end{eqnarray}
Our previous result show that $A_+$ has the same argument as $A_-$. Therefore
\beq
\frac{\pi}{2}+\frac{\chi_-}{2}+\phi_-=-\frac{\pi}{2}-\frac{\chi_+}{2}+\phi_+
\eeq
so that the total scattering angle $\chi^{\rm (tot)}\equiv \phi_+-\phi_--\pi$ (including radiation-reaction) is simply given by
\begin{widetext}
\beq
\chi^{\rm (tot)}=\frac12(\chi_-+\chi_+)\equiv \frac12 \left[\chi^{\rm (conserv)}(\tilde E_-,j_-)+\chi^{\rm (conserv)}(\tilde E_+,j_+)\right]\,,
\eeq
\end{widetext}
which is the relation that we have indicated above.

Let us briefly indicate why this result extends to the case where the unperturbed, conservative dynamics is treated, say, at the 2PN accuracy. 
In that case one cannot use the Laplace vector because of periastron precession. Instead one can use the version of the method of variation of constants used in Refs. \cite{Damour:1983tz,Damourtorino1985,Damour:2004bz}, and adapt it to the hyperbolic case.
Then the crucial quantities which encode the effect of radiation-reaction on the scattering angle are the \lq\lq varying constants" $c_1(t)$, $c_2(t)$ and $c_\lambda(t)$ that enter the expression for $\phi(t)$ given in Eqs. (32b) and (33b) of Ref. \cite{Damour:2004bz}, namely
\begin{widetext}
\beq
\phi(t)=\int_{t_0}^t dt [1+k(c_1(t),c_2(t))]  n(c_1(t),c_2(t))+c_\lambda (t)+W(\ell; c_1(t),c_2(t)]\,.
\eeq
\end{widetext}
Here, $c_1(t)$ and $c_2(t)$ denote $\tilde E(t)$ and $j(t)$, while the third quantity $c_\lambda(t)$ corresponds to a possible additional angular displacement of the apsidal line, beyond the effect linked to the radiation-reaction-driven adiabatic variations of $\tilde E(t)$ and $j(t)$. The quantity $c_\lambda(t)$ corresponds in our above simplified treatment to the direction of the vector ${\mathbf A}(t)$. We found above that the direction of ${\mathbf A}(t)$ did not include a secular change under the influence of ${\pmb {\mathcal F}}$, because of symmetry reasons linked, finally, to the time-odd character of ${\pmb {\mathcal F}}$.
This fact has a correspondant in $c_\lambda(t)$. Indeed, Ref.  \cite{Damour:2004bz} found that there were no secular changes in $c_\lambda(t)$ (and $c_\ell (t)$)
precisely because $dc_\lambda(t)/dt$ is an odd function of $\phi$, around the periastron, and remarked that this was linked to the time-odd character of 
${\pmb {\mathcal F}}$. When applying this result to a scattering situation, one again finds that the total scattering angle will be given by the average of the conservative $\chi^{\rm (conserv)}(\tilde E,j)$ over the incoming $(\tilde E_-,j_-)$ and outgoing $(\tilde E_+,j_+)$ values of the two secularly-evolving \lq\lq constants," $\tilde E(t)$ and $j(t)$ (i.e., $c_1(t)$
 and $c_2(t)$ in the notation of \cite{Damour:2004bz}).

Let us finally give an explicit estimate of the modification
\begin{widetext}
\beq
\delta^{\rm (RR)} \chi= \frac12 \left(
\frac{\partial \chi^{\rm (conserv)} (\tilde E,j)}{\partial \tilde E}\delta^{\rm (RR)} E
+\frac{\partial \chi^{\rm (conserv)} (\tilde E,j)}{\partial j}\delta^{\rm (RR)} j
\right)
\eeq
\end{widetext}
of the scattering angle entailed by radiation-reaction. We will do this calculation at the leading PN order in the value of ${\pmb {\mathcal F}}$, i.e., at the $O(1/c^5)$ order only.
We therefore need the values of the losses of energy and angular momentum during a hyperbolic encounter.
From the (Newtonian-order) energy flux at infinity
\beq
\Phi_E=\frac{8}{5G c^5}\nu^2 \left(\frac{GM}{r} \right)^4 \left( 4v^2-\frac{11}{3}\dot r^2\right)\,,
\eeq
we compute the integral
\beq
\int_{-\infty}^{+\infty} dt \Phi_E(t)
\eeq
along the unperturbed motion, using $\phi$, rather than $t$, as integration variable, i.e.,
\beq
\delta^{\rm (RR)} \tilde E=\tilde E_+-\tilde E_-=-\int_{-\phi_+}^{\phi_+} d\phi \frac{r^2}{GM\mu j}\Phi_E\,.
\eeq
Computing this integral, we find
\begin{widetext}
\beq
\delta^{\rm (RR)} \tilde E=-\frac{2\nu }{15c^5 j_-^7}\left[\frac{1}{3}(673 e_-^2+602)\sqrt{e_-^2-1}+(37 e_-^4+292 e_-^2+96)  \phi^0_+(e_-)\right]\,,
\eeq
\end{widetext}
where $\phi^0_+(e_-)$ is defined (in keeping with Eq. (\ref{phi0+})) as
\beq
\phi^0_+(e_-)\equiv {\rm arccos}\left(-\frac{1}{e_-}\right)=\frac{\pi}{2}+{\rm arcsin}\left(\frac{1}{e_-}\right)\,.
\eeq 
This result agrees with Eq. (2.10) in \cite{blanchetschafer}.
Similarly, from the Newtonian angular momentum flux at infinity,
\beq
\Phi_J=\frac{8}{5c^5}\nu^2 (jGM) \left( \frac{GM}{r}\right)^3 \left(2v^2-3\dot r^2+\frac{2}{r}\right)\,
\eeq
we computed 
\begin{widetext}
\beq
\delta^{\rm (RR)} j= j_+- j_-=-\int_{-\phi_+}^{\phi_+} d\phi \frac{r^2}{(GM)^2 \nu j}\frac{\Phi_J}{M }\,.  
\eeq
\end{widetext}
We find
\begin{eqnarray}
\delta^{\rm (RR)} j &=& -\frac{8\nu}{5c^5 j_-^4}\left[(2e_-^2+13)\sqrt{e_-^2-1} \right.\nonumber\\
&& \left.+(7 e_-^2+8) \phi^0_+(e_-)\right]\,,
\end{eqnarray}
where $\phi^0_+(e_-)$ is the same function as above. 

As the (conservative) scattering angle is a function of the eccentricity, i.e., the combination
\beq
e(\tilde E,j)=\sqrt{1+2\tilde Ej^2}
\eeq
of $\tilde E$ and $j$, we are mainly interested in the radiation-reaction-driven change in the eccentricity, namely
\beq
\delta^{\rm (RR)} e=\frac{\partial e(\tilde E,j)}{\partial \tilde E}\delta^{\rm (RR)} E+
\frac{\partial e(\tilde E,j)}{\partial j}\delta^{\rm (RR)} j\,.
\eeq
Using the results above for $\delta^{\rm (RR)} E$ and $\delta^{\rm (RR)} j$ we find
\beq \label{dRRe}
\delta^{\rm (RR)} e=-\frac{2}{15\nu}\frac{e_-}{c^5 j_-^5} Q(e_-)\,,
\eeq
where
\begin{widetext}
\beq
\label{Qe_def}
Q(e_-)=\frac{\sqrt{e_-^2-1}}{3e_-^2}(72e_-^4+1069 e_-^2+134)+(304+121e_-^2)
\phi^0_+(e_-)\,.
\eeq
\end{widetext}
We have also checked this result by computing the change in the Laplace vector ${\mathbf A}$.
We find that the $\phi$-derivative of the associated complex quantity $A=A_x+iA_y$ reads
\begin{eqnarray}
\frac{d A}{d\phi}&=&\frac{8}{15c^5}\nu j e^{i\phi}\left[ -3i j^2 \hat r'{}^3+6j^2 \hat r \hat r'{}^2\right.\nonumber\\
&& \left. +i(7\hat r-15j^2)\hat r^2\hat r'-12(\hat r+j^2)\hat r^3  \right]\,, 
\end{eqnarray}
where the prime denotes a $\phi$-derivative. Inserting the Newtonian orbit 
$\hat r=j^2/(1+e\cos \phi)$, and integrating between $\phi_-$ and $\phi_+$ yields
\beq
\delta^{\rm (RR)} A=A_+-A_-= -\frac{2}{15}\nu \frac{e_-}{c^5j_-^5} Q(e_-)\,,
\eeq
in agreement with Eq. (\ref{dRRe})
Finally, as
\begin{eqnarray}
\chi^{\rm (conserv)}(e)&=&2{\rm arccos}\left(-\frac{1}{e}\right)-\pi \nonumber\\
&=& 2{\rm arcsin} \left(\frac{1}{e} \right)
\end{eqnarray}
we have ${\partial \chi^{\rm (conserv)}}/{\partial e} =-2/(e\sqrt{e^2-1})$ so that
\begin{eqnarray}
\delta^{\rm (RR)}\chi &=& \frac12  \frac{\partial \chi^{\rm (conserv)}(e)}{\partial e} \delta^{\rm (RR)}e\nonumber\\
&=&- \frac{ \delta^{\rm (RR)}e}{e\sqrt{e^2-1}}\,.
\end{eqnarray}
Finally, the radiation-reaction contribution to the scattering angle is given by
\beq
\label{deltachi_fin}
\delta^{\rm (RR)}\chi=\frac{1}{c^5}\, \frac{2\nu}{15}\frac{1}{j_-^5\sqrt{e_-^2-1}} Q(e_-)+O\left(\frac{1}{c^7} \right)\,,
\eeq
where $Q(e_-)$ is defined in Eq. (\ref{Qe_def}).

\section{Summary and outlook}\label{summary}
Let us summarize the main results of our work:
\begin{enumerate}
  \item We have introduced a new approach to the computation of the gravitational radiation-reaction, based on the identities (\ref{n11}), (\ref{n11bis}) satisfied by the combined energy and angular momentum flux function $\Phi_{EJ}$, Eq. (\ref{n11}).

  \item We have computed some \lq\lq minimal" version of the 2PN accurate radiation-reaction force ${\pmb {\mathcal F}}({\mathbf x},{\mathbf p})$ which must be added on the rhs of the Hamiltonian EOB equations of motion when describing general orbits (elliptic-like or hyperbolic-like). The radial, ${\mathcal F}_r$, and azimuthal, ${\mathcal F}_\phi$, components of the radiation-reaction force are explicitly given as functions of the EOB position and momenta by Eqs. (\ref{Fr_EOB}) and (\ref{Fphieob}). Our calculations were based on the transformation properties of the three basic scalars $X_1\sim {\mathbf p}^2/\mu^2\sim {\mathbf v^2}$, $X_2\sim p_r^2/\mu^2\sim \dot r^2$ and $X_3\sim GM/r$ between the various coordinate systems used in PN theory (harmonic, ADM and EOB).

  \item We have also computed the \lq\lq Schott" contribution to the energy, corresponding to the above minimal construction of ${\pmb {\mathcal F}}$. It is given as a function of the EOB position and momenta by Eq. (\ref{minimalschottenergy2}). In particular, we pointed out that $E_{\rm (schott)}$ does not vanish during quasi-inspiral but is proportional to $p_r$ and is given by Eq. (\ref{minimalschottenergy2_circ}).
  
\item We provided a new understanding of the gauge freedom in the construction of the radiation-reaction. It is linked to the arbitrary choice of (i) the Schott contribution to the angular momentum, and (ii) the part of the Schott energy which is proportional to the cube of the radial momentum $p_r$. This explains very simply why there exist $2 \times 1$ arbitrary parameters in ${\pmb {\mathcal F}}$ at the Newtonian order,  $2 \times 3$ at the 1PN order  and $2 \times 6$ at the 2PN order [and then $(n+1) (n+2) $ at $n$ PN order]. 
  
\item We pointed out that there is an inconsistency between the assumptions that are standardly used in current implementations of the radiation-reaction force in the EOB formalism, namely Eqs. (\ref{b1}) and (\ref{b2}). We showed that if one adopts the assumption ${\mathcal F}_\phi=-\Phi_J$ (which is convenient, and always possible) this essentially determines (during inspiral) a nonzero value for the radial component of the radiation-reaction force, given by Eqs. (\ref{Fr_long}), (\ref{Fr_inspiral}) and (\ref{Fphi_inspiral}).
  
\item We introduced a new way of parametrizing (conservative) hyperbolic orbits in PN theory, by the simple quasi-conic equation (at 2PN)
\begin{widetext}
\beq
\frac{{\sf p}}{r}+\frac{1}{c^2}\alpha_2 \left(\frac{{\sf p}}{r}  \right)^2+\frac{1}{c^4}\alpha_3 \left(\frac{{\sf p}}{r}  \right)^3+\frac{1}{c^4}\alpha_4 \left(\frac{{\sf p}}{r}  \right)^4=1+\bar e \cos \frac{\phi-\phi_0}{K}\,,
\eeq
\end{widetext}
and emphasized that the two quantities $\bar e$ (\lq\lq eccentricity") and $K$ (\lq\lq periastron advance") are gauge invariant. The gauge-invariant eccentricity $\bar e$ is related to the scattering angle $\chi$ and to $K$ via Eq. (\ref{c7}). Moreover, $K$ and $\chi$ are given, in EOB theory, by simple (complete or incomplete) integrals over the inverse-radius $u=GM/r$, Eqs. (\ref{eq_per_K}), (\ref{eq_per_chi}).
\item We have showed how the effect of radiation-reaction on the scattering angle  can be computed (modulo correction $O({\pmb {\mathcal F}}^2)=O(1/c^{10})$) from the sole knowledge of the losses of energy and angular momentum at infinity, see Eqs. (\ref{deltachi}), (\ref{deltachi2}) and (\ref{deltachi3}). This result might be used to subtract the effect of radiation-reaction on the scattering angle obtained in  numerical simulations, by using only numerical data in the asymptotic domain at infinity. We also gave an explicit expression, at leading order in $1/c$, for the additional contribution to the scattering angle due to radiation-reaction, see Eq. (\ref{deltachi_fin}). 
\end{enumerate}

Finally, let us point out some of the future research directions that would complete our results:
\begin{itemize}
  \item[{(a)}] In the present work we have not included the effects of tails on the radiation-reaction. We plan to treat this issue in a future publication.
  \item[{(b)}] Here we obtained the components of the radiation-reaction force ${\pmb {\mathcal F}}$ in the form of a standard, non-resummed PN expansion. However, the current most successful implementations of the EOB formalism make a crucial use of efficient re-summations of ${\mathcal F}_\phi$, in the circular limit. It would be interesting to concoct resummation schemes in the more general context considered here. For instance, in the case of slightly elliptic orbits one might hope to improve the numerical validity of our PN-expanded ${\mathcal F}_i$'s by first factorizing the \lq\lq circular part" of these components, and re-summing them by the method introduced in \cite{Damour:2008gu}. We gave some partial results towards this goal in Sec. IV.
  \item[(c)]   Let us finally mention that, in order to have a complete EOB formalism for general orbits, there remains the problem of expressing the emitted gravitational waveforms in terms of the EOB phase-space variables. The transformation formulas we provided should be also useful in this respect.
\end{itemize}
\vglue 1cm

\begin{acknowledgments}
We thank ICRANet for partial support. D.B. is grateful to IHES for
  hospitality during different stages of this project.
\end{acknowledgments}

\vglue 1cm

\appendix

\section{The 2PN energy and angular momentum fluxes in the far zone in EOB coordinates}

\subsection{Energy flux}
\label{en_flux}

The 2PN energy flux (excluding tail terms), scaled as in Eq. (\ref{scaledfluxes}), can be written as
\begin{eqnarray}
\widehat \Phi_E^{\rm (eob)}(X_A^e)&=&[X_3^e]^4 \left(C_AX_A^e+\epsilon^2 C_{AB}X_A^eX_B^e\right. \nonumber\\
&& \left.+\epsilon^4 C_{ABC}X_A^eX_B^eX_C^e \right)
\end{eqnarray}
where
\beq
C_A=\frac{8}{5}\nu\left[4,-\frac{11}{3},0  \right]\,,
\eeq
\begin{widetext}
\beq
\label{Cab_rels}
\begin{array}{lll}
C_{11}=\frac{2}{315}\nu (1347-2556\nu)\,\quad &C_{12}=\frac{2}{105}\nu (-1333+412\nu)\,\quad  &C_{13}=-\frac{2}{2835}\nu(36720-6696\nu)\,\quad \cr
C_{22}=-\frac{2}{945}\nu (-18549+108\nu)\,\quad &C_{23}=\frac{2}{945}\nu (12960-2304\nu)\,\quad  &C_{33}=\frac{2}{2835}\nu (432-1728\nu)\,\quad
\end{array}
\eeq
and
\beq
\label{Cabc_rels}
\begin{array}{ll}
C_{111} = \frac{4}{315}\nu(-159-838\nu+1874\nu^2)&
C_{112} = \frac{4}{315}\nu (490\nu^2+1523\nu+1101)\cr
C_{113} = -\frac{8}{945}\nu (1390\nu^2-6362\nu+4761)&
C_{122} = -\frac{4}{315}\nu (-372\nu+2034\nu^2+43)\cr
C_{123} = \frac{4}{315}\nu (-2242\nu+8111+368\nu^2)&
C_{133} = \frac{2}{8505}\nu (22836\nu^2-61596\nu+198961)\cr
C_{222} = \frac{4}{315}\nu (-4498\nu+2828\nu^2-2501)&
C_{223}= \frac{8}{2835}\nu (590\nu-58611+974\nu^2)\cr
C_{233} = -\frac{2}{2835}\nu (7644\nu^2-32052\nu+81263)&
C_{333} = -\frac{16}{945}\nu (145-648\nu+272\nu^2)\,.
\end{array}
\eeq
\end{widetext}
Note that we are listing here and below the independent components of the symmetric \lq\lq tensor" $C_{A_1\ldots A_n}$. When explicitly effecting the multisummations present in the contractions $C_{A_1\ldots A_n}X_{A_1\ldots A_n}$ (with $X_{A_1\ldots A_n}=X_{A_1}\ldots X_{A_n}$) they appear multiplied by the symmetry factors of Eq. (\ref{symfactor}), namely
\begin{widetext}
\begin{eqnarray}
C_{AB}X_{AB}&=& C_{11}X_{11}+C_{22}X_{22}+C_{33}X_{33}+2C_{12}X_{12}+2C_{13}X_{13}+2C_{23}X_{23}\nonumber\\
C_{ABC}X_{ABC}&=& C_{111}X_{111}+C_{222}X_{222}+C_{333}X_{333}+3 C_{112}X_{112}+3 C_{113}X_{113}\nonumber\\
&& +3 C_{122}X_{122}+3 C_{133}X_{133}
+3 C_{223}X_{223}+3 C_{233}X_{233}+6 C_{123}X_{123}
\,.
\end{eqnarray}
\end{widetext}

\subsection{Angular momentum flux}
\label{am_flux}

The 2PN angular momentum flux (excluding tail terms), scaled as in Eq. (\ref{scaledfluxes}), can be written as
\begin{eqnarray}
\widehat \Phi_J^{\rm (eob)}(X_A^e)&=& j\, [X_3^e]^3 \left(B_AX_A^e+\epsilon^2 B_{AB}X_A^eX_B^e\right. \nonumber\\
&& \left.+\epsilon^4 B_{ABC}X_A^eX_B^eX_C^e \right)
\end{eqnarray}
where
\beq
B_A=\frac{8}{5}\nu\left[2,-3,2  \right]\,,
\eeq
\begin{widetext}
\beq
\begin{array}{lll}
B_{11}= \frac{1}{315}\nu (330-1272\nu)\,\quad &B_{12}= \frac{1}{630}\nu (-396+900\nu)\,\quad  &B_{13}= \frac{1}{630}\nu (-5928-3288\nu)\,\quad \cr
B_{22}= -\frac{1}{315}\nu (-1710-1080\nu)\,\quad &B_{23}=-\frac{1}{630}\nu (-11520-600\nu) \,\quad  &B_{33}= -\frac{2}{945}\nu (11898-1548\nu)\,\quad
\end{array}
\eeq
and
\beq
\begin{array}{ll}
B_{111} =  \frac{1}{315}\nu (-1051\nu-50+750\nu^2) &
B_{112} =   \frac{1}{315}\nu (2051\nu^2+3347\nu-971)\cr
B_{113} =  \frac{1}{945}\nu (7802\nu^2-6057+6238\nu)&
B_{122} =  -\frac{1}{63}\nu (872\nu^2-430+1489\nu)\cr
B_{123} = -\frac{1}{1890}\nu (5516\nu^2-10392\nu-26869)&
B_{133} =  -\frac{2}{945}\nu (4312+4448\nu^2-21843\nu)\cr
B_{222} = \frac{2}{9}\nu (76\nu^2-78+155\nu) &
B_{223}= \frac{8}{945}\nu (272\nu^2-2725\nu-5255) \cr
B_{233} = \frac{4}{945}\nu (-2997+1098\nu^2-6566\nu) &
B_{333} =  \frac{4}{2835}\nu (46085+3042\nu^2-6741\nu)\,. 
\end{array}
\eeq
\end{widetext}

\subsection{Combined energy and angular momentum flux ${\widehat \Phi}_{EJ}^{\rm (eob)}$}
\label{enandam_flux}

The  combined   flux ${\widehat \Phi}_{EJ}^{\rm (eob)}={\widehat \Phi}_{E}^{\rm (eob)}-GM \dot \phi_e {\widehat \Phi}_{J}^{\rm (eob)}$ (excluding tail terms) can be written as
\beq
{\widehat \Phi}_{EJ}^{\rm (eob)}(X_A^e)= [X_3^e]^3 Q_{2,4} (X_A^e)
\eeq
\begin{widetext}
where
\beq
\begin{array}{lll}
Q_{11}=-\frac{16}{5}\nu  \,\quad &Q_{12}=  4\nu\,\quad  &Q_{13}= \frac{8}{5}\nu\,\quad \cr
Q_{22}= -\frac{24}{5}\nu\,\quad &Q_{23}=-\frac{4}{3}\nu  \,\quad  &Q_{33}=0 
\end{array}
\eeq
and
\beq
\begin{array}{ll}
Q_{111} = -\frac{1}{315}\nu (-174-1776\nu)   &
Q_{112} =  -\frac{1}{945}\nu (534+3432\nu)   \cr
Q_{113} =  \frac{1}{945}\nu (10134-2328\nu) &
Q_{122} =  -\frac{1}{945}\nu (534+3432\nu) \cr
Q_{123} = \frac{1}{945}\nu (-18234+1536\nu)&
Q_{133} =  \frac{2}{945}\nu (-3690+468\nu)\cr
Q_{222} =  -\frac{1}{315}\nu (-1710-1080\nu)&
Q_{223}=  -\frac{1}{2835}\nu (-76194+2952\nu)\cr
Q_{233} = -\frac{2}{2835}\nu (-12510+1548 \nu) &
Q_{333} =  \frac{2}{2835}\nu (432-1728\nu)\,;
\end{array}
\eeq
finally, the $15$ independent components of $Q_{ABCD}$ are
\beq
\begin{array}{ll}
Q_{1111} =  -\frac{1}{315}\nu (163-202\nu+1764\nu^2)   &
Q_{1112} =  -\frac{1}{1260}\nu (3372\nu^2+9424\nu-3445)   \cr
Q_{1113} = \frac{1}{1260}\nu (456\nu^2-16682\nu+1905)   &
Q_{1122} =   \frac{1}{630}\nu (10984\nu+6252\nu^2-2959)\cr
Q_{1123} = \frac{1}{1890}\nu (14159\nu-5543+7784\nu^2) &
Q_{1233} = -\frac{1}{1890}\nu (3296\nu^2-2134\nu-108839)\cr
Q_{1222} = -\frac{1}{252}\nu (6916\nu+3788\nu^2-2211)&
Q_{1223}= -\frac{1}{3780}\nu (-30434\nu-66503+30312\nu^2) \cr
Q_{1133} = \frac{1}{945}\nu (790\nu^2+3200\nu-19679)  &
Q_{2222} = \frac{2}{9}\nu (76\nu^2-78+155\nu)\cr
Q_{2223} =  \frac{1}{210}\nu (-6779\nu-9211+2428\nu^2) &
Q_{2233} =\frac{2}{2835}\nu (3991\nu^2-9005\nu-133530) \cr
Q_{3333} =   \frac{1}{1890}\nu (1768\nu^2-1836\nu+24455)&
Q_{3331} = \frac{1}{1890}\nu (1768\nu^2-1836\nu+24455) \cr
Q_{3332} =-\frac{1}{5670}\nu (5400\nu^2-40068\nu+118193)\,.  & \cr
\end{array}
\eeq
Similarly to Eqs. (\ref{Cab_rels}) and (\ref{Cabc_rels}) above, the symmetry factors multiplying the independent components of the symmetric tensor
$Q_{ABCD}$ are given by
\begin{eqnarray}
Q_{ABCD}X_{ABCD}&= & Q_{1111}X_{1111}+Q_{2222}X_{2222}+Q_{3333}X_{3333} \nonumber\\
&& +4 Q_{1112}X_{1112}+4 Q_{1113}X_{1113}+4 Q_{1222}X_{1222}\nonumber\\
&& +4 Q_{1333}X_{1333}+4 Q_{2223}X_{2223}+4 Q_{2333}X_{2333}\nonumber\\
&& 6 Q_{1122}X_{1122}+6 Q_{1133}X_{1133}+6Q_{2233}X_{2233}\nonumber\\
&& +12 Q_{1123}X_{1123}+12 Q_{1223}X_{1223}+12 Q_{1233}X_{1233}
\end{eqnarray}
\end{widetext}

\section{Hamilton equations in EOB coordinates: expansion at 2PN}
\label{eq_re}

$\bullet$ Equation for   $\dot r_e$ \noindent

From Hamilton's equations we have
\beq
\dot r_e= \frac{\partial \tilde {\mathcal H}_{\rm (eob)}}{\partial \tilde p_r^{(e)}}=\tilde C ({\mathbf n}_e \cdot \tilde {\mathbf p}_e)=\tilde C\tilde p_r^{\rm (e)}
\eeq
with
\begin{eqnarray}
\tilde C&=&1+\epsilon^2 \tilde C_A X_A^e+\epsilon^4 \tilde C_{AB}X_A^eX_B^e\,,
\end{eqnarray}
where
\beq
\tilde C_A=\left[-\frac{1+\nu}{2} , 0, \nu-3\right]
\eeq
and
\begin{widetext}
\beq
\begin{array}{lll}
\tilde C_{11}=\frac38 (1+\nu+\nu^2) \,\quad &\tilde C_{12}=0 \,\quad  &\tilde C_{13}=\frac34(1+\nu-\nu^2) \,\quad \cr
\tilde C_{22}=0 \,\quad &\tilde C_{23}= \frac12(1+\nu) \,\quad  &\tilde C_{33}=\frac12(3\nu^2+7\nu+3)\,.
\end{array}
\eeq
\end{widetext}

$\bullet$ Equation for   $\dot {\tilde p}_r^{(e)}$ \noindent

From Hamilton's equations we have
\beq
\dot {\tilde p}_r^{(e)}= -\frac{\partial \tilde {\mathcal H}_{\rm (eob)}}{\partial r_e}
\eeq
In $X_A^e$ variables we have
\begin{eqnarray}
\dot {\tilde p}_r^{(e)} 
&=& X_3^e \hat  C_{1,3}(X_A^e)
\end{eqnarray}
where  
\beq
\hat C_A=\left[1 ,-1,-1\right]
\eeq
and
\beq
\begin{array}{ll}
\hat C_{11}=-\frac12(1+\nu)\,\quad &\hat C_{12}=\frac14(1+\nu) \cr
\hat C_{13}=-\frac34(1-\nu)\,\quad & \hat C_{22}=0 \cr
\hat C_{23}= -\frac12\nu \,\quad  &\hat C_{33}=-1-\nu\,.
\end{array}
\eeq
and
\beq
\begin{array}{ll}
\hat C_{111} = \frac38(1+\nu+\nu^2)&
\hat C_{112} =-\frac18(1+\nu+\nu^2) \cr
\hat C_{113} = -\frac5{24} (3\nu^2- \nu -1) &
\hat C_{122} =0\cr
\hat C_{123} =\frac{1}{12}(2+2\nu +3\nu^2)  &
\hat C_{133} =\frac13 (3\nu^2-\nu-1)  \cr
\hat C_{222} = 0&
\hat C_{223}=-\frac13 (1+\nu) \cr
\hat C_{233} =-\frac16 (3\nu^2 -9\nu -5 )&
\hat C_{333} =-\frac32 (\nu^2 -\nu +1)\,. 
\end{array}
\eeq

\section{Schott energy at 2PN in EOB coordinates: minimal gauge expression}
\label{Schott_energy}
\begin{widetext}
The minimal gauge expression for the Schott energy is given by the first of Eqs. (\ref{minimalschottenergy2}), that is
\begin{eqnarray}
\label{Eschott_appendix}
\tilde E_{\rm schott}^{\rm (min)}&=& \frac{1}{c^5}\sqrt{X_2^e} [X_3^e]^2 C_{1,3}(X_A^e)
\end{eqnarray}
where $\sqrt{X_2^e}$ denotes $\tilde p_r$ (with its sign),
\beq
\label{C1app}
C_A=\frac{16}{5}\nu \left[1 ,-1,0\right]
\eeq
and
\beq
\begin{array}{lll}
C_{11}=\nu \left(\frac{22}{21}-\frac{424}{105}\nu\right)\,\quad & C_{12}=\nu \left( \frac{68}{21}\nu-\frac{194}{105}\right)\,\quad  & C_{13}=\nu \left(-\frac{1382}{105}-\frac{76}{105}\nu\right)\,\quad \cr
C_{22}=\nu \left(-\frac{256}{105}\nu+\frac{278}{105}\right)\,\quad & C_{23}=\nu \left(\frac{226}{21}+\frac{32}{21}\nu\right)\,\quad  & C_{33}=\nu \left(\frac{32}{105}-\frac{128}{105}\nu\right)\,.
\end{array}
\eeq
and
\beq
\begin{array}{ll}
 C_{111} = \nu \left(-\frac{1051}{315}\nu+\frac{50}{21}\nu^2-\frac{10}{63}\right)&
 C_{112} = \nu \left(\frac{8}{105}+\frac{61}{15}\nu-\frac{242}{315}\nu^2\right)\cr
 C_{113} = \nu \left(\frac{184}{45}\nu^2-\frac{6236}{945}+\frac{8224}{945}\nu\right)&
 C_{122} = \nu \left(-\frac{4}{9}\nu^2-\frac{277}{63}\nu+\frac{128}{315}\right)\cr
 C_{123} = \nu \left(-\frac{326}{63}\nu^2-\frac{3334}{945}\nu+\frac{7814}{945}\right)&
 C_{133} = \nu \left(-\frac{304}{189}\nu^2+\frac{100}{7}\nu+\frac{4414}{189}\right)\cr
 C_{222} = \nu \left( \frac{1363}{315}\nu+\frac{44}{35}\nu^2-\frac{58}{45}\right)&
 C_{223} = \nu \left(\frac{1846}{315}\nu^2-\frac{926}{945}\nu-\frac{8762}{945}\right)\cr
 C_{233} = \nu \left(\frac{2048}{945}\nu^2-\frac{694}{63}\nu-\frac{1484}{135}\right)&
 C_{333} = \nu \left(\frac{416}{35}\nu-\frac{640}{189}\nu^2-\frac{2608}{945}\right)\,.
\end{array}
\eeq
\end{widetext}

\section{Radiation reaction force at 2PN in EOB coordinates: Minimal gauge expressions}
\label{Fr_and_phi_eob}

The minimal gauge expression for the radial component of the  radiation reaction force is given by the second of Eqs. (\ref{minimalschottenergy2}), that is

\begin{eqnarray}
\tilde {\mathcal F}_r^{\rm (eob)}(X_A^e)&=& \frac{1}{GM c^5}\sqrt{X_2^e} [X_3^e]^3 R_{1,3}(X_A^e)
\end{eqnarray}
where
\beq
R_A=\frac{8}{5}\nu \left[7 ,-7,-\frac13\right]
\eeq
and
\begin{widetext}
\beq
\begin{array}{lll}
R_{11}=\nu \left(\frac{1252}{105}-\frac{2588}{105}\nu\right)\,\quad & R_{12}=\nu \left(\frac{158}{7}\nu-\frac{438}{35}\right) \,\quad  & R_{13}=\nu \left(-\frac{248}{7}-\frac{634}{105}\nu\right)\,\quad \cr
R_{22}= \nu \left(-\frac{2152}{105}\nu+\frac{1376}{105}\right)\,\quad & R_{23}=\nu \left(\frac{3106}{105}+\frac{304}{35}\nu\right)\,\quad  & R_{33}=\nu \left( -\frac{624}{35}+\frac{16}{21}\nu\right)\,.
\end{array}
\eeq
and
\beq
\begin{array}{ll}
R_{111} =\nu \left(  -\frac{3229}{315}+\frac{3277}{105}\nu^2-\frac{718}{63}\nu \right)&
R_{112} =\nu \left( \frac{289}{35}+\frac{282}{35}\nu-\frac{769}{35}\nu^2 \right)\cr
R_{113} =\nu \left(  \frac{25217}{945}\nu^2-\frac{6103}{135}+\frac{16418}{315}\nu \right)&
R_{122} =\nu \left( -\frac{1276}{189}+\frac{12688}{945}\nu^2-\frac{5441}{945}\nu \right)\cr
R_{123} =\nu \left( -\frac{1097}{35}\nu^2-\frac{40184}{945}\nu+\frac{30532}{945} \right)&
R_{133} =\nu \left( -\frac{584}{35}\nu^2+\frac{66032}{945}\nu+\frac{218}{189} \right)\cr
R_{222} =\nu \left( -\frac{1756}{315}\nu^2+\frac{1417}{315}\nu+\frac{86}{15} \right)&
R_{223} =\nu \left(  \frac{10916}{315}\nu^2+\frac{715}{21}\nu-\frac{3604}{189} \right)\cr
R_{233} =\nu \left(  \frac{6602}{405}\nu^2-\frac{28034}{567}\nu+\frac{8336}{189} \right)&
R_{333} =\nu \left( -\frac{3548}{315}\nu^2+\frac{9526}{105}\nu+\frac{33338}{567} \right)\,.
\end{array}
\eeq
\end{widetext}

The minimal gauge expression for the azimuthal component of the  radiation reaction force is given by the third of Eqs. (\ref{minimalschottenergy2}), that is

\begin{eqnarray}
{\mathcal F}_\phi^{\rm (eob)}&=& \frac{j}{c^5}  [X_3^e]^3 S_{1,3}(X_A^e)\,,
\end{eqnarray}
where
\beq
S_A=-\frac{8}{5}\nu \left[2 ,-3,2\right]
\eeq
and
\begin{widetext}
\beq
\begin{array}{lll}
S_{11}=\nu \left(-\frac{22}{21}+\frac{424}{105}\nu\right)\,\quad & S_{12}=\nu \left(-\frac{10}{7}\nu+\frac{22}{35}\right) \,\quad  & S_{13}=\nu \left(\frac{988}{105}+\frac{548}{105}\nu \right)\,\quad \cr
S_{22}=\nu \left(-\frac{24}{7}\nu-\frac{38}{7}\right) \,\quad & S_{23}=\nu \left( -\frac{128}{7}-\frac{20}{21}\nu \right)\,\quad  & S_{33}=\nu \left(\frac{2644}{105}-\frac{344}{105}\nu\right)\,,
\end{array}
\eeq
and
\beq
\begin{array}{ll}
\label{D8app}
S_{111} =\nu \left(\frac{1051}{315}\nu-\frac{50}{21}\nu^2+\frac{10}{63}\right) &
S_{112} =\nu \left( \frac{971}{315}-\frac{3347}{315}\nu-\frac{293}{45}\nu^2\right) \cr
S_{113} =\nu \left(  -\frac{7802}{945}\nu^2+\frac{673}{105}-\frac{6238}{945}\nu \right)&
S_{122} =\nu \left( \frac{872}{63}\nu^2+\frac{1489}{63}\nu-\frac{430}{63} \right)\cr
S_{123} =\nu \left( \frac{394}{135}\nu^2-\frac{1732}{315}\nu-\frac{26869}{1890} \right) &
S_{133} =\nu \left( \frac{8896}{945}\nu^2-\frac{1618}{35}\nu+\frac{1232}{135} \right)\cr
S_{222} =\nu \left( -\frac{152}{9}\nu^2+\frac{52}{3}-\frac{310}{9}\nu \right)&
S_{223} =\nu \left( -\frac{2176}{945}\nu^2+\frac{4360}{189}\nu+\frac{8408}{189}\right) \cr
S_{233} =\nu \left( -\frac{488}{105}\nu^2+\frac{3752}{135}\nu+\frac{444}{35} \right)&
S_{333} =\nu \left(  -\frac{1352}{315}\nu^2+\frac{428}{45}\nu-\frac{36868}{567}\right)\,.
\end{array}
\eeq

Finally, for the expression Eq. (\ref{Fr_X234}) of ${\mathcal F}_r(X_2^e,X_3^e,X_4^e)$ in terms of the EOB variables $X_2^e,X_3^e,X_4^e$ we have the following coefficients
\beq
\label{TI_coeffs}
T_I=\nu\left[ 0, \frac{32}{3} , -\frac{56}{5} \right]\,,
\eeq
and
\beq
\label{TIJ_coeffs}
\begin{array}{lll}
T_{23}= \nu \left(\frac{4}{7}\nu +\frac{100}{21} \right)\,\quad & 
T_{24}= \nu \left( -\frac{76}{105}\nu -\frac{232}{105}\right)\,\quad  & 
T_{33}= \nu \left(-\frac{3776}{105}\nu -\frac{4532}{105}\right) \,\quad \cr
T_{34}= \nu \left(\frac{1172}{35}\nu +\frac{998}{105}\right) \,\quad & 
T_{44}= \nu \left(-\frac{400}{21}\nu+\frac{368}{21}\right) \,. &  
\end{array}
\eeq
and
\beq
\label{TIJK_coeffs}
\begin{array}{ll}
 T_{223} = \nu \left( -\frac{206}{315}\nu^2-\frac{94}{35}\nu-\frac{14}{15}\right)  &
 T_{224} = \nu \left( \frac{88}{189}\nu^2+\frac{1382}{945}\nu +\frac{1088}{945}\right)\cr
 T_{233} = \nu \left(-\frac{1312}{2835}\nu^2-\frac{17678}{567}\nu-\frac{1024}{135}\right) &
 T_{234} = \nu \left( -\frac{263}{945}\nu^2+\frac{550}{27}\nu +\frac{209}{21} \right) \cr
 T_{244} = \nu \left( \frac{104}{315}\nu^2-\frac{1786}{315}\nu +\frac{562}{315}\right)&
 T_{333} = \nu \left( \frac{1748}{35}\nu^2 \frac{1138}{5}\nu -\frac{351098}{2835}\right)\cr
 T_{334} = \nu \left(-\frac{73384}{945}\nu^2-\frac{63173}{945}\nu +\frac{16148}{189} \right)&
 T_{344} = \nu \left( \frac{7544}{135}\nu^2 -\frac{5584}{315}\nu -\frac{33976}{945} \right)\cr
 T_{444} = \nu \left(-\frac{836}{105}\nu^2+\frac{5393}{315}\nu  -\frac{968}{315} \right) \,.&
\end{array}
\eeq

Similarly, for the expression Eq. (\ref{Fphi_X234}) of ${\mathcal F}_\phi(X_2^e,X_3^e,X_4^e)$ in terms of the EOB variables $X_2^e,X_3^e,X_4^e$ (having also used the expression (\ref{j234}) for $j$), we have the following coefficients
\beq
\label{TI_coeffs}
V_I=\nu\left[ \frac85 ,-\frac{32}{5} ,\frac{16}{5}  \right]\,,
\eeq
and
\beq
\label{TIJ_coeffs}
\begin{array}{lll}
V_{22}=\nu\left( -\frac{236}{105}\nu-\frac{548}{105}\right) \,\quad & V_{23}= \nu\left(\frac{722}{105}\nu-\frac{1312}{105}\right) \,\quad  & V_{24}=\nu\left(-\frac{38}{21}\nu +\frac{128}{105} \right) \,\quad \cr
V_{33}=\nu\left(\frac{56}{5}\nu +\frac{3502}{105}\right)  \,\quad & V_{34}=\nu\left(-\frac{352}{35}\nu -\frac{458}{105}  \right)\,\quad 
& V_{44}=\nu\left(\frac{256}{105}\nu -\frac{278}{105}\right)
\end{array}
\eeq
and
\beq
\label{TIJK_coeffs}
\begin{array}{ll}
 V_{222} =\nu\left(  \frac{857}{315}\nu^2 +\frac{499}{63}\nu + \frac{1973}{315}\right) &
 V_{223} =\nu\left(  -\frac{5938}{945}\nu^2 +\frac{13864}{945}\nu  +\frac{6742}{315} \right)\cr
 V_{224} =\nu\left(   \frac{176}{315}\nu^2 -\frac{398}{63}\nu + \frac{32}{63}  \right)&
 V_{233} =\nu\left(  -\frac{13987}{945}\nu^2-\frac{2726}{105}\nu +\frac{16648}{945} \right) \cr
 V_{234} =\nu\left( \frac{12526}{945}\nu^2 +\frac{11008}{945}\nu +\frac{8731}{1890}\right) &
 V_{244} =\nu\left(   -\frac{701}{105}\nu^2-\frac{668}{105}\nu  +\frac{289}{105} \right)\cr
 V_{333} =\nu\left(    -\frac{16}{5}\nu^2-\frac{25357}{315}\nu+\frac{1261}{405} \right)&
 V_{334} =\nu\left(  \frac{3916}{315}\nu^2+\frac{32549}{945}\nu -\frac{4828}{189}  \right)\cr
 V_{344} =\nu\left(   -\frac{8408}{945}\nu^2 +\frac{1579}{189}\nu+\frac{115}{21}  \right)&
 V_{444} =\nu\left(  -\frac{44}{35}\nu^2 -\frac{1363}{315}\nu+\frac{58}{45} \right)\,.
\end{array}
\eeq
\end{widetext}

\section{Coordinate transformations in phase space: harmonic, ADM, EOB}
\label{coord_trans}

In PN theory there exist at least three different coordinate systems that are largely used: harmonic (h), ADM (a) and EOB (e).
Each of these systems has its own utility and we shall discuss here their transformation laws at the 2PN order.
We work with the scaled position variables ${\mathbf q}_h={\mathbf x}_h/(GM)$, ${\mathbf q}_a={\mathbf x}_a/(GM)$, ${\mathbf q}_e={\mathbf x}_e/(GM)$ and similarly for velocity or momentum (per unit reduced mass) variables, which are simply denoted by ${\mathbf p}_h$, ${\mathbf p}_a$, ${\mathbf p}_e$ without recalling the tilde notation.
Phase space variables associated with harmonic coordinates are only $({\mathbf q}_h,{\mathbf v}_h)$ (no ordinary Hamiltonian exits in this case), whereas for the ADM and EOB cases one has either $({\mathbf q}_a,{\mathbf v}_a)$ and $({\mathbf q}_e,{\mathbf v}_e)$, respectively or $({\mathbf q}_a,{\mathbf p}_a)$ and $({\mathbf q}_e,{\mathbf p}_e)$.
With each choice of phase space variables (h,a, or e) is associated a family of fundamental scalars, that is for example
$$
({\mathbf q}_h,{\mathbf v}_h)\,\rightarrow \, X_1^h=v_h^2\,, X_2^h=({\mathbf n}_h \cdot {\mathbf v}_h)^2\,,X_3^h=\frac{1}{q_h}
$$
where ${\mathbf n}_h={\mathbf q}_h/q_h$, etc. We list below the main transformation laws among phase space vectors as well as fundamental scalars.

\begin{widetext}

\subsection{ADM vs harmonic coordinates}
\noindent 
ADM vs harmonic phase space vector 2PN-transformations are the following:\\

\noindent 1) $({\mathbf q}_a,{\mathbf v}_a)\rightarrow ({\mathbf q}_h,{\mathbf v}_h)$\\
\begin{eqnarray}
{\mathbf q}_h &=& {\mathbf q}_a+\epsilon^4 \left\{\left[\left(3\nu+\frac14\right)\frac{1}{q_a}+\frac58 \nu p_a^2-\frac18 \nu({\mathbf n}_a\cdot {\mathbf p}_a)^2\right]{\mathbf n}_a-\frac94  \nu ({\mathbf n}_a\cdot {\mathbf p}_a){\mathbf p}_a\right\}\,\nonumber\\
{\mathbf v}_h&=&{\mathbf v}_a+\epsilon^4 \left\{-\frac{({\mathbf n}_a\cdot {\mathbf v}_a)}{8q_a}
\left[ \nu\left(7v_a^2-3({\mathbf n}_a\cdot {\mathbf v}_a)^2+38 \frac{1}{q_a}\right)+4\frac{1}{q_a} \right]{\mathbf n}_a\right.\nonumber\\
&&\left.-\frac{1}{8q_a}\left[ \nu\left(-17 ({\mathbf n}_a\cdot {\mathbf v}_a)^2 +13  v_a^2-42 \frac{1}{q_a}\right) -2\frac{1}{q_a}\right]{\mathbf v}_a\right\}\,.
\end{eqnarray}

\noindent 2) ${\mathbf v}_a \leftrightarrow {\mathbf p}_a$\\ 
\begin{eqnarray}
{\mathbf v}_a&=&{\mathbf p}_a+\epsilon^2 \left[-\frac{1}{q_a}\nu ({\mathbf n}_a\cdot {\mathbf p}_a){\mathbf n}_a+\left(\frac{(3\nu-1)}{2}p_a^2-(3+\nu)\frac{1}{q_a}\right){\mathbf p}_a \right]\nonumber\\
&& +\epsilon^4 \left[\frac{1}{q_a}({\mathbf n}_a\cdot {\mathbf p}_a)\left(3\nu \frac{1}{q_a}-\frac{\nu^2}{2} p_a^2-\frac32\nu^2 ({\mathbf n}_a\cdot {\mathbf p}_a)^2 \right){\mathbf n}_a \right.\nonumber\\
&& \left. +\left(\frac38(5\nu^2-5\nu+1) p_a^4 -\frac12(3\nu^2+20\nu-5) \frac{1}{q_a}p_a^2+ (8\nu+5)\frac{1}{q_a^2}-\frac{\nu^2}{2}\frac{1}{q_a}({\mathbf n}_a\cdot {\mathbf p}_a)^2\right){\mathbf p}_a\right]\,,\nonumber\\
{\mathbf p}_a&=&{\mathbf v}_a+\epsilon^2 \left[\frac{1}{q_a}\nu ({\mathbf n}_a\cdot {\mathbf v}_a){\mathbf n}_a+\left(\frac{(1-3\nu)}{2}v_a^2+(3+\nu)\frac{1}{q_a}\right){\mathbf v}_a \right]\nonumber\\
&& +\epsilon^4\left\{({\mathbf n}_a\cdot {\mathbf v}_a){\mathbf n}_a\ \left[\frac32\nu^2 \frac{({\mathbf n}_a\cdot {\mathbf v}_a)^2}{q_a}+\frac{\nu(2-5\nu)}{2}\frac{v_a^2}{q_a}+3\nu(1+\nu)\frac{1}{q_a^2}\right]\right.\nonumber\\
&& \left. {\mathbf v}_a\left[\frac{\nu(2-5\nu)}{2}\frac{({\mathbf n}_a\cdot {\mathbf v}_a)^2}{q_a}+\frac{39\nu^2-21\nu+3}{8}v_a^4  
-\frac{9\nu^2+12\nu-7}{2}\frac{v_a^2}{q_a}+(\nu^2-2\nu+4)\frac{1}{q_a^2}
\right]\right\}\,.
\end{eqnarray}

\noindent 3) ${\mathbf n}_h \leftrightarrow {\mathbf n}_a$\\ 
\begin{eqnarray}
{\mathbf n}_h&=&{\mathbf n}_a+\frac{9}{4q_a}\nu\epsilon^4 \left[({\mathbf n}_a\cdot {\mathbf v}_a){\mathbf n}_a-{\mathbf v}_a \right]\,,\quad 
{\mathbf n}_a={\mathbf n}_h-\frac{9}{4q_h}\nu\epsilon^4 \left[({\mathbf n}_h\cdot {\mathbf v}_h){\mathbf n}_h-{\mathbf v}_h \right]\,. 
\end{eqnarray}

\noindent 4) ${\mathbf v}_h \rightarrow {\mathbf v}_a$\\ 
\begin{eqnarray}
{\mathbf v}_a&=&{\mathbf v}_h-\epsilon^4 \left\{\left[\frac{3\nu}{8q_h}({\mathbf n}_h\cdot {\mathbf v}_h)^2-\frac{19\nu+2}{4q_h^2}-\frac{7\nu}{8q_h}v_h^2  \right]({\mathbf n}_h\cdot {\mathbf v}_h){\mathbf n}_h \right.\nonumber\\
&& \left.+\left[\frac{17\nu}{8} \frac{({\mathbf n}_h\cdot {\mathbf v}_h)^2}{q_h} +\frac{1+21\nu}{4q_h^2}-\frac{13\nu}{8q_h}v_h^2\right]{\mathbf v}_h\right\}\,.
\end{eqnarray}

\noindent 5) ${\mathbf v}_h \rightarrow {\mathbf p}_a$\\ 
\begin{eqnarray}
{\mathbf p}_a&=&{\mathbf v}_h +\epsilon^2 \left[ {\mathbf v}_h \left( \frac{1-3\nu}{2}v_h^2+\frac{\nu+3}{q_h} \right)+  \frac{\nu}{q_h}({\mathbf n}_h\cdot {\mathbf v}_h){\mathbf n}_h\right]\nonumber\\
&& + \epsilon^4 \left[{\mathbf v}_h \left( \frac{39 \nu^2-21\nu+3}{8}v_h^4-\frac{36\nu^2 +35\nu-28}{8}\frac{v_h^2}{q_h} -\frac{\nu(20 \nu+9)}{8} \frac{({\mathbf n}_h\cdot {\mathbf v}_h)^2}{q_h}+\frac{4\nu^2-29\nu+15}{4q_h^2}\right)  \right. \nonumber\\
&& \left. +({\mathbf n}_h\cdot {\mathbf v}_h) {\mathbf n}_h \left(\frac{3\nu (4\nu-1)}{8q_h} ({\mathbf n}_h\cdot {\mathbf v}_h)^2 -\frac{5\nu (4\nu-3)}{8q_h}v_h^2+\frac{12\nu^2+31\nu+2}{4q_h^2} \right)  \right]\,.
\end{eqnarray}

Concerning the transformation of fundamental scalar quantities, we recall once more the notation introduced in Sec. II, namely
$X_1^h=v_h^2$, $X_2^h=({\mathbf n}_h\cdot {\mathbf v}_h)^2$, $X_3^h=\frac{1}{q_h}$,.
The same notation for the ADM variables leads, as explained before, to the two possible choices
\beq
X_1^a=p_a^2\,,\quad X_2^a=({\mathbf n}_a\cdot {\mathbf p}_a)^2\,,\quad X_3^a=\frac{1}{q_a}
\eeq
and
\beq
Y_1^a=v_a^2\,,\quad Y_2^a=({\mathbf n}_a\cdot {\mathbf v}_a)^2\,,\quad Y_3^a=\frac{1}{q_a}=X_3^a\,.
\eeq
We find explicitly
\begin{eqnarray}
\label{Xh_ADM}
X_1^h&=&Y_1^a+\epsilon^4Y_3^a \left[ Y_2^a \left(  \frac34 \nu Y_2^a  -\frac{2+19\nu}{2}Y_3^a\right)+Y_1^a\left(  \frac{5}2 \nu Y_2^a -\frac{13}{4}\nu Y_1^a +\frac{1+21\nu}{2}Y_3^a\right) \right]\nonumber\\
X_2^h&=& Y_2^a\left[1+\epsilon^4 Y_3^a\left( \frac{19}{2}\nu Y_2^a-\frac{19}{2}\nu Y_1^a +\frac{(2\nu-1)}{2}Y_3^a\right)  \right]
\nonumber\\
X_3^h &=&Y_3^a+\epsilon^4[Y_3^a]^2\left[
\frac{\nu}{8}(19Y_2^a -5Y_1^a )-\frac{1}{4}(1+12\nu)Y_3^a
\right]\,,
\end{eqnarray}
and
\begin{eqnarray}
\label{Yadm_h}
Y_1^a&=&X_1^h+\epsilon^4 X_3^h \left[ X_2^h \left( - \frac34 \nu X_2^h  +\frac{2+19\nu}{2}X_3^h\right)+X_1^h\left(  -\frac{5}2 \nu X_2^h +\frac{13}{4}\nu X_1^h -\frac{1+21\nu}{2}X_3^h\right) \right]\nonumber\\
Y_2^a&=& X_2^h \left[1-\epsilon^4 X_3^h\left( \frac{19}{2}\nu X_2^h-\frac{19}{2}\nu X_1^h +\frac{(2\nu-1)}{2}X_3^h\right)  \right]\nonumber\\
Y_3^a&=&  X_3^h \left[1- \epsilon^4 X_3^h \left(
\frac{\nu}{8}(19X_2^h -5X_1^h )-\frac{1}{4}(1+12\nu)X_3^h\right)\right]\,.
\end{eqnarray}
Equivalently, using our ``tensorial'' notation  Eqs. (\ref{Yadm_h}) are summarized by
\begin{enumerate}
  \item   $Y_1^a={}_1 Q_{1,3}^{ah}(X_A^h)$, 
with ${}_1Q_B^{ah}=\delta_{1B}$
and the ${}_1Q^{ah}_{AB}$ all vanishing, while
\beq
\begin{array}{lll}
{}_1C^{ah}_{111} =0  \quad &
{}_1C^{ah}_{113} =  \frac{13}{12}\nu \quad  &
{}_1C^{ah}_{123} = -\frac{5}{12}\nu  \cr
{}_1C^{ah}_{133} =  -\frac{7}{2}\nu-\frac16  \quad &
{}_1C^{ah}_{223} =  -\frac{\nu}{4}\quad &
{}_1C^{ah}_{233} =  \frac{19}{6}\nu+\frac13\,.
\end{array}
\eeq

  \item   $Y_2^a={}_2 Q_{1,3}^{ah}(X_A^h)$, 
with ${}_2Q_B^{ah}=\delta_{2B}$
and the ${}_2Q^{ah}_{AB}$  all vanishing, while
\beq
\begin{array}{llll}
{}_2C^{ah}_{112} =   0\quad &
{}_2C^{ah}_{123} =  \frac{19}{12}\nu   \quad &
{}_2C^{ah}_{223} = -\frac{19}{6}\nu   &
{}_2C^{ah}_{233} = \frac16 -\frac{\nu}{3} \,.
\end{array}
\eeq

  \item   $Y_3^a={}_3 Q_{1,3}^{ah}(X_A^h)$, 
with ${}_3Q_B^{ah}=\delta_{3B}$
and the ${}_3Q^{ah}_{AB}$ all vanishing, while
\beq
\begin{array}{lll}
{}_3Q^{ah}_{133} = \frac{5}{24}\nu   \quad &
{}_3Q^{ah}_{233} = -\frac{19}{24}\nu   \quad &
{}_3Q^{ah}_{333} =  \frac14 +3\nu\,.
\end{array}
\eeq
\end{enumerate}
Similarly, we may summarize the relations $X_A^a={}_AC_{1,3}^{ah}(X_A^h)$ as indicated below

\begin{enumerate}
  \item   $X_1^a={}_1 C_{1,3}^{ah}(X_A^h)$, with
${}_1C_B^{ah}=\delta_{1B}$, 
\beq
\begin{array}{lll}
{}_1C^{ah}_{11}=1-3\nu\,\quad & {}_1C^{ah}_{13}=\nu+3 \,\quad  & {}_1C^{ah}_{23}=\nu \,,
\end{array}
\eeq
and finally
\beq
\begin{array}{lll}
{}_1C^{ah}_{111} =  12\nu^2-\frac{27}{4}\nu +1\quad &
{}_1C^{ah}_{113} =  -4\nu^2 -\frac{67}{12}\nu+\frac{10}{3}\quad  &
{}_1C^{ah}_{123} =   -\frac{13}{6}\nu^2+\frac{5}{12}\nu \cr
{}_1C^{ah}_{133} =   \nu^2-\frac{17}{6}\nu +\frac{11}{2}\quad &
{}_1C^{ah}_{223} =  \nu^2-\frac14 \nu\quad &
{}_1C^{ah}_{233} =  3\nu^2+\frac{43}{6}\nu+\frac13\,.
\end{array}
\eeq

  \item   $X_2^a={}_2 C_{1,3}^{ah}(X_A^h)$, with ${}_2C_B^{ah}=\delta_{2B}$, 
\beq
\begin{array}{ll}
{}_2C^{ah}_{12}= \frac{1-3\nu}{2} \,\quad &   {}_2C^{ah}_{23}=2\nu+3  \,,
\end{array}
\eeq
and finally
\beq
\begin{array}{lll}
{}_2C^{ah}_{112} =4\nu^2-\frac{9}{4}\nu+\frac13  \quad &
{}_2C^{ah}_{123} = -\frac{10}{3}\nu^2-\frac{5}{4}\nu+\frac53    \quad &
{}_2C^{ah}_{223} =  -\frac{2}{3}\nu^2-\frac{5}{2}\nu  \cr
{}_2C^{ah}_{233} =  4\nu^2+\frac{13}{3}\nu+\frac{35}{6}  \,.\quad &
 \quad &
\end{array}
\eeq

  \item   $X_3^a={}_3 C_{1,3}^{ah}(X_A^h)$, with 
${}_3C_B^{ah}=\delta_{3B}$; 
here all the ${}_3C^{ah}_{AB}$ vanish while
\beq
\begin{array}{lll}
{}_3C^{ah}_{133} = \frac{5}{24}\nu  \quad &
{}_3C^{ah}_{233} = -\frac{19}{24}\nu    \quad &
{}_3C^{ah}_{333} =   3\nu+\frac14\,.
\end{array}
\eeq
\end{enumerate}

\subsection{Harmonic vs EOB coordinates}
\noindent 
Harmonic vs EOB phase space vector 2PN-transformations are the following:\\
 
\noindent 1) ${\mathbf p}_e\leftrightarrow  {\mathbf v}_e $\\
\begin{eqnarray}
{\mathbf v}_e=\dot {\mathbf q}_e&=&{\mathbf p}_e +\epsilon^2\left[{\mathbf p}_e\left(-\frac{\nu+1}{2}p_e^2-(\nu-1)\frac{1}{q_e}  \right)-2\frac{1}{q_e}({\mathbf n}_e \cdot {\mathbf p}_e){\mathbf n}_e\right]\nonumber\\
&& +\epsilon^4 \left\{{\mathbf n}_e \frac{1}{q_e}({\mathbf n}_e \cdot {\mathbf p}_e)\left[(\nu+1)p_e^2+2(1+2\nu)\frac{1}{q_e}\right]+{\mathbf p}_e \left[(\nu+1)({\mathbf n}_e \cdot {\mathbf p}_e)^2\frac{1}{q_e}\right.\right.\nonumber\\
&& \left.\left. +\frac38(1+\nu+\nu^2)p_e^4-\frac12 \frac{1}{q_e}(-1-\nu+3\nu^2)\left(p_e^2-\frac{1}{q_e}\right)\right] \right\}\,.
\end{eqnarray}

\begin{eqnarray}
{\mathbf p}_e&=&{\mathbf v}_e +\epsilon^2\left[2{\mathbf n}_e\frac{({\mathbf n}_e \cdot {\mathbf v}_e)}{q_e}+\frac{{\mathbf v}_e}{2}\left(v_e^2(1+\nu)-2(\nu-1)\frac{1}{q_e}  \right)
\right]\nonumber\\
&& \quad +\epsilon^4  \left[{\mathbf n}_e\frac{({\mathbf n}_e \cdot {\mathbf v}_e)}{q_e} \left( -2\frac{(4\nu-3)}{q_e} +(1+\nu)v_e^2\right) +\frac{{\mathbf v}_e}{8} \left( \frac{-4\nu^2-12\nu+12}{q_e^2} \right.\right. \nonumber\\
&& \left.\left.\qquad +\frac{-4\nu^2-4\nu-12}{q_e}v_e^2+\frac{8(1+\nu)({\mathbf n}_e \cdot {\mathbf v}_e)^2}{q_e}+3v_e^2(\nu^2+3\nu+1)\right) \right]\,.
\end{eqnarray}
We recall that ${\mathbf p}_e$ denote  momenta per unit  reduced mass (indicated without the tilde, for convenience).

\noindent 2) $({\mathbf q}_h, {\mathbf v}_h)\leftrightarrow  ({\mathbf q}_e,{\mathbf p}_e)$\\

\begin{eqnarray}
{\mathbf q}_e&=& {\mathbf q}_h+\epsilon^2\left[\left(\frac{2+\nu}{2q_h}-\frac{\nu}{2}v_h^2\right){\mathbf q}_h
-\nu({\mathbf q}_h \cdot {\mathbf v}_h){\mathbf v}_h  \right] \nonumber\\
&&+ \epsilon^4 \left[ \left(\frac{\nu(11\nu-3)}{8}v_h^4-\frac{\nu(19+9\nu)}{8q_h}v_h^2-\frac{3\nu (5\nu-3)}{8q_h}({\mathbf n}_h \cdot {\mathbf v}_h)^2+\frac{\nu(\nu-19)}{4q_h^2}\right) {\mathbf q}_h \right.\nonumber\\
&& \left.+ 
\left(\frac{\nu(7\nu-1)}{2}v_h^2 -\frac{3\nu (3+5\nu)}{4q_h}\right) ({\mathbf q}_h \cdot {\mathbf v}_h){\mathbf v}_h\right]\nonumber\\
{\mathbf p}_e&=& {\mathbf v}_h+\epsilon^2\left[\frac{3\nu+2}{2q_h}({\mathbf n}_h \cdot {\mathbf v}_h){\mathbf n}_h
+\left(\frac{1-2\nu}{2}v_h^2 -\frac{3\nu (3+5\nu)}{4q_h}  \right){\mathbf v}_h  \right] \nonumber\\
&&+ \epsilon^4 \left\{ 
({\mathbf n}_h \cdot {\mathbf v}_h){\mathbf n}_h \left(\frac{21\nu (1+\nu)}{8q_h} ({\mathbf n}_h \cdot {\mathbf v}_h)^2 -\frac{23\nu^2-15\nu-4}{8q_h}v_h^2+\frac{15\nu^2+23\nu +12}{4q_h^2} \right)
\right.\nonumber\\
&& \left.+ 
\left[\left(\frac38 -2\nu+3\nu^2\right)v_h^4-\frac{21\nu^2+7\nu-24}{8q_h}v_h^2-\frac{\nu (17\nu+25)}{8q_h}({\mathbf n}_h \cdot {\mathbf v}_h)^2\right.\right. \nonumber\\
&& \left.\left. +\frac{\nu^2-14\nu+3}{2q_h^2}   \right] {\mathbf v}_h\right\}\,.
\end{eqnarray}

\begin{eqnarray}
{\mathbf q}_h&=& {\mathbf q}_e+\epsilon^2\left[\left( \frac{\nu}{2}p_e^2-\frac{\nu+2}{2q_e} \right) {\mathbf q}_e+\nu ({\mathbf q}_e \cdot {\mathbf p}_e){\mathbf p}_e\right]\nonumber\\
&& +\epsilon^4 \left[ {\mathbf q}_e \left(-\frac{\nu(5\nu+17)}{8q_e}({\mathbf n}_e \cdot {\mathbf p}_e)^2-\frac{\nu(1+\nu)}{8}p_e^4+\frac{\nu(3\nu-1)}{8q_e}p_e^2-\frac{\nu(\nu-19)}{4q_e^2}  \right)\right.\nonumber\\
&& \left.
+\nu ({\mathbf q}_e \cdot {\mathbf p}_e)\left( \frac{\nu-1}{2}p_e^2+\frac{\nu-19}{4q_e} \right){\mathbf p}_e  \right]\nonumber\\
{\mathbf v}_h&=& {\mathbf p}_e+\epsilon^2\left[
-\frac{3\nu +2}{2q_e}({\mathbf n}_e \cdot {\mathbf p}_e){\mathbf n}_e
+\left( -\frac{\nu+4}{2q_e}+\frac{2\nu-1}{2}p_e^2 \right){\mathbf p}_e
\right]\nonumber\\
&& +\epsilon^4 \left[ 
\left(\frac{3\nu (5\nu+1)}{8q_e} ({\mathbf n}_e \cdot {\mathbf p}_e)^2 -\frac{7\nu^2 +23\nu-4}{8q_e}p_e^2-\frac{3\nu^2-9\nu-4}{4q_e^2} \right)({\mathbf n}_e \cdot {\mathbf p}_e){\mathbf n}_e 
\right.\nonumber\\
&& \left.
+ \left(  -\frac{15\nu^2 -29\nu -8}{8q_e}({\mathbf n}_e \cdot {\mathbf p}_e)^2+\frac{3-8\nu}{8}p_e^4+\frac{7\nu^2-41\nu+8}{8q_e}p_e^2-\frac{\nu^2-15\nu-1}{2q_e^2}\right){\mathbf p}_e  \right]\,.
\end{eqnarray}

Let us consider now the transformation law of the fundamental scalars $X_A^e$ and $X_A^h$ as represented by  $X_A^h={}_A C_{1,3}^{he}(X_A^e)$, with
\begin{enumerate}
  \item   $X_1^h={}_1 C_{1,3}^{he}(X_A^e)$, with ${}_1C_B^{he}=\delta_{1B}$, 
\beq
\begin{array}{lll}
{}_1C^{he}_{11}= -1+2\nu\,\quad & {}_1C^{he}_{13}= -\frac{\nu}{2}-2 \,\quad  & {}_1C^{he}_{23}=-\frac{3}{2}\nu -1  \,,
\end{array}
\eeq
and finally
\beq
\begin{array}{lll}
{}_1C^{he}_{111} = \nu^2 -3\nu+1 \quad &
{}_1C^{he}_{113} = \frac{\nu^2}{4}-\frac{55}{12}\nu+\frac43 \quad  &
{}_1C^{he}_{123} =  -\frac{17}{12}\nu^2+\frac{1}{6}\nu +\frac23\cr
{}_1C^{he}_{133} =  -\frac{\nu^2}{4}+\frac{17}{3}\nu+\frac53\quad &
{}_1C^{he}_{223} =  \frac{5}{4}\nu^2+\frac14 \nu  \quad &
{}_1C^{he}_{233} = \frac34 \nu^2 +\frac{29}{6}\nu+\frac73\,.
\end{array}
\eeq

  \item   $X_2^h={}_2 C_{1,3}^{he}(X_A^e)$, with ${}_2C_B^{he}=\delta_{2B}$, 
\beq
\begin{array}{lll}
{}_2C^{he}_{12}= -\frac12+2\nu\,\quad & {}_2C^{he}_{22}= -2\nu \,\quad  & {}_2C^{he}_{23}=-3-2\nu  \,,
\end{array}
\eeq
and finally
\beq
\begin{array}{lll}
{}_2C^{he}_{112} = 2\nu^2 -2\nu+\frac13 \quad &
{}_2C^{he}_{123} = -\frac54 \nu^2 -\frac{25}{4}\nu+1   \quad &
{}_2C^{he}_{223} = \frac76 \nu^2+\frac{17}{2}\nu+\frac23  \cr
{}_2C^{he}_{122} =-3\nu^2+\nu   \quad &
{}_2C^{he}_{222} = 4\nu^2    \quad &
{}_2C^{he}_{233} = \frac{\nu^2}{2}+\frac{21}{2}\nu+4 \,.
\end{array}
\eeq

  \item   $X_3^h={}_3 C_{1,3}^{he}(X_A^e)$, with 
${}_3C_B^{he}=\delta_{3B}$, 
\beq
\begin{array}{lll}
{}_3C^{he}_{13}=-\frac{\nu}{4}\,\quad & {}_3C^{he}_{23}=-\frac{\nu}{2}  \,\quad  & {}_3C^{he}_{33}=1+\frac{\nu}{2}   \,,
\end{array}
\eeq
and finally
\beq
\begin{array}{lll}
{}_3C^{he}_{113} = \frac{\nu^2}{8}+\frac{\nu}{24}  \quad &
{}_3C^{he}_{123} =    \frac{\nu}{12}\quad &
{}_3C^{he}_{133} =  -\frac{7}{24}\nu(1+\nu)\cr
{}_3C^{he}_{223} =   \frac{\nu^2}{2}\quad &
{}_3C^{he}_{233} =  -\frac{5}{24}\nu^2 +\frac{13}{8}\nu \quad &
{}_3C^{he}_{333} = \frac{\nu^2}{2}-\frac{15}{4}\nu+1 \,.
\end{array}
\eeq
\end{enumerate}

Similarly, for the transformation $X_A^e={}_AC_{1,3}^{eh}(X_A^h)$ we have
\begin{enumerate}
  \item   $X_1^e={}_1 C_{1,3}^{eh}(X_A^h)$, with ${}_1C_B^{eh}=\delta_{1B}$, 
\beq
\begin{array}{lll}
{}_1C^{eh}_{11}= 1-2\nu \,\quad & {}_1C^{eh}_{13}=\frac{\nu}{2}+2   \,\quad  & {}_1C^{eh}_{23}=\frac{3}{2}\nu+1   \,,
\end{array}
\eeq
and finally
\beq
\begin{array}{lll}
{}_1C^{eh}_{111} =  7\nu^2-5\nu+1 \quad &
{}_1C^{eh}_{113} =  -\frac{25}{12}\nu^2-\frac{7}{4}\nu+\frac{8}{3}  \quad &
{}_1C^{eh}_{123} =  -\frac{13}{6}\nu^2 -\frac12 \nu +\frac13 \cr
{}_1C^{eh}_{133} =   \frac{5}{12}\nu^2-4\nu+\frac73\quad &
{}_1C^{eh}_{223} =    \frac{7}{4}\nu (1+\nu)\quad &
{}_1C^{eh}_{233} =  \frac{15}{4}\nu^2+\frac{43}{6}\nu+\frac{11}{3}\,.
\end{array}
\eeq

  \item   $X_2^e={}_2 C_{1,3}^{eh}(X_A^h)$, with ${}_2C_B^{eh}=\delta_{2B}$, 
\beq
\begin{array}{lll}
{}_2C^{eh}_{12}=  \frac12 -2\nu\,\quad & {}_2C^{eh}_{22}= 2\nu  \,\quad  & {}_2C^{eh}_{23}= 3+2\nu  \,,
\end{array}
\eeq
and finally
\beq
\begin{array}{lll}
{}_2C^{eh}_{112} = 6\nu^2-\frac83 \nu +\frac13 \quad &
{}_2C^{eh}_{123} = -\frac{53}{12}\nu^2-\frac{29}{12}\nu+\frac53   \quad &
{}_2C^{eh}_{223} =   \frac{25}{6}\nu^2+\frac{23}{6}\nu\cr
{}_2C^{eh}_{122} =   -5\nu^2+\nu\quad &
{}_2C^{eh}_{222} =   4\nu^2\quad &
{}_2C^{eh}_{233} =   \frac{25}{6}\nu^2+\frac{19}{6}\nu+6\,.
\end{array}
\eeq

  \item   $X_3^e={}_3 C_{1,3}^{eh}(X_A^h)$, with 
${}_3C_B^{eh}=\delta_{3B}$ 
\beq
\begin{array}{lll}
{}_3C^{eh}_{13}= \frac{\nu}{4}\,\quad & {}_3C^{eh}_{23}=\frac{\nu}{2}   \,\quad  & {}_3C^{eh}_{33}= -\frac{\nu}{2}-1   \,,
\end{array}
\eeq
and finally
\beq
\begin{array}{lll}
{}_3C^{eh}_{113} =  -\frac{3}{8}\nu^2 +\frac{\nu}{8} \quad &
{}_3C^{eh}_{123} =   -\frac{\nu^2}{2}+\frac{1}{12}\nu  \quad &
{}_3C^{eh}_{133} =  \frac{5}{24}\nu^2+\frac{11}{24}\nu \cr
{}_3C^{eh}_{223} =   \frac{\nu^2}{2} \quad&
{}_3C^{eh}_{233} =   \frac{37}{24}\nu^2-\frac{7}{24}\nu\quad &
{}_3C^{eh}_{333} =  \frac{23}{4}\nu +1\,.
\end{array}
\eeq
\end{enumerate}

\subsection{EOB  vs ADM coordinates}
\noindent 
EOB vs ADM phase space vector 2PN-transformations (see e.g., sec. VI, Eqs. (6.22) of ref. \cite{Buonanno:1998gg}) are the following:\\

\noindent 1) $({\mathbf q}_a, {\mathbf v}_a)\leftrightarrow  ({\mathbf q}_e,{\mathbf p}_e)$\\

\begin{eqnarray}
{\mathbf q}_e&=& {\mathbf q}_a +\epsilon^2\left [\left(-\frac{\nu}{2}p_a^2+\frac{1}{q_a}\left(1+\frac{\nu}{2} \right)\right){\mathbf q}_a -({\mathbf q}_a\cdot {\mathbf p}_a)\nu{\mathbf p}_a\right]\nonumber\\
&& +\epsilon^4 \left[\left( \frac{\nu}{8}(1-\nu)p_a^4 +\frac{\nu}{4}\left(5-\frac{\nu}{2}\right) \frac{p_a^2}{q_a} + 
\nu\left(1+\frac{\nu}{8}\right)\frac{({\mathbf q}_a\cdot {\mathbf p}_a)^2}{q_a^3} +\frac14(1-7\nu+\nu^2) \frac{1}{q_a^2}\right){\mathbf q}_a
\right.\nonumber\\
&& \left.
+({\mathbf q}_a\cdot {\mathbf p}_a)\left[\frac{\nu}{2} (1+\nu) p_a^2 +\frac32 \nu \left(1-\frac{\nu}{2}\right) \frac{1}{q_a} \right]{\mathbf p}_a\right]
\nonumber\\
{\mathbf p}_e&=& {\mathbf p}_a+\epsilon^2 \left[({\mathbf q}_a\cdot {\mathbf p}_a)\frac{1}{q_a^3}\left(1+\frac{\nu}{2} \right){\mathbf q}_a+\frac{\nu}{2}p_a^2-\frac{1}{q_a}\left(1+\frac{\nu}{2} \right) {\mathbf p}_a\right]\nonumber\\
&&+\epsilon^4 \left\{ ({\mathbf q}_a\cdot {\mathbf p}_a)\frac{1}{q_a^3}\left[\frac{\nu}{8}(10-\nu) p_a^2 +\frac38\nu(8+3\nu) \frac{({\mathbf q}_a\cdot {\mathbf p}_a)^2}{q^2}+ \frac14(-2-18\nu+\nu^2)\frac{1}{q_a}  \right]{\mathbf q}_a\right.\nonumber\\
&& \left.+\left(\frac{\nu}{8}(-1+3\nu) p_a^4-\frac{3}{4}\nu\left(3+\frac{\nu}{2}\right)\frac{p_a^2}{q_a} 
-\frac{\nu}{8}(16+5\nu)\frac{({\mathbf q}_a\cdot {\mathbf p}_a)^2}{q_a^3}+ \frac14(3+11\nu)\frac{1}{q_a^2}\right){\mathbf p}_a 
\right\}
\end{eqnarray}
\begin{eqnarray}
{\mathbf q}_a&=& {\mathbf q}_e +\epsilon^2 \left[\left(\frac{\nu}{2}p_e^2-\frac{1}{q_e}\left(1+\frac{\nu}{2} \right)\right){\mathbf q}_e+\nu({\mathbf q}_e\cdot {\mathbf p}_e){\mathbf p}_e\right]\nonumber\\
&&
+\epsilon^4 \left\{\left[ -\frac{\nu}{8}(1+\nu)p_e^4 + \frac{3}{4}\nu\left(\frac{\nu}{2}-1\right)\frac{p_e^2}{q_e} - 
 \nu\left(2+\frac{5}{8}\nu\right)\frac{({\mathbf q}_e\cdot {\mathbf p}_e)^2}{q_e^3} +  \frac{-\nu^2+7\nu -1}{4} \frac{1}{q_e^2}\right]{\mathbf q}_e \right.\nonumber\\
&& \left.+({\mathbf q}_e\cdot {\mathbf p}_e)\left[\frac{\nu(\nu-1}{2}  p_e^2 +\frac{\nu}{2} \left(-5+\frac{\nu}{2}\right)\frac{1}{q_e} \right]{\mathbf p}_e\right\}\,\nonumber\\
{\mathbf p}_a&=& {\mathbf p}_e +\epsilon^2 \left[-\left(1+\frac{\nu}{2} \right)({\mathbf q}_e\cdot {\mathbf p}_e)\frac{1}{q_e^3}{\mathbf q}_e+\left(-\frac{\nu}{2}p_e^2+\frac{1}{q_e}\left(1+\frac{\nu}{2} \right)  \right){\mathbf p}_e\right]\nonumber\\
&&+
\epsilon^4 \left\{ ({\mathbf q}_e\cdot {\mathbf p}_e)\frac{1}{q_e^3}\left[ \frac{3}{4}\nu \left(\frac{\nu}{2}-1\right) p_e^2 + \frac{3}{8}\nu^2\frac{({\mathbf q}_e\cdot {\mathbf p}_e)^2}{q_e^2}+\left(-\frac32 +\frac52\nu-\frac34 \nu^2\right)  \frac{1}{q_e}  \right]{\mathbf q}_e \right. \nonumber\\
&& \left.+\left[\frac{\nu(1+3\nu)}{8}p_e^4 -\frac{\nu}{4}\left(1+\frac72 \nu\right)\frac{p_e^2}{q_e} 
 +\nu\left(1+\frac{\nu}{8}\right)\frac{({\mathbf q}_e\cdot {\mathbf p}_e)^2}{q_e^3}+\left(\frac54 -\frac34\nu +\frac{\nu^2}{2}\right)  \frac{1}{q_e^2}\right]{\mathbf p}_e\right\}
\end{eqnarray}
Let us consider now the transformation law of the fundamental scalars $X_A^e$ and $X_A^a$ as represented by  $X_A^a={}_A T_{1,3}^{ae}(X_A^e)$, with
\begin{enumerate}
  \item   $X_1^a={}_1 T_{1,3}^{ae}(X_A^e)$, with ${}_1T_B^{ae}=\delta_{1B}$, 
\beq
\begin{array}{lll}
{}_1T^{ae}_{11}= -\nu\,\quad & {}_1T^{ae}_{13}= 1+\frac{\nu}{2} \,\quad  & {}_1T^{ae}_{23}=-\frac{\nu}{2}-1   \,,
\end{array}
\eeq
and finally
\beq
\begin{array}{lll}
{}_1T^{ae}_{111} =\nu^2+\frac14 \nu  \quad &
{}_1T^{ae}_{113} =-\frac{3}{4}\nu^2-\frac12 \nu   \quad  &
{}_1T^{ae}_{123} = \frac14 \nu (1+\nu)  \cr
{}_1T^{ae}_{133} = \frac{5}{12}\nu^2-\frac16 \nu +\frac76  \quad &
{}_1T^{ae}_{223} =\frac{\nu^2}{4}   \quad &
{}_1T^{ae}_{233} = -\frac{7}{12}\nu^2+\frac43 \nu -\frac43 \,.
\end{array}
\eeq

  \item   $X_2^a={}_2 T_{1,3}^{ae}(X_A^e)$, with ${}_2T_B^{ae}=\delta_{2B}$, 
\beq
\begin{array}{ll}
{}_2T^{ae}_{12}= \frac{\nu}{2}\,\quad & {}_2T^{ae}_{22}=-2\nu   \,,
\end{array}
\eeq
and finally
\beq
\begin{array}{lll}
{}_2T^{ae}_{112} =  -\frac{\nu}{4}\quad &
{}_2T^{ae}_{122} = -\nu^2+\frac13 \nu    \quad &
{}_2T^{ae}_{123} =\frac14 \nu^2-\frac12 \nu    \cr
{}_2T^{ae}_{223} = -\frac12 \nu^2 +\nu \quad &
{}_2T^{ae}_{222} =4\nu^2      \quad &
{}_2T^{ae}_{233} =  -\frac16 \nu^2+\frac76 \nu-\frac16 \,.
\end{array}
\eeq

  \item   $X_3^a={}_3 T_{1,3}^{ae}(X_A^e)$, with 
${}_3T_B^{ae}=\delta_{3B}$, 
\beq
\begin{array}{lll}
{}_3T^{ae}_{13}= -\frac{\nu}{4}\,\quad & {}_3T^{ae}_{23}= -\frac{\nu}{2} \,\quad  & {}_3T^{ae}_{33}=\frac{\nu}{2}+1    \,,
\end{array}
\eeq
and finally
\beq
\begin{array}{lll}
{}_3T^{ae}_{113} =  \frac{\nu^2}{8}+\frac{\nu}{24} \quad &
{}_3T^{ae}_{123} =  \frac{\nu}{12}  \quad &
{}_3T^{ae}_{133} =  -\frac{7}{24}\nu^2-\frac{\nu}{12} \cr
{}_3T^{ae}_{223} =\frac{\nu^2}{2}   \quad &
{}_3T^{ae}_{233} =  -\frac{5}{24}\nu^2+\frac{5}{6}\nu \quad &
{}_3T^{ae}_{333} =  \frac{\nu^2}{2}-\frac{3}{4}\nu+\frac{5}{4}\,.
\end{array}
\eeq
\end{enumerate}

Similarly, for the transformation law    $X_A^e={}_A H_{1,3}^{ea}(X_A^a)$ we have
\begin{enumerate}
  \item   $X_1^e={}_1 H_{1,3}^{ea}(X_A^a)$, with ${}_1H_B^{ea}=\delta_{1B}$, 
\beq
\begin{array}{lll}
{}_1H^{ea}_{11}= \nu \,\quad & {}_1H^{ea}_{13}=-\frac{\nu}{2}-1   \,\quad  & {}_1H^{ea}_{23}=\frac{\nu}{2}+1    \,,
\end{array}
\eeq
and finally
\beq
\begin{array}{lll}
{}_1H^{ea}_{111} = \nu^2-\frac{\nu}{4}  \quad &
{}_1H^{ea}_{113} = -\frac{5}{12}\nu^2-\frac{11}{6}\nu   \quad  &
{}_1H^{ea}_{123} =-\frac{\nu^2}{6}-\frac{\nu}{12}  \cr
{}_1H^{ea}_{133} = \frac{\nu^2}{12}+\frac{13}{6}\nu+\frac56   \quad &
{}_1H^{ea}_{223} =\frac{3}{4}\nu^2+2\nu   \quad &
{}_1H^{ea}_{233} =  \frac{\nu^2}{12}-\frac{10}{3}\nu-\frac{2}{3}\,.
\end{array}
\eeq

  \item   $X_2^e={}_2 H_{1,3}^{ae}(X_A^a)$, with ${}_2H_B^{ae}=\delta_{2B}$, 
\beq
\begin{array}{ll}
{}_2H^{ea}_{12}= -\frac{\nu}{2} \,\quad & {}_2H^{ea}_{22}= 2\nu   \,,
\end{array}
\eeq
and finally
\beq
\begin{array}{lll}
{}_2H^{ea}_{112} = \frac{\nu}{4} \quad &
{}_2H^{ea}_{122} = -\nu^2-\frac{\nu}{3}    \quad &
{}_2H^{ea}_{123} =  -\frac{1}{12}\nu^2+\frac56 \nu  \cr
{}_2H^{ea}_{223} =  \frac16 \nu^2-\frac53 \nu\quad &
{}_2H^{ea}_{222} =   4\nu^2\quad &
{}_2H^{ea}_{233} =   \frac16 \nu^2-\frac76 \nu+\frac16\,.
\end{array}
\eeq

  \item   $X_3^e={}_3 H_{1,3}^{ae}(X_A^a)$, with 
${}_3T_B^{ae}=\delta_{3B}$, 
\beq
\begin{array}{lll}
{}_3H^{ea}_{13}=  \frac{\nu}{4}\,\quad & {}_3H^{ea}_{23}=\frac{\nu}{2} \,\quad  & {}_3H^{ea}_{33}= -\frac{\nu}{2}-1    \,,
\end{array}
\eeq
and finally
\beq
\begin{array}{lll}
{}_3H^{ea}_{113} =   \frac{\nu^2}{8}-\frac{\nu}{24}\quad &
{}_3H^{ea}_{123} =   -\frac{\nu}{12}  \quad &
{}_3H^{ea}_{133} =    -\frac{\nu^2}{8}-\frac34 \nu\cr
{}_3H^{ea}_{223} =  \frac{\nu^2}{2}\quad &
{}_3H^{ea}_{233} =   -\frac{\nu^2}{8}-\frac32 \nu\quad &
{}_3H^{ea}_{333} =  \frac{11}{4}\nu+\frac34 \,.
\end{array}
\eeq
\end{enumerate}

\end{widetext}

\subsection{Transformation of the angular momentum variables}

While the conserved angular momentum of the system, $\bf J$, has its usual, simple expression in ADM
and EOB variables, namely (in reduced form $\bf j= \bf J/ \mu M$)
\beq
\bf j=  \bf q_a \times \bf p_a =  \bf q_e \times \bf p_e  \, ,
\eeq
its expression in harmonic variables involves an extra PN-correcting factor $f_h=1 + O(1/c^2)$,
namely
\beq
{\bf j}=  f_h \, \bf q_h \times \bf v_h   \, ,
\eeq
where
\begin{eqnarray}
f_h &=& 1 + \epsilon^2\left(-\frac12 (3\nu-1) X_1^h+(\nu+3)X_3^h \right)\nonumber\\
&&+\epsilon^4\left[\frac14 (4\nu^2+14-41\nu) X_{33}^h  \right. \nonumber\\
&& \left. -\frac12(9\nu^2 +10\nu -7) X_{13}^h-\frac12(5\nu^2 +2\nu) X_{23}^h   \right. \nonumber\\
&& \left.
+\frac{3}{8}  (1+13\nu^2-7\nu)X_{11}^h\right]\,.
\end{eqnarray}

\section{Some reminders of Newtonian theory}
\label{newtonian}

The relative motion  of two bodies with masses $m_1$ and $m_2$ can be treated as that  of a single body with effective mass $\mu=m_1m_2/(m_1+m_2)$.
Indeed, after separation of the motion of the center of mass (with $M=m_1+m_2$)
\beq
 {\bf R}=\frac{m_1 {\bf x}_1+m_2{\bf x}_2}{M}\, 
\eeq
one gets the following Lagrangian for the dynamics of the relative motion
\beq
{\mathcal L}_0=\mu\left(\frac12  \dot{\bf r}^2+\frac{GM}{r}\right)\,,
\eeq
where ${\bf r}\equiv  {\bf x}_1- {\bf x}_2$ and $r=|{\bf r}|$, from which follow the momenta
\beq
{\bf p}=\mu \dot {\mathbf r}=\mu {\mathbf v}\,
\eeq
and then the Hamiltonian
\beq
{\mathcal H}_0=\mu\left(\frac{{\bf p}^2}{2\mu^2} -\frac{GM}{r}\right)\,.
\eeq
We systematically use a \lq\lq tilde notation" for quantities per unit reduced mass; for example
\beq
\tilde {\mathcal L}_0={\mathcal L}_0/\mu\,,\quad \tilde {\mathcal H}_0={\mathcal H}_0/\mu  \,.
\eeq
The conservation of the angular momentum 
\beq
{\mathbf J}={\mathbf r} \times {\mathbf p}=\mu {\mathbf r} \times {\mathbf v}\equiv \mu \tilde {\mathbf J}\,,
\eeq
allows one to study the motion in the $x-y$ orbital plane (orthogonal to ${\mathbf J}=J{\mathbf e}_z$).
Using polar coordinates $x^i=(r,\phi)$ leads to the 
 Lagrangian per unit reduced mass 
\beq
\tilde {\mathcal L}_0(r, \dot r, \phi , \dot \phi )= \frac12  (\dot r^2 +r^2 \dot \phi^2)+\frac{GM}{r}\,,
\eeq
so that
\beq
p_r=\frac{\partial {\mathcal L}_0}{\partial \dot r}=\mu \dot r\,,\qquad
p_\phi=\frac{\partial {\mathcal L}_0}{\partial \dot \phi}=\mu r^2 \dot \phi
\eeq
and
\beq
\tilde {\mathcal H}_0(p_r,r,p_\phi,  \phi)= \frac1{2\mu^2}   \left(p_r^2 +\frac{p_\phi^2}{r^2}  \right)- \frac{GM}{r}\,.
\eeq
The dynamics simplifies if we use 
the following rescaled  variables
\beq
\hat r=\frac{r}{GM}\,,\quad \tilde p_r=\frac{p_r}{\mu}\,,\quad j=\frac{{\tilde p}_\phi}{GM}=\frac{J}{GM\mu}\,,\quad \hat t =\frac{t}{GM}\,.
\eeq
The Hamiltonian corresponding to these scaled variables is
\beq
\tilde {\mathcal H}_0(\tilde p_r,\hat r,j,  \phi)= \frac1{2}   \left(\tilde p_r^2 +\frac{j^2}{\hat r^2}  \right)- \frac{1}{\hat r}\,,
\eeq
and the equations of motion read
\beq
\label{rddot2}
\frac{d \hat r}{d\hat t }= \tilde p_r\,,\quad \frac{d \phi}{d\hat t}=\frac{j}{\hat r^2}\,,\quad
\frac{d \tilde p_r}{d\hat t }=\frac{j^2}{{\hat r}^3}- \frac{1}{{\hat r}^2} \,.
\eeq

The integration of the radial equation fully determines the orbit
\beq
\hat r(\phi)=\frac{\sf p}{1+e_0\cos \phi }\,,\qquad {\sf p}=j^2\,,
\eeq
or
\beq
\label{r_di_phi_equation}
\frac{j^2}{\hat r(\phi)}={1+e_0\cos \phi }\,,
\eeq
also implying
\beq
\label{dot_r_equation}
\frac{d \hat r}{d\hat t}  =\frac{e_0}{j}\sin \phi \,.
\eeq
$e_0$ being the eccentricity of the orbit given by 
\beq
e_0(\tilde E, j)=1+2\tilde E j^2\,;
\eeq
where $\tilde E=\tilde {\mathcal H}_0$ is the conserved energy per unit reduced mass.

One has now to distinguish among the various types of orbits: elliptic ($0\le e_0<1$; $e_0=0$ in the circular case), parabolic ($e_0=1$) and hyperbolic ($e_0>1$).

\begin{itemize}
  \item \underline{Elliptic orbits} 
The solution of the equations of motion can be given in terms of  the eccentric anomaly $u$  as follows
\begin{eqnarray}
\label{elliptic1}
\hat n( \hat t-\hat t_0)&=& u-e_0 \sin  u \,, \nonumber\\
\hat r&=& \hat a_0 (1-e_0 \cos  u )\,,\nonumber\\
\phi-\phi_0&=&  
2{\rm arctan}\left[\sqrt{\frac{1+e_0}{1-e_0}}\tan \left(\frac{u}{2}\right)  \right]
\end{eqnarray}
where
\beq
\hat n 
=\sqrt{\frac{1}{\hat a_0^3}}\,,\quad \hat a_0= -\frac{1}{2\tilde E}\,,  
\eeq
$\hat a_0$ being the scaled semimajor axis of the ellipse, $\hat a_0=a_0/(GM)$. Other useful relations are
\beq
j=\frac{\sqrt{1-e_0^2}}{\sqrt{-2\tilde E}}\,,
\eeq
implying
\beq
{\sf p}=j^2=
\hat a_0 (1-e_0^2)\,,
\eeq
and
\begin{eqnarray}
\hat x&=&\hat r\cos (\phi-\phi_0)= \hat a_0 (\cos  u -e_0)\nonumber\\
\hat y&=&\hat r\sin (\phi-\phi_0)=\hat a_0 \sqrt{1-e_0^2}\sin u \,.
\end{eqnarray}

The circular orbit case,i.e. $\dot r=0=\ddot r$, corresponds   to $e_0=0$, 
that is
\beq
j^2={\hat r}\,,\qquad \tilde E=-\frac1{2{\hat r}} \,, 
\eeq 
or
\beq
1+2\tilde E j^2=0\,.
\eeq

\item \underline{Parabolic orbits}
The parabolic case ($\tilde E=0$) is obtained from the elliptic one  taking \lq\lq consistently" the limit $\tilde E\to 0$.
For instance, in Eq. (\ref{elliptic1}) one poses
\beq
u= \sqrt{-2\tilde E}  x 
\eeq
and takes the limit $\tilde E\to 0^-$ keeping $x$ fixed.
The result is
\begin{eqnarray}
\hat t-\hat t_0&=&\frac{x^3}{6} \nonumber\\
\hat r &=&\frac{x^2}{2 }=\frac{j^2 (\phi-\phi_0)^2}{8}\nonumber\\
\phi-\phi_0&=&\frac{2x}{j}\,.
\end{eqnarray}

  \item \underline{Hyperbolic orbits}
 Transition to the hyperbolic case is accomplished by the substitution
\beq
u=i \bar u\,,
\eeq
in the elliptic case relations, so that
\begin{eqnarray}
\bar  n (\hat t-\hat t_0)&=& -\bar u+e_0 \sinh  (\bar u)\, \nonumber\\
\hat r&=&\hat   a_0  (1-e_0 \cosh (\bar u))\nonumber\\
\phi-\phi_0&=& 2{\rm arctan}\left[\sqrt{\frac{e_0+1}{e_0-1}}\tanh \left(\frac{\bar u}{2}\right)  \right]\,,
\end{eqnarray}
with
\beq
\bar n=\sqrt{-\frac{1}{\hat a_0^3}}\,,\qquad \hat a_0=- \frac{1}{2\tilde E}\,.
\eeq
The ``parameter'' ${\sf p}$
entering the polar form of the orbits is still given by
\beq
 {\sf p}=j^2=\hat a_0 (1-e_0^2)\,.
\eeq
The scattering angle is given by \cite{LL} 
\beq
\label{F29}
\tan \frac{\chi}{2}=\frac{1}{\sqrt{e_0^2-1}}=\frac{1}{\sqrt{2\tilde E j^2}}\,,
\eeq
where $e_0\equiv \sqrt{1+2\tilde E j^2}$.
Note also the equivalent relations (whose 2PN analogs we often use in the main text)
\begin{eqnarray}
\frac{\chi}{2}&=&{\rm arccos}\left( -\frac{1}{e_0} \right)-\frac{\pi}{2}\nonumber\\
&=&{\rm arcsin}\left( \frac{1}{e_0} \right)\,.
\end{eqnarray}
The scattering angle can also be expressed in terms of $\hat r_{\rm (min)}$ and $\tilde p_{\rm (max)}$. Indeed, at the point of minimal distance (periastron) $r=r_{\rm (min)}$ one has $\tilde p_r=0$ and  $\tilde p_{\rm (max)}=\tilde p_\phi /r_{\rm min}=j GM/r_{\rm min}=j/\hat r_{\rm (min)}$. Hence, 
\begin{eqnarray}
\label{EJrmin}
j&=&\hat r_{\rm (min)}\tilde p_{\rm (max)}\,,\nonumber\\
\tilde E &=&\frac12 \tilde p_{\rm (max)}^2 -\frac{1}{\hat r_{\rm min}}\,;
\end{eqnarray}
so that
\beq
1+2\tilde E j^2=(\tilde p_{\rm (max)}^2 \hat r_{\rm (min)}-1)^2\,,
\eeq
which can be replaced in Eq. (\ref{F29}) if one wishes to express $\tan \chi/2$ in terms of $\tilde p^2_{\rm (max)} \hat r_{\rm (min)}$.
\end{itemize}

Anticipating applications of our framework to numerical relativity simulations of hyperbolic encounters, let us indicate an estimate of the simulation time 
$t_{\rm (stop)}$ (counted from the periastron passage) necessary for extracting from the corresponding polar angle $\phi_+(t_{\rm (stop)})$ (counted from the periastron) the scattering angle $\chi$ with some prescribed accuracy $\varepsilon=10^{-N}\ll 1$.

Consider the Newtonian relations for hyperbolic motion with 
$t_0=0=\phi_0$, i.e.,
\begin{eqnarray}
\label{time_eq}
&& e_0 \sinh \bar u -\bar u= \bar n \hat t\nonumber\\
&& \tan \left(\frac{\phi}{2}\right)=\sqrt{\frac{e_0+1}{e_0-1}}\tanh \left(\frac{\bar u}{2}\right)\,,
\end{eqnarray}
where $\bar n = |\hat a_0|^{-3/2}.$
The asymptotic value for $\phi$ corresponds to $\bar u \to +\infty$, that is
\beq
\label{F33}
\tan \left(\frac{\phi_+}{2}\right)=\sqrt{\frac{e_0+1}{e_0-1}}\,. 
\eeq
From Eq. (\ref{F33}) also follows
\beq
\tan \phi_+=-\sqrt{e_0^2-1}\,, 
\eeq
so that
\beq
\tan  \left(\frac{\chi}{2}\right)=-\frac{1}{\tan \phi_+}\,.
\eeq
Let us define an \lq\lq incompleted" instantaneous scattering angle $\chi(t)$ by
\beq
\phi(t)=\frac{\pi}{2}+\frac{\chi (t)}{2}.
\eeq
From Eq. (\ref{time_eq}), $\chi(t)$ satisfies (when it is large and positive)
\begin{eqnarray}
\cot \left(\frac{\chi}{2}\right)&=&\cot \left(\frac{\chi_\infty}{2}\right)\tanh \frac{\bar u}{2}=\cot \left(\frac{\chi_\infty}{2}\right) \frac{1-e^{-\bar u}}{1+e^{-\bar u}}\nonumber\\
&\approx &
\cot \left(\frac{\chi_\infty}{2} \right)(1-2e^{-\bar u})
\end{eqnarray}
or
\beq
\frac{\cot \left(\frac{\chi}{2}\right)}{\cot \left(\frac{\chi_\infty}{2}\right)}\approx 1-2e^{-\bar u}\,.
\eeq
From the \lq\lq time" equation (\ref{time_eq})$_1$ evaluated for large $\bar u$, i.e.,
\beq
e_0 \frac{e^{\bar u}}{2}\approx \bar n \hat t
\eeq 
we have
\beq
e^{-\bar u}\approx \frac{e_0}{2\bar n \hat t}
\eeq
so that
\beq
\label{F39}
\frac{\cot \left(\frac{\chi}{2}\right)}{\cot \left(\frac{\chi_\infty}{2}\right)}\approx 1-\frac{e_0}{\bar n \hat t}\,.
\eeq
The condition for the left hand side of Eq. (\ref{F39}) to differ from $1$ only within some precision $\varepsilon=10^{-N}$ is then
\beq
10^{-N}\approx \frac{e_0}{\bar n \hat t_{\rm (stop)}}
\eeq
that is
\beq
\hat t_{\rm (stop)}\approx \frac{\sqrt{1+2\tilde E j^2}}{(2\tilde E)^{3/2}}10^N.
\eeq

\end{document}